\documentclass[11pt]{article}
\usepackage{amsmath,amssymb}
\usepackage[usenames]{color}
\usepackage{pstricks}
\numberwithin{equation}{section}

\newcommand{\nc}{\newcommand}
\def\vvdots{\mathinner{\mkern1mu\raise1pt\vbox{\kern7pt\hbox{.}}\mkern2mu
  \raise4pt\hbox{.}\mkern2mu\raise7pt\hbox{.}\mkern1mu}}
\nc{\fh}{\hat{f}}
\nc{\muh}{\hat{\mu}}
\nc{\nuh}{\hat{\nu}}
\nc{\bib}{\bibitem}
\nc{\al}{\alpha}
\nc{\g}{\gamma}
\nc{\G}{\Gamma}
\nc{\D}{\Delta}
\nc{\eps}{\epsilon}
\nc{\la}{\lambda}
\nc{\La}{\Lambda}
\nc{\var}{\varphi}
\nc{\pa}{\partial}
\nc{\nn}{\nonumber \\ }
\nc{\hf}{\frac{1}{2}}
\nc{\dz}{\frac{dz}{2\pi i}}
\nc{\bin}[2]{\left(\!\!\!\begin{array}{c} {#1}\\ {#2} \end{array}\!\!\!\right)}
\nc{\be}{\begin{equation}}
\nc{\ee}{\end{equation}}
\nc{\bea}{\begin{eqnarray}}
\nc{\eea}{\end{eqnarray}}
\nc{\bra}[1]{\langle {#1}|}
\nc{\ket}[1]{|{#1}\rangle}
\nc{\ketw}[1]{({#1})^{\phantom{a}}_{{\cal W}}}
\nc{\chit}{\raisebox{0.25ex}{$\chi$}}
\nc{\chih}{\raisebox{0.25ex}{$\hat\chi$}}
\nc{\Db}{\mbox{\boldmath $D$}}
\nc{\Hb}{\mbox{\boldmath $H$}}
\nc{\Lc}{{\cal L}}
\nc{\Vc}{{\cal V}}
\nc{\Ib}{\mbox{\boldmath $I$}}
\nc{\qb}{\bar{q}}
\nc{\Ac}{\mathcal{A}}
\nc{\Bc}{\mathcal{B}}
\nc{\Cc}{\mathcal{C}}
\nc{\Dc}{\mathcal{D}}
\nc{\Ec}{\mathcal{E}}
\nc{\Hc}{\mathcal{H}}
\nc{\Ic}{\mathcal{I}}
\nc{\Jc}{\mathcal{J}}
\nc{\Nc}{\mathcal{N}}
\nc{\Oc}{\mathcal{O}}
\nc{\Rc}{\mathcal{R}}
\nc{\Xc}{\mathcal{X}}
\nc{\Yc}{\mathcal{Y}}
\nc{\Zc}{\mathcal{Z}}
\nc{\fus}{\mbox{}\,\hat\otimes\,\mbox{}}
\nc{\Xt}{\tilde{X}}
\nc{\Yt}{\tilde{Y}}
\nc{\ch}{{\rm ch}}
\nc{\R}{{\cal R}}
\nc{\dkk}{\delta_{j,\{k,k'\}}^{(2)}}
\nc{\ddkk}{\delta_{j,\{k,k'\}}^{(4)}}
\nc{\dddkk}{\delta_{j,\{k,k'\}}^{(8)}}
\nc{\dnn}{\delta_{j,\{n,n'\}}^{(2)}}
\nc{\ddnn}{\delta_{j,\{n,n'\}}^{(4)}}
\nc{\dddnn}{\delta_{j,\{n,n'\}}^{(8)}}
\def\vvdots{\mathinner{\mkern1mu\raise1pt\vbox{\kern7pt\hbox{.}}\mkern2mu
  \raise4pt\hbox{.}\mkern2mu\raise7pt\hbox{.}\mkern1mu}}
\nc{\gauss}[2]{\left[\!\!\begin{array}{c} {#1}\\ {#2} \end{array}\!\!\right]}
\nc{\sbin}[2]{\left\{\!\!\!\begin{array}{c} {#1}\\ {#2} 
\end{array}\!\!\!\right\}}
\nc{\sbinlr}[2]{\Big\langle\!\!\begin{array}{c} {#1}\\ {#2} 
\end{array}\!\!\Big\rangle}
\nc{\bino}[2]{\left(\!\!\begin{array}{c} {#1}\\ {#2} \end{array}\!\!\right)}

\definecolor{lightblue}{rgb}{.61,.61,1}
\definecolor{midblue}{rgb}{.7,.7,1}
\definecolor{lightlightblue}{rgb}{.85,.85,1}
\definecolor{lightestblue}{rgb}{.96,.96,1}

\begin{document}

\topmargin -5mm
\oddsidemargin 5mm

\setcounter{page}{1}

\mbox{}\vspace{-16mm}
\thispagestyle{empty}

\begin{center}
{\huge {\bf Polynomial fusion rings}}\\[.3cm]
{\huge {\bf of ${\cal W}$-extended logarithmic minimal models}}

\vspace{7mm}
{\Large J{\o}rgen Rasmussen}
\\[.3cm]
{\em Department of Mathematics and Statistics, University of Melbourne}\\
{\em Parkville, Victoria 3010, Australia}\\[.4cm]
{\tt J.Rasmussen\!\!\ @\!\!\ ms.unimelb.edu.au}

\end{center}

\vspace{8mm}
\centerline{{\bf{Abstract}}}
\vskip.4cm
\noindent
The countably infinite number of Virasoro representations of the logarithmic minimal model
${\cal LM}(p,p')$ can be reorganized into a finite number of ${\cal W}$-representations with respect
to the extended Virasoro algebra symmetry \!${\cal W}$.
Using a lattice implementation of fusion, we recently determined the fusion algebra of 
these representations and found that it closes, albeit without an identity for $p>1$.
Here, we provide a fusion-matrix realization of this fusion algebra and identify a fusion ring 
isomorphic to it.
We also consider various extensions of it and quotients thereof, and introduce and analyze commutative
diagrams with morphisms between the involved fusion algebras and the corresponding
quotient polynomial fusion rings.
One particular extension is reminiscent of the fundamental fusion algebra of ${\cal LM}(p,p')$
and offers a natural way of introducing the missing identity for $p>1$.
Working out explicit fusion matrices is facilitated by a further enlargement based on a pair
of mutual Moore-Penrose inverses intertwining between the ${\cal W}$-fundamental and enlarged
fusion algebras. 
\vskip1cm
%
%
\renewcommand{\thefootnote}{\arabic{footnote}}
\setcounter{footnote}{0}

\section{Introduction}
\label{SecIntro}

The fusion algebras of the infinite series of logarithmic minimal models ${\cal LM}(p,p')$, 
introduced in~\cite{PRZ0607}, are discussed in~\cite{RP0706,RP0707}.
In these works, it is found that closure of the {\em fundamental} fusion algebra 
of ${\cal LM}(p,p')$ requires an infinite set of indecomposable representations of rank 1, 2 or 3.
The ones of rank 1 are so-called Kac representations of which some, but in general only some, 
are irreducible (highest-weight) representations. The fundamental fusion
algebra $\langle (2,1),(1,2)\rangle_{p,p'}$ 
is so named since it is generated from the two fundamental Kac representations $(2,1)$ and $(1,2)$.
A fusion-matrix realization of this fusion algebra is presented in~\cite{RP0709} 
along with a quotient polynomial fusion ring isomorphic with it. 
This extends part of Gepner's work~\cite{Gep91} on {\em rational} conformal field theories 
to a class of {\em irrational} conformal field theories, 
the latter being the series of logarithmic minimal models.
Both the matrix realizations and the fusion rings in~\cite{RP0709} 
are described in terms of Chebyshev polynomials.

A central question of much current interest~\cite{WLogCFT,GK9606} is whether an 
extended symmetry algebra ${\cal W}$~\cite{Walgebra} exists for logarithmic 
conformal field theories~\cite{LogCFT} like the logarithmic minimal models. 
Such a symmetry should allow the countably {\em infinite} number of Virasoro representations 
to be reorganized into a {\em finite} number of ${\cal W}$-extended representations closing 
under fusion. In the case of the logarithmic minimal models ${\cal LM}(1,p')$, the existence of 
such an extended ${\cal W}$-symmetry and the associated fusion rules are by now 
well established~\cite{GK9606,WLM1p,PRR0803}. 

The extension to $p>1$, however, proved itself rather difficult. A significant breakthrough 
came with the works~\cite{FGST0606a,FGST0606b} which strongly indicate the existence of a  
${\cal W}_{p,p'}$ symmetry algebra for {\em general} augmented minimal models,
but offer only very limited insight into the associated fusion algebras.
Recently, a detailed description of these fusion algebras has been 
provided in~\cite{RP0804,Ras0805}. 
In these papers, we generalized the approach of~\cite{PRR0803} using a strip-lattice
implementation of fusion to obtain the fusion rules of the entire series of 
logarithmic minimal models in the ${\cal W}$-extended picture.
Contrary to the situation in the Virasoro picture, there is in general (for $p>1$) no identity nor a pair
of ${\cal W}$-fundamental representations in this ${\cal W}$-extended picture.

In~\cite{PRR0803}, it was shown that symplectic fermions~\cite{Symplectic}
is just critical dense polymers ${\cal LM}(1,2)$~\cite{Polymers} 
viewed in the ${\cal W}$-extended picture. 
Likewise in the general case~\cite{Ras0805}, including critical percolation~\cite{RP0804}, 
the extended picture is described
by the {\em same} lattice model as the Virasoro picture~\cite{PRZ0607,RP0706,RP0707}. 
It is nevertheless useful to distinguish between the two pictures by denoting
the logarithmic minimal models viewed in the
extended picture by ${\cal WLM}(p,p')$ and reserve the notation ${\cal LM}(p,p')$ for 
the logarithmic minimal models in the non-extended Virasoro picture. 

An objective of the present work is to determine fusion-matrix realizations and the corresponding
polynomial fusion rings of the ${\cal W}$-extended fusion algebras mentioned above. 
To achieve this, we first revisit the logarithmic minimal models
in the Virasoro picture to identify a particular algebraic ideal of the fundamental fusion algebra.
Considering its strong resemblance with the
${\cal W}$-extended fusion algebra of~\cite{Ras0805}, 
we are led to propose an extension of this ${\cal W}$-extended fusion algebra
reminiscent of the fundamental fusion algebra in the Virasoro picture.
We thus call this larger algebra the ${\cal W}$-{\em fundamental} fusion algebra and posit
the complete set of supplementary fusion rules which are subsequently shown to
yield an associative and commutative fusion algebra.
It also offers a canonical way of introducing the elusive identity and pair of ${\cal W}$-fundamental
representations. For $p>1$, this identity is a ${\cal W}$-reducible yet ${\cal W}$-indecomposable
representation of rank~1. The dimension of the ${\cal W}$-fundamental fusion algebra is
$7pp'-3p-3p'+1$, while the dimension of the empirically obtained ${\cal W}$-extended fusion 
algebra of~\cite{Ras0805} is $6pp'-2p-2p'$. 

To assist in the construction of {\em explicit} fusion matrices, we introduce a further enlargement
of the ${\cal W}$-extended fusion algebra. The dimension of this enlarged fusion algebra 
is $9pp'-3p-3p'+1$, and a pair of rectangular matrices is introduced to intertwine between the 
fundamental and enlarged bases. These matrices can be chosen as a pair of
mutual Moore-Penrose inverses (see~\cite{Penrose}, for example), one of which is a sparse
binary matrix.

As a means to study and describe how the many fusion algebras and fusion rings are
interrelated, we introduce and discuss commutative diagrams with morphisms between 
the various algebras and rings. This provides, in particular, a convenient framework for representing
the ${\cal W}$-fundamental fusion algebra as a {\em quotient} of the fundamental fusion
algebra in the Virasoro picture. Two additional quotient constructions are examined. 
According to~\cite{PR0812}, the first of these
plays an important role as the algebra of the fusion matrices obtained from a Verlinde formula
applied to the modular $S$ matrix of the set of so-called projective characters 
in the ${\cal W}$-extended picture~\cite{FGST0606b,Ras0805}.
This algebra is also related to the fusion algebra arising from lattice considerations 
when omitting the so-called disentangling procedure~\cite{Ras0805}, as we will discuss.
The other example allows us to view the usual fusion algebra of the
{\em rational} minimal model ${\cal M}(p,p')$~\cite{BPZ84,DiFMS} as a quotient of the fundamental
fusion algebra of ${\cal LM}(p,p')$ in the Virasoro picture or as a quotient of the 
${\cal W}$-fundamental fusion algebra of ${\cal WLM}(p,p')$.

The layout of this paper is as follows. 
In Section~\ref{SecLM}, we review the logarithmic 
minimal model ${\cal LM}(p,p')$ and its fusion algebra~\cite{PRZ0607,RP0707}, and consider 
a particular ideal of the fundamental fusion algebra. 
Their fusion-matrix realizations and the corresponding polynomial fusion rings are also discussed.
These are expressed in terms of Chebyshev polynomials reviewed in Appendix~\ref{AppCheb},
while Appendix~\ref{AppQuo} settles our conventions for quotient polynomial rings.
Section~\ref{SecWLM} summarizes some main results~\cite{Ras0805} on 
${\cal WLM}(p,p')$ as obtained from lattice considerations,
with an exhaustive list of the explicit fusion rules deferred to Appendix~\ref{AppFus}.
The ${\cal W}$-fundamental fusion algebras are introduced in Section~\ref{SecFundWLM}.
The complete set of supplementary fusion rules is posited and we present a conjecture for
the embedding diagrams and ensuing characters of the proposed representations.
There are $(p-1)(p'-1)$ such representations and they are all
${\cal W}$-reducible yet ${\cal W}$-indecomposable representations of rank 1.
One of these is the identity of the ${\cal W}$-fundamental fusion algebra.
Section~\ref{SecFundWLM} is concluded with a description of the fusion-matrix
realizations and corresponding polynomial fusion rings of the 
empirically obtained ${\cal W}$-extended fusion algebras given in Appendix~\ref{AppFus}
as well as the (for $p>1$) larger ${\cal W}$-fundamental fusion algebras.
In Section~\ref{SecEnlarged}, we introduce the enlarged system to facilitate the
construction of explicit fusion matrices.
In Section~\ref{SecQuo}, we introduce the commutative diagrams with morphisms
between fusion algebras and fusion rings. We describe the ${\cal W}$-fundamental
fusion algebras as quotients of the fundamental fusion algebras in the Virasoro picture.
We also discuss the aforementioned quotient construction with links to modular transformations
and the disentangling procedure. Section~\ref{SecQuo} is concluded with the
discussion of the fusion algebras of the rational minimal models as quotients of
the fusion algebras of the logarithmic minimal models.
%
Section~\ref{SecDisc} contains some concluding remarks. It also presents a fusion-ring
description of the fusion algebra of~\cite{RS0701} and offers a proposal for
${\cal W}$-extensions of this fusion algebra and generalizations thereof.



\subsection*{Notation}
\vskip.1cm 
For $n,m\in\mathbb{Z}$,
\be
 \mathbb{Z}_{n,m}\ =\ \mathbb{Z}\cap[n,m]
\ee 
denotes the set of integers from $n$ to $m$, both included, 
while the set of non-negative integers is written $\mathbb{N}_0$.
Certain properties of integers modulo 2 are 
\be
 \eps(n)\ =\ \frac{1-(-1)^n}{2},\qquad\quad
   n\cdot m\ =\ \frac{3-(-1)^{n+m}}{2},\qquad\quad n,m\in\mathbb{Z}
\label{eps}
\ee
where the dot product is seen to be associative.
By a direct sum of representations $A_n$
with unspecified lower summation bound, we mean a direct sum in steps of 2
whose lower bound is given by the parity $\eps(N)$ of the upper bound, that is,
\be
 \bigoplus_{n}^{N}A_n\ =\ \bigoplus_{n=\eps(N),\ \!{\rm by}\ \!2}^{N}\!\! A_n,\qquad\quad N\in\mathbb{Z}
\label{sum2}
\ee
This direct sum vanishes for negative $N$. For ordinary sums of entities $f_n$, we likewise introduce
\be
 \sum_{n}^{N}f_n\ =\ \sum_{n=\eps(N),\ \!{\rm by}\ \!2}^{N}\!\! f_n,\qquad\quad N\in\mathbb{Z}
\label{sum2a}
\ee
Unless otherwise specified, we let
\be
 \kappa,\kappa'\in\mathbb{Z}_{1,2},\quad r\in\mathbb{Z}_{1,p},\quad
  s\in\mathbb{Z}_{1,p'},\quad a,a'\in\mathbb{Z}_{1,p-1},\quad b,b'\in\mathbb{Z}_{1,p'-1},
  \quad\al\in\mathbb{Z}_{0,p-1},\quad\beta\in\mathbb{Z}_{0,p'-1}
\ee
and $k,k',n\in\mathbb{N}$. $T_n(x)$ and $U_n(x)$ are Chebyshev polynomials
of the first and second kind, respectively, see Appendix~\ref{AppCheb}.

\section{Logarithmic minimal model ${\cal LM}(p,p')$}
\label{SecLM}

A logarithmic minimal model ${\cal LM}(p,p')$ is defined~\cite{PRZ0607} for every coprime pair of
positive integers $p<p'$.
The model ${\cal LM}(p,p')$ has central charge
\be
 c\ =\  1-6\frac{(p'-p)^2}{pp'}
\label{c}
\ee
and conformal weights
\be
 \D_{r,s}\ =\ \frac{(rp'-sp)^{2}-(p'-p)^2}{4pp'},\hspace{1.2cm} r,s\in\mathbb{N}
\label{D}
\ee
The fundamental fusion algebra~\cite{RP0706,RP0707}
\be
  \big\langle(2,1), (1,2)\big\rangle_{p,p'}\ =\ \big\langle(a,b), (pk,b), (a,p'k), (pk,p'),
   \R_{pk,s}^{a,0}, \R_{r,p'k}^{0,b},  
  \R_{pk,p'}^{a,b}\big\rangle_{p,p'}
\label{A2112}
\ee
of the logarithmic
minimal model ${\cal LM}(p,p')$ is generated by the two fundamental Kac representations
$(2,1)$ and $(1,2)$ and contains a countably infinite number of inequivalent, indecomposable 
representations of rank 1, 2 or 3. The set of these representations is given by
\be
 \Jc^{\mathrm{Fund}}_{p,p'}\ =\ \big\{\R_{a,b}^{0,0}, \R_{pk,b}^{0,0}, \R_{a,p'k}^{0,0}, \R_{pk,p'}^{0,0},
   \R_{pk,s}^{a,0}, \R_{r,p'k}^{0,b}, \R_{pk,p'}^{a,b}\big\}
\label{Jfund}
\ee
where we have introduced the convenient notation
\be
 \R_{r,s}^{0,0}\ \equiv\ (r,s),\qquad\quad r,s\in\mathbb{N}
\label{R00}
\ee
Their rudimentary properties are summarized in the following.
To gain transparency when discussing distinctions and relations to other fusion algebras,
we sometimes use the alternative notation
\be
 \mathrm{Fund}[{\cal LM}(p,p')]\ =\ \big\langle(2,1), (1,2)\big\rangle_{p,p'}
\label{F2112}
\ee
for the fundamental fusion algebra of ${\cal LM}(p,p')$, 
see also the discussion in Section~\ref{SecWfunRep}.

For $r,s\in\mathbb{N}$, the character of the Kac representation $(r,s)$ is
\be
 \chit_{r,s}(q)\ =\ \frac{q^{\frac{1-c}{24}+\D_{r,s}}}{\eta(q)}\big(1-q^{rs}\big)
  \ =\ \frac{1}{\eta(q)}\big(q^{(rp'-sp)^2/4pp'}-q^{(rp'+sp)^2/4pp'}\big)
\label{chikac}
\ee
where the Dedekind eta function is given by
\be
  \eta(q)\ =\ q^{\frac{1}{24}} \prod_{n=1}^\infty (1-q^n)
\label{eta}
\ee
Such a representation is of rank 1 and is irreducible if $r\in\mathbb{Z}_{1,p}$ and $s\in p'\mathbb{N}$
or if $r\in p\mathbb{N}$ and $s\in\mathbb{Z}_{1,p'}$. It is a reducible yet indecomposable
representation if $r\in\mathbb{Z}_{1,p-1}$ and $s\in\mathbb{Z}_{1,p'-1}$,
while it is a fully reducible representation if $r\in p\mathbb{N}$ and $s\in p'\mathbb{N}$ where
\be
 (kp,k'p')\ =\ (k'p,kp')\ =\ \bigoplus_{j=|k-k'|+1,\ \!{\rm by}\ \!2}^{k+k'-1}(jp,p')\ =\ 
   \bigoplus_{j=|k-k'|+1,\ \!{\rm by}\ \!2}^{k+k'-1}(p,jp')
\label{kpkp}
\ee
These are the only Kac representations appearing in the fundamental fusion algebra (\ref{A2112}).
The characters of the reducible yet indecomposable Kac representations just mentioned
can be written as sums of two irreducible Virasoro characters
\be
 \chit_{a,b}(q)\ =\ \ch_{a,b}(q)+\ch_{2p-a,b}(q)\ =\ \ch_{a,b}(q)+\ch_{a,2p'-b}(q)
\ee
In general, the irreducible Virasoro characters read~\cite{FSZ87}
\bea
 {\rm ch}_{a+(k-1)p,b}(q)&=&K_{2pp',(a+(k-1)p)p'-bp;k}(q)-K_{2pp',(a+(k-1)p)p'+bp;k}(q)\nn
 {\rm ch}_{a+kp,p'}(q)&=&
  \frac{1}{\eta(q)}\big(q^{((k-1)p+a)^2p'/4p}-q^{((k+1)p-a)^2p'/4p}\big) \nn
 {\rm ch}_{kp,s}(q)&=&
  \frac{1}{\eta(q)}\big(q^{(kp'-s)^2p/4p'}-q^{(kp'+s)^2p/4p'}\big)
\label{laq}
\eea
where $K_{n,\nu;k}(q)$ is defined as
\be
 K_{n,\nu;k}(q)\ =\ \frac{1}{\eta(q)}\sum_{j\in\mathbb{Z}\setminus\mathbb{Z}_{1,k-1}}q^{(\nu-jn)^2/2n}
\label{Kk}
\ee

The representations denoted by $\R_{kp,s}^{a,0}$ and $\R_{r,kp'}^{0,b}$ are indecomposable
representations of rank 2, while $\R_{kp,p'}^{a,b}\equiv\R_{p,kp'}^{a,b}$ 
is an indecomposable representation of rank 3. Their characters read
\bea
 \chit[\R_{kp,s}^{a,0}](q)&=&
    \big(1-\delta_{k,1}\delta_{s,p'}\big)\ch_{kp-a,s}(q)+2\ch_{kp+a,s}(q)+\ch_{(k+2)p-a,s}(q)\nn
 \chit[\R_{r,kp'}^{0,b}](q)&=&
    \big(1-\delta_{k,1}\delta_{r,p}\big)\ch_{r,kp'-b}(q)+2\ch_{r,kp'+b}(q)+\ch_{r,(k+2)p'-b}(q)\nn
 \chit[\R_{kp,p'}^{a,b}](q)&=&\big(1-\delta_{k,1}\big)\ch_{(k-1)p-a,b}(q)+2\ch_{(k-1)p+a,b}(q)
   +2\big(1-\delta_{k,1}\big)\ch_{kp-a,p'-b}(q)\nn
   &&+\ 4\ch_{kp+a,p'-b}(q)+\big(2-\delta_{k,1}\big)\ch_{(k+1)p-a,b}(q)
  +2\ch_{(k+1)p+a,b}(q)\nn
   &&+\ 2\ch_{(k+2)p-a,p'-b}(q)+\ch_{(k+3)p-a,b}(q)
\label{chiR}
\eea
Indecomposable representations of rank 3 appear for $p>1$ only.
A decomposition similar to (\ref{kpkp}) also applies to the higher-rank {\em decomposable}
representations $\R_{kp,k'p'}^{\al,\beta}$ as we have 
\be
 \R_{kp,k'p'}^{\al,\beta}\ =\ \R_{k'p,kp'}^{\al,\beta}
   \ =\ \bigoplus_{j=|k-k'|+1,\ \!{\rm by}\ \!2}^{k+k'-1}\R_{jp,p'}^{\al,\beta}
   \ =\ \bigoplus_{j=|k-k'|+1,\ \!{\rm by}\ \!2}^{k+k'-1}\R_{p,jp'}^{\al,\beta}
\ee

The fusion rules governing the fundamental fusion algebra
(\ref{A2112}) are discussed in~\cite{RP0706,RP0707}.
Here, we merely note that fusion in the fundamental fusion algebra 
decomposes into `horizontal' and `vertical' components 
\be
 \R_{p,kp'}^{\al,\beta}\ =\ \R_{p,1}^{\al,0}\otimes\R_{1,kp'}^{0,\beta}
   \ =\ \R_{kp,1}^{\al,0}\otimes\R_{1,p'}^{0,\beta}
\label{decomp}
\ee
The Kac representation $(1,1)$ is the identity of the fundamental fusion algebra.
For $p>1$, this is a reducible yet indecomposable representation, while for $p=1$, it
is an irreducible representation.

\subsection{Outer fusion algebra}

Following the detailed description of the underlying fusion rules in~\cite{RP0707},
many fusion subalgebras of the fundamental fusion algebra (\ref{A2112}) are readily identified.
One of these is generated by all $\R_{r,s}^{\al,\beta}\in\Jc^{\mathrm{Fund}}_{p,p'}$ but
the $(p-1)(p'-1)$
reducible yet indecomposable Kac representations $(a,b)$. Since this corresponds to leaving
out or omitting the bottom-left corner of the infinitely-extended Kac table, we will refer
to this fusion subalgebra as the {\em outer fusion algebra}. We denote it by
\be
 \mathrm{Out}[{\cal LM}(p,p')]\ =\ \big\langle(pk,b), (a,p'k), (pk,p'),
   \R_{pk,s}^{a,0}, \R_{r,p'k}^{0,b}, \R_{pk,p'}^{a,b}\big\rangle_{p,p'}
\label{LMout}
\ee
and introduce
\be
 \Jc^{\mathrm{Out}}_{p,p'}\ =\ \Jc^{\mathrm{Fund}}_{p,p'}\setminus\big\{(a,b)\big\}
   \ =\ \big\{\R_{pk,b}^{0,0}, \R_{a,p'k}^{0,0}, \R_{pk,p'}^{0,0},
   \R_{pk,s}^{a,0}, \R_{r,p'k}^{0,b}, \R_{pk,p'}^{a,b}\big\}
\label{Jout}
\ee
as the set of its generators. 
The decomposition $\Jc^{\mathrm{Fund}}_{p,p'}=\Jc^{\mathrm{Out}}_{p,p'}\cup\big\{(a,b)\big\}$ is
thus a disjoint union. This outer fusion algebra is not discussed elsewhere in the literature.
The main reason for doing it here is that its relationship with the fundamental fusion algebra 
provides a scenario whose ${\cal W}$-extended analogue plays an important role, as we will see.

It is noted that the identity $(1,1)$ of the fundamental fusion algebra is among the representations
{\em not} partaking in the outer fusion algebra for $p>1$. For $p>1$, the outer fusion algebra
has no identity element.
For $p=1$, on the other hand, the subtracted set in (\ref{Jout}) is empty implying that
\be
 \mathrm{Out}[{\cal LM}(1,p')]\ =\ \mathrm{Fund}[{\cal LM}(1,p')]
\label{OF1p}
\ee
It is also observed that the outer fusion algebra is an (non-negative integer)
{\em ideal} of the fundamental fusion algebra
in the sense that, for all $A\in\Jc^{\mathrm{Fund}}_{p,p'}$ and $B\in\Jc^{\mathrm{Out}}_{p,p'}$,
\be
 A\otimes B\ \in\ \mathrm{Span}_{\mathbb{N}_0}\big(\Jc^{\mathrm{Out}}_{p,p'}\big)
\label{ABout}
\ee
as a fusion multiplication in $\mathrm{Fund}[{\cal LM}(p,p')]$.

\subsection{Fusion matrices and fusion rings}
\label{FusMat}

The fusion algebra, see~\cite{DiFMS} for example,
\be
 \phi_i\otimes\phi_j\ =\ \bigoplus_{k\in\mathcal{J}}{{\cal N}_{i,j}}^k\phi_k,\hspace{1cm}i,j\in\mathcal{J}
\ee
of a {\em rational} conformal field theory is finite and
can be represented by a commutative matrix algebra $\langle N_i;\ i\in\mathcal{J}\rangle$ 
where the entries of the $|\mathcal{J}|\times|\mathcal{J}|$ matrix $N_i$ are
\be
 {(N_i)_j}^k\ =\ {{\cal N}_{i,j}}^k,\hspace{1cm}i,j,k\in\mathcal{J}
\ee
and where the fusion multiplication $\otimes$ has been replaced by ordinary matrix multiplication.
In~\cite{Gep91}, Gepner found that every such algebra is isomorphic to a ring
of polynomials in a finite set of variables modulo an ideal defined as the vanishing conditions
of a finite set of polynomials in these variables. He also conjectured that this ideal of constraints 
corresponds to the local extrema of a potential, see~\cite{FusionPotential} for
further elaborations on this conjecture.

Since the fundamental fusion algebra of the logarithmic minimal model
${\cal LM}(p,p')$ has infinitely many
elements, the associated fusion matrices are infinite-dimensional. The corresponding
conformal field theory is {\em irrational} (in this case {\em logarithmic}~\cite{PRZ0607}) and the results
of Gepner~\cite{Gep91} do not necessarily apply. 
Nevertheless, one of the main results of~\cite{RP0709}, Proposition~\ref{SecLM}.2 below, shows that a
(quotient) polynomial fusion ring can be identified after all.
It is based on a matrix realization of the fundamental fusion algebra where the 
fusion matrices corresponding to the various representations appearing in (\ref{A2112}) 
are denoted by $N_{\R_{r,s}^{\al,\beta}}$.
The commuting fusion matrices associated to the fundamental
representations $(2,1)$ and $(1,2)$ are denoted also by $X=N_{(2,1)}$ and $Y=N_{(1,2)}$,
where it is recalled that $\R_{r,s}^{0,0}\equiv (r,s)$.
Since we are considering a countably infinite number 
of representations, $X$ and $Y$ are infinite-dimensional.
According to the following proposition from~\cite{RP0709}, every fusion matrix $N_{\R_{r,s}^{\al,\beta}}$
can be written in terms of Chebyshev polynomials (see Appendix~\ref{AppCheb}) 
in $X$ and $Y$.
First, we follow~\cite{RP0709} and introduce the polynomial
\bea
  P_{n,n'}(x,y)&=&\Big(T_n\big(\frac{x}{2}\big)-T_{n'}\big(\frac{y}{2}\big)\Big)
   U_{n-1}\big(\frac{x}{2}\big)U_{n'-1}\big(\frac{y}{2}\big)\nn
  &=&\frac{1}{2}\Big(U_{2n-1}\big(\frac{x}{2}\big)U_{n'-1}\big(\frac{y}{2}\big)
    -U_{n-1}\big(\frac{x}{2}\big)U_{2n'-1}\big(\frac{y}{2}\big)\Big)
\label{Pnn}
\eea
where the second equality is a simple consequence of (\ref{2TU}). 
\\[.2cm]
{\bf Proposition \ref{SecLM}.1} (\cite{RP0709})\ \ \ Modulo the polynomial $P_{p,p'}(X,Y)$
defined in (\ref{Pnn}), the matrices
\be
 N_{\R_{i,j}^{\al,\beta}}(X,Y)\ =\ 
  \big(2-\delta_{\al,0}\big)T_{\al}\big(\frac{X}{2}\big)U_{i-1}\big(\frac{X}{2}\big)
  \big(2-\delta_{\beta,0}\big)T_{\beta}\big(\frac{Y}{2}\big)U_{j-1}\big(\frac{Y}{2}\big),\qquad
 \R_{i,j}^{\al,\beta}\in\Jc^{\mathrm{Fund}}_{p,p'}
\label{NR}
\ee
constitute a fusion-matrix realization of the fundamental fusion algebra 
$\mathrm{Fund}[{\cal LM}(p,p')]$ with the fusion multiplication $\otimes$ and direct summation
$\oplus$ replaced by matrix multiplication and addition, respectively.
\\[.2cm]
{\bf Proposition \ref{SecLM}.2} (\cite{RP0709})\ \ \ 
The fundamental fusion algebra is isomorphic to the polynomial
ring generated by $X$ and $Y$ modulo the ideal $(P_{p,p'}(X,Y))$, that is,
\be
 \mathrm{Fund}[{\cal LM}(p,p')]\ \simeq\ \mathbb{C}[X,Y]/\big(P_{p,p'}(X,Y)\big)
\ee
The isomorphism reads
\be
 \R_{i,j}^{\al,\beta}\ \leftrightarrow\ 
  \big(2-\delta_{\al,0}\big)T_{\al}\big(\frac{X}{2}\big)U_{i-1}\big(\frac{X}{2}\big)
  \big(2-\delta_{\beta,0}\big)T_{\beta}\big(\frac{Y}{2}\big)U_{j-1}\big(\frac{Y}{2}\big),\qquad
 \R_{i,j}^{\al,\beta}\in\Jc^{\mathrm{Fund}}_{p,p'}
\label{RCP}
\ee

It is noted that $X$ and $Y$ in Proposition \ref{SecLM}.2 are formal entities and hence need 
not be identified with the fusion matrices $X$ and $Y$ of Proposition \ref{SecLM}.1.
A similar comment applies in the following as well when discussing other fusion-matrix realizations
and their associated polynomial fusion rings.

Due to the fusion property (\ref{ABout}), propositions analogous to the ones above apply to the
outer fusion algebra as well. 
We stress, though, that the fundamental representations $(2,1)$ and $(1,2)$
do {\em not} belong to $\Jc^{\mathrm{Out}}_{p,p'}$ for $p,p'>2$, respectively. 
In these cases, a natural matrix realization of the outer fusion algebra 
is described in a basis different from any of the ones used in Proposition \ref{SecLM}.1
since $\Jc^{\mathrm{Out}}_{p,p'}\subsetneq\Jc^{\mathrm{Fund}}_{p,p'}$.
\\[.2cm]
{\bf Proposition \ref{SecLM}.3} \ \ \ 
Let $\{\tilde{N}_{\R};\ \R\in\Jc_{p,p'}^{\mathrm{Fund}}\}$ be a 
fusion-matrix realization of the fundamental fusion algebra 
$\mathrm{Fund}[{\cal LM}(p,p')]$ in some basis, that is, some ordering of the elements
of $\Jc_{p,p'}^{\mathrm{Fund}}$. 
For every $\R\in\Jc_{p,p'}^{\mathrm{Out}}$, the matrix $N_{\R}$ 
is constructed from the fusion matrix
$\tilde{N}_{\R}\in\{\tilde{N}_{\R};\ \R\in\Jc_{p,p'}^{\mathrm{Out}}\}
\subseteq\{\tilde{N}_{\R};\ \R\in\Jc_{p,p'}^{\mathrm{Fund}}\}$
by deleting the rows and columns corresponding to
the $(p-1)(p'-1)$ Kac representations $(a,b)$.
The set $\{N_{\R};\ \R\in\Jc_{p,p'}^{\mathrm{Out}}\}$
constitutes a fusion-matrix realization of the outer fusion algebra $\mathrm{Out}[{\cal LM}(p,p')]$.
\\[.2cm]
{\bf Proof}\ \ \ This follows from the fact that the elements of the subset 
$\Jc_{p,p'}^{\mathrm{Out}}\subseteq\Jc_{p,p'}^{\mathrm{Fund}}$
generate the fusion algebra $\mathrm{Out}[{\cal LM}(p,p')]$ which is (an ideal and hence)
a subalgebra of the fusion algebra $\mathrm{Fund}[{\cal LM}(p,p')]$.
\\
$\Box$
\\[.2cm]
{\bf Proposition \ref{SecLM}.4} \ \ \ For the set of labels $(\al,\beta;i,j)$ characterizing the elements
$\R_{i,j}^{\al,\beta}\in\Jc_{p,p'}^{\mathrm{Out}}$, the elements of the set
$\big\{ \big(2-\delta_{\al,0}\big)T_{\al}\big(\frac{X}{2}\big)U_{i-1}\big(\frac{X}{2}\big)
  \big(2-\delta_{\beta,0}\big)T_{\beta}\big(\frac{Y}{2}\big)U_{j-1}\big(\frac{Y}{2}\big)\big\}$
generate an ideal of the quotient polynomial ring
$\mathbb{C}[X,Y]/(P_{p,p'}(X,Y))$. The outer fusion algebra
$\mathrm{Out}[{\cal LM}(p,p')]$ is isomorphic to this ideal.
\\[.2cm]
{\bf Proof}\ \ \ The elements of the proposed ideal are here indicated by their lifts to the ambient
quotient polynomial ring $\mathbb{C}[X,Y]/(P_{p,p'}(X,Y))$. Continuing this practice
and recalling the fusion property (\ref{ABout}), the proposition follows 
from Proposition \ref{SecLM}.2. In short, the elements of $\Jc_{p,p'}^{\mathrm{Out}}$ generate
an ideal with respect to the fusion multiplication of $\mathrm{Fund}[{\cal LM}(p,p')]$. Since this 
fusion algebra is isomorphic to a quotient polynomial ring, the isomorphic image of
$\Jc_{p,p'}^{\mathrm{Out}}$ in the ring must form an ideal of the ring itself.
The fusion algebra generated by $\Jc_{p,p'}^{\mathrm{Out}}$ is thus isomorphic to
this ideal.
\\
$\Box$
\\[.2cm]
We emphasize that, for $p,p'>2$ respectively, 
the two generators $X$ and $Y$ will {\em not} themselves be part of
the quotient polynomial fusion ring of the outer fusion algebra $\mathrm{Out}[{\cal LM}(p,p')]$. 
The generators are merely `borrowed' from the similar description of the 
bigger fundamental fusion algebra $\mathrm{Fund}[{\cal LM}(p,p')]$
and thus only act as building blocks.

\section{${\cal W}$-extended logarithmic minimal models}
\label{SecWLM}

In this section, we summarize some of our findings~\cite{Ras0805} on
${\cal WLM}(p,p')$ as obtained from lattice considerations. We will extend these
results in Section~\ref{SecFundWLM} where we introduce a fundamental extension of the
fusion algebra obtained empirically in~\cite{Ras0805}.

\subsection{Representation content}
\label{SecRepCont}

There are $2p+2p'-2$ ${\cal W}$-indecomposable rank-1 representations
\be
 \big\{\ketw{\kappa p,s},\ketw{r,\kappa p'}\big\}
 \hspace{1.2cm}\mathrm{subject\ to}\ \ \ \ketw{p,\kappa p'}\equiv\ketw{\kappa p,p'}
\label{r1}
\ee
where $\ketw{p,p'}$ is listed twice,
$4pp'-2p-2p'$ ${\cal W}$-indecomposable rank-2 representations
\be
 \big\{\ketw{\R_{\kappa p,s}^{a,0}}, \ketw{\R_{r,\kappa p'}^{0,b}}\big\}
\label{r2}
\ee
and $2(p-1)(p'-1)$ ${\cal W}$-indecomposable rank-3 representations
\be
 \big\{\ketw{\R_{\kappa p,\kappa' p'}^{a,b}}\big\}
 \hspace{1.2cm}\mathrm{subject\ to}\ \ \ 
  \ketw{\R_{p,2p'}^{a,b}}\equiv\ketw{\R_{2p,p'}^{a,b}}\quad
   \mathrm{and}\quad \ketw{\R_{2p,2p'}^{a,b}}\equiv\ketw{\R_{p,p'}^{a,b}}
\label{r3}
\ee
The total number of ${\cal W}$-indecomposable representations obtained from the lattice
is thus $6pp'-2p-2p'$.
Compactly, the various ${\cal W}$-indecomposable representations satisfy 
\be
 \ketw{\R_{(\kappa\cdot\kappa')p,p'}^{\al,\beta}}
  \ \equiv\ \ketw{\R_{\kappa p,\kappa' p'}^{\al,\beta}}
  \ \equiv\ \ketw{\R_{p,(\kappa\cdot\kappa')p'}^{\al,\beta}}
\label{identi}
\ee
and it is convenient to mimic (\ref{R00}) by 
\be
 \ketw{\R_{r,s}^{0,0}}\ \equiv\ \ketw{r,s},  \qquad\quad  r,s\in\mathbb{N}
\ee

In terms of Virasoro-indecomposable representations, the ${\cal W}$-indecomposable rank-1 representations decompose as
\be
 \ketw{\kappa p,s}\ =\ \bigoplus_{k\in\mathbb{N}}(2k-2+\kappa)((2k-2+\kappa)p,s),\qquad
 \ketw{r,\kappa p'}\ =\ \bigoplus_{k\in\mathbb{N}}(2k-2+\kappa)(r,(2k-2+\kappa)p')
\label{r1Vir}
\ee
where the two expressions for $\ketw{p,p'}$ agree and where the identity
$\ketw{p,2p'}\equiv\ketw{2p,p'}$ is verified explicitly.
Similarly, the ${\cal W}$-indecomposable rank-2 representations decompose as
\be
 \ketw{\R_{\kappa p,s}^{a,0}}\ =\ 
   \bigoplus_{k\in\mathbb{N}}(2k-2+\kappa)\R_{(2k-2+\kappa)p,s}^{a,0},\qquad\quad
 \ketw{\R_{r,\kappa p'}^{0,b}}\ =\ \bigoplus_{k\in\mathbb{N}}(2k-2+\kappa)\R_{r,(2k-2+\kappa)p'}^{0,b}
\label{r2Vir}
\ee
while the ${\cal W}$-indecomposable rank-3 representations decompose as
\be
 \ketw{\R_{\kappa p,p'}^{a,b}}  
   \ =\ \bigoplus_{k\in\mathbb{N}}(2k-2+\kappa)\R_{p,(2k-2+\kappa)p'}^{a,b}
   \ =\ \bigoplus_{k\in\mathbb{N}}(2k-2+\kappa)\R_{(2k-2+\kappa)p,p'}^{a,b}
\label{r3Vir}
\ee

\subsection{${\cal W}$-irreducible characters}

It is recalled that Virasoro-irreducible characters are denoted by $\ch_{\rho,\sigma}(q)$ 
where $\rho,\sigma\in\mathbb{N}$ as we reserve the notation
$\chit_{\rho,\sigma}(q)$ for the characters of the (in general reducible) 
Kac representations $(\rho,\sigma)$. Only if the Kac representation happens
to be Virasoro-irreducible, cf. the discussion following (\ref{eta}), do we use both notations.
In the ${\cal W}$-extended picture, on the other hand, we will denote the character of 
a ${\cal W}$-irreducible representation of conformal weight $\D_{\rho,\sigma}$ simply by 
$\chih_{\rho,\sigma}(q)$.

It is believed~\cite{Ras0805} that the $2(p+p'-1)$ ${\cal W}$-indecomposable
representations (\ref{r1}) are in fact ${\cal W}$-{\em irreducible},
and that there are an additional $\frac{5}{2}(p-1)(p'-1)$ ${\cal W}$-irreducible rank-1
representations appearing as {\em subfactors} of the ${\cal W}$-indecomposable
rank-2 and -3 representations. 
This brings the total number of ${\cal W}$-irreducible characters to 
\be
 N_{\mathrm{irr}}(p,p') \ =\ 2pp' +\frac{1}{2}(p-1)(p'-1)
\ee
The characters of the ${\cal W}$-irreducible representations (\ref{r1}) are
\bea
 \chih_{\kappa p,s}(q)&=&\sum_{k\in\mathbb{N}}(2k-2+\kappa)\ch_{(2k-2+\kappa)p,s}(q)
   \ =\ \frac{1}{\eta(q)}\sum_{k\in\mathbb{Z}}(2k-2+\kappa)q^{((2k-2+\kappa)p'-s)^2p/4p'}\nn
 \chih_{r,\kappa p'}(q)&=&\sum_{k\in\mathbb{N}}(2k-2+\kappa)\ch_{r,(2k-2+\kappa)p'}(q)
   \ =\ \frac{1}{\eta(q)}\sum_{k\in\mathbb{Z}}(2k-2+\kappa)q^{((2k-2+\kappa)p-r)^2p'/4p} 
\label{r1char}
\eea
while $\frac{1}{2}(p-1)(p'-1)$ of the additional ${\cal W}$-irreducible representations 
simply correspond to Virasoro-irreducible representations and have characters given by
\be
  \chih_{a,b}(q)\ =\ \chih_{p-a,p'-b}(q)\ =\ \ch_{a,b}(q)
   \ =\ \frac{1}{\eta(q)}\sum_{k\in\mathbb{Z}}
    \Big(q^{(ap'-bp+2kpp')^2/4pp'}-q^{(ap'+bp+2kpp')^2/4pp'}\Big)
\label{chihab}
\ee
The remaining $2(p-1)(p'-1)$ ${\cal W}$-irreducible representations have characters
\bea
 \!\!\!\!\!\!\chih_{\kappa p+a,b}(q)&=&\sum_{k\in\mathbb{N}}(2k-2+\kappa)\ch_{(2k-2+\kappa)p+a,b}(q) \nn
  &=& \frac{1}{\eta(q)}\sum_{k\in\mathbb{Z}}k(k-1+\kappa)\Big(
    q^{(ap'-bp+(2k-2+\kappa)pp')^2/4pp'}-q^{(ap'+bp+(2k-2+\kappa)pp')^2/4pp'}\Big) 
\label{chih}
\eea
satisfying
\be
    \chih_{\kappa p+a,p'-b}(q)\ =\ \chih_{p-a,\kappa p'+b}(q)
\ee
The characters of the higher-rank ${\cal W}$-indecomposable representations can then be
expressed in terms of ${\cal W}$-irreducible characters
\bea
 \chit\big[\ketw{\R_{\kappa p,s}^{a,0}}\big](q)
   &=& \delta_{\kappa,1}\big(1-\delta_{s,p'}\big)\chih_{p-a,s}(q)+2\chih_{(4-\kappa)p-a,s}(q)
    +2\chih_{\kappa p+a,s}(q)\nn
 \chit\big[\ketw{\R_{r,\kappa p'}^{0,b}}\big](q)&=&
    \delta_{\kappa,1}\big(1-\delta_{r,p}\big)\chih_{r,p'-b}(q)+2\chih_{r,(4-\kappa)p'-b}(q)
    +2\chih_{r,\kappa p'+b}(q)\nn
 \chit\big[\ketw{\R_{\kappa p,p'}^{a,b}}\big](q)
   &=& 2\delta_{\kappa,1}\chih_{a,b}(q)+2\delta_{\kappa,2}\chih_{p-a,b}(q)
    +4\chih_{\kappa p+a,p'-b}(q)\nn
   &+&4\chih_{\kappa p+p-a,b}(q)+4\chih_{(3-\kappa)p+a,b}(q)
    +4\chih_{a,(3-\kappa)p'+b}(q)
\eea

\subsection{Fusion algebra}

We denote the fusion multiplication in the ${\cal W}$-extended picture by $\!\fus\!$ and
reserve the symbol $\otimes$ for the fusion multiplication in the Virasoro picture.
The fusion rules underlying the fusion algebra
in the ${\cal W}$-extended picture ${\cal WLM}(p,p')$
are summarized in Appendix \ref{AppFus}. 
Here, we denote this fusion algebra by 
\be
 \mathrm{Out}[{\cal WLM}(p,p')]\ =\ 
  \big\langle\ketw{\kappa p,b},\ketw{a,\kappa p'},\ketw{\kappa p,p'},\ketw{\R_{\kappa p,s}^{a,0}},
  \ketw{\R_{r,\kappa p'}^{0,b}},\ketw{\R_{\kappa p,p'}^{a,b}}  \big\rangle_{p,p'}
\label{WfusOut}
\ee
foreshadowing the existence of its extension $\mathrm{Fund}[{\cal WLM}(p,p')]$ to be
discussed in Section~\ref{SecFundWLM}.
This pairing of the two ${\cal W}$-extended (`outer' and `fundamental') fusion algebras is of course
reminiscent of the analogous situation in the Virasoro picture.
We also introduce
\be
 \ketw{\Jc^{\mathrm{Out}}_{p,p'}}\ =\ \big\{
  \ketw{\R_{\kappa p,b}^{0,0}},\ketw{\R_{a,\kappa p'}^{0,0}},\ketw{\R_{\kappa p,p'}^{0,0}},
   \ketw{\R_{\kappa p,s}^{a,0}},\ketw{\R_{r,\kappa p'}^{0,b}},\ketw{\R_{\kappa p,p'}^{a,b}}\big\}
\label{JWout}
\ee
as the set of generators of (\ref{WfusOut}). Its cardinality is
\be
 |\ketw{\Jc^{\mathrm{Out}}_{p,p'}}|\ =\ 6pp'-2p-2p'
\ee
The fusion algebra (\ref{WfusOut})
is both associative and commutative, while there is no identity for $p>1$.
For $p=1$, the ${\cal W}$-irreducible representation $\ketw{1,1}$ is the identity.

\section{Fundamental fusion algebra of ${\cal WLM}(p,p')$}
\label{SecFundWLM}

Considering the similarity between $\mathrm{Out}[{\cal LM}(p,p')]$ 
and $\mathrm{Out}[{\cal WLM}(p,p')]$, 
we here propose a particular and in some sense
canonical extension of the ${\cal W}$-extended algebra as the analogue of 
the fundamental fusion algebra $\mathrm{Fund}[{\cal LM}(p,p')]$
in the Virasoro picture. We will refer to this extension
as the ${\cal W}$-{\em fundamental} fusion algebra and denote it by
$\mathrm{Fund}[{\cal WLM}(p,p')]$. It is the topic of the present section.

We emphasize, that Yang-Baxter integrable boundary 
conditions corresponding to the new
${\cal W}$-representations are yet to be constructed. For now, we can 
therefore only claim that the proposed extension  
is {\em algebraic} in nature.

\subsection{Representation content of fundamental extension}

We extend the set of generators of the outer fusion algebra (\ref{WfusOut}) by introducing
\be
 \ketw{\Jc^{\mathrm{Fund}}_{p,p'}}\ =\ \big\{\ketw{\R_{a,b}^{0,0}},
  \ketw{\R_{\kappa p,b}^{0,0}},\ketw{\R_{a,\kappa p'}^{0,0}},\ketw{\R_{\kappa p,p'}^{0,0}},
  \ketw{\R_{\kappa p,s}^{a,0}},
  \ketw{\R_{r,\kappa p'}^{0,b}},\ketw{\R_{\kappa p,p'}^{a,b}}\big\}
\label{JWfund}
\ee
where
\be
 \ketw{\Jc^{\mathrm{Fund}}_{p,p'}}\ =\ \big\{\ketw{a,b}\big\}\cup\ketw{\Jc^{\mathrm{Out}}_{p,p'}}
\label{JWdisj}
\ee
is a disjoint union.
The cardinality of $\ketw{\Jc^{\mathrm{Fund}}_{p,p'}}$ is thus
\be
 |\ketw{\Jc^{\mathrm{Fund}}_{p,p'}}|\ =\ 
  (p-1)(p'-1)+|\ketw{\Jc^{\mathrm{Out}}_{p,p'}}|\ =\ 7pp'-3p-3p'+1
\label{CardJFund}
\ee
As we will argue below, one can extend the domain of the fusion multiplication $\!\fus\!$
from $\ketw{\Jc^{\mathrm{Out}}_{p,p'}}$ to $\ketw{\Jc^{\mathrm{Fund}}_{p,p'}}$, 
thereby obtaining the fusion algebra
\be
 \mathrm{Fund}[{\cal WLM}(p,p')]\ =\ 
  \big\langle\ketw{a,b},\ketw{\kappa p,b},\ketw{a,\kappa p'},\ketw{\kappa p,p'},
  \ketw{\R_{\kappa p,s}^{a,0}},
  \ketw{\R_{r,\kappa p'}^{0,b}},\ketw{\R_{\kappa p,p'}^{a,b}}  \big\rangle_{p,p'}
\label{WfusFund}
\ee
Furthermore, we will demonstrate that this fusion algebra
is generated from repeated fusions of the two `fundamental' representations $\ketw{2,1}$
and $\ketw{1,2}$ (strictly speaking,
in addition to the identity $\ketw{1,1}$, cf. the discussion in Section~\ref{SecWfunRep})
\be
 \mathrm{Fund}[{\cal WLM}(p,p')]\ =\ \big\langle \ketw{1,1}, \ketw{2,1},\ketw{1,2}\big\rangle_{p,p'}
\label{Fund2112}
\ee
much akin to the situation in the Virasoro picture (\ref{F2112}). In hindsight, this is of course
the reason we refer to (\ref{WfusFund}) as the {\em fundamental} fusion algebra of
${\cal WLM}(p,p')$. We also note that the extension (\ref{JWdisj}) is trivial for $p=1$ since the set 
$\{\ketw{a,b}\}$ is empty in this case. This implies that 
\be
 \mathrm{Fund}[{\cal WLM}(1,p')]\ =\ \mathrm{Out}[{\cal WLM}(1,p')]
\ee
again reminiscent of the Virasoro picture (\ref{OF1p}).

The embedding patterns, conformal dimensions and characters of the new ${\cal W}$-representations
$\{\ketw{a,b}\}$ are discussed in the following.

\subsubsection{Embedding patterns and characters}

Just as the reducible yet indecomposable Kac representation $(a,b)$ has conformal weight
$\D_{a,b}$ and appears as a
subrepresentation of the indecomposable rank-2 representations $\R_{p,b}^{p-a,0}$
and $\R_{a,p'}^{0,p'-b}$~\cite{Embedding,RP0707},
we find it natural to expect that the new ${\cal W}$-representation 
$\ketw{a,b}$ has conformal weight $\D_{a,b}$ and appears as a subrepresentation of the 
${\cal W}$-indecomposable rank-2 
representations $\ketw{\R_{p,b}^{p-a,0}}$ and $\ketw{\R_{a,p'}^{0,p'-b}}$
(and hence of the indecomposable rank-3 representation $\ketw{\R_{p,p'}^{p-a,p'-b}}$, 
cf. the discussion of embedding patterns in~\cite{Ras0805}).
Also, the embedding patterns of $\ketw{\R_{p,b}^{p-a,0}}$ and $\ketw{\R_{a,p'}^{0,p'-b}}$
are believed~\cite{Ras0805} to be
\be
 \mbox{
 \begin{picture}(100,100)(-25,0)
    \unitlength=0.8cm
  \thinlines
\put(-4,2){$\ketw{\R_{p,b}^{p-a,0}}:$}
\put(1.5,3.7){$\ketw{\D_{2p+a,b}}$}
\put(-1.3,2){$\ketw{\D_{2p-a,b}}$}
\put(3.9,2){$\ketw{\D_{2p-a,b}}$}
\put(1.5,0.3){$\ketw{\D_{2p+a,b}}$}
\put(3.5,2.1){\vector(-1,0){2.1}}
\put(1.6,3.4){\vector(-4,-3){1.2}}
\put(4.6,1.7){\vector(-4,-3){1.2}}
\put(1.6,0.8){\vector(-4,3){1.2}}
\put(4.6,2.5){\vector(-4,3){1.2}}
\put(1.85,1.25){$\ketw{\D_{a,b}}$}
\put(3.7,1.9){\vector(-2,-1){0.5}}
\put(1.8,1.6){\vector(-2,1){0.5}}
 \end{picture}
}
\hspace{5cm}
 \mbox{
 \begin{picture}(100,100)(-25,0)
    \unitlength=0.8cm
  \thinlines
\put(-4.1,2){$\ketw{\R_{a,p'}^{0,p'-b}}:$}
\put(1.5,3.7){$\ketw{\D_{a,2p'+b}}$}
\put(-1.3,2){$\ketw{\D_{a,2p'-b}}$}
\put(3.9,2){$\ketw{\D_{a,2p'-b}}$}
\put(1.5,0.3){$\ketw{\D_{a,2p'+b}}$}
\put(3.5,2.1){\vector(-1,0){2.1}}
\put(1.6,3.4){\vector(-4,-3){1.2}}
\put(4.6,1.7){\vector(-4,-3){1.2}}
\put(1.6,0.8){\vector(-4,3){1.2}}
\put(4.6,2.5){\vector(-4,3){1.2}}
\put(1.85,1.25){$\ketw{\D_{a,b}}$}
\put(3.7,1.9){\vector(-2,-1){0.5}}
\put(1.8,1.6){\vector(-2,1){0.5}}
 \end{picture}
}
\label{Remb}
\ee 
where the horizontal arrows indicate the non-diagonal action of the Virasoro mode $L_0$,
while $\ketw{\D}$ is a shorthand for the ${\cal W}$-irreducible representation of conformal
weight $\D$.
Since
\be
 \D_{2p-a,b}\ =\ \D_{a,2p'-b},\qquad\quad \D_{2p+a,b}\ \neq\ \D_{a,2p'+b}
\ee
we are thus led to conjecture that the embedding pattern of $\ketw{a,b}$ is given by
\be
 \mbox{
 \begin{picture}(100,60)(-30,0)
    \unitlength=0.8cm
  \thinlines
\put(-3.3,1){$\ketw{a,b}:$}
\put(-0.7,1.9){$\ketw{\D_{2p-a,b}}$}
\put(2.7,0){$\ketw{\D_{a,b}}$}
\put(2.6,0.6){\vector(-4,3){1.2}}
\put(5,1){$=$}
 \end{picture}
}
\hspace{2cm}
 \mbox{
 \begin{picture}(100,60)(-30,0)
    \unitlength=0.8cm
  \thinlines
\put(-0.7,1.9){$\ketw{\D_{a,2p'-b}}$}
\put(2.7,0){$\ketw{\D_{a,b}}$}
\put(2.6,0.6){\vector(-4,3){1.2}}
 \end{picture}
}
\label{abemb}
\ee
implying that $\ketw{a,b}$ is a 
{\em ${\cal W}$-reducible yet ${\cal W}$-indecomposable representation of rank 1}.
Its conformal dimension is indeed
\be
 \D[\ketw{a,b}]\ =\ \D_{a,b}
\ee
while the associated character reads
\bea
 \chit[\ketw{a,b}](q)&=&\chih_{a,b}(q)+\chih_{2p-a,b}(q)\ =\ \chih_{a,b}(q)+\chih_{a,2p'-b}(q)\nn
  &=&\frac{1}{\eta(q)}\sum_{k\in\mathbb{Z}}(k^2-1)
    \Big(q^{(ap'+bp+2kpp')^2/4pp'}-q^{(ap'-bp+2kpp')^2/4pp'}\Big)
\label{abchar}
\eea

Combined with the result for $p=1$ (\ref{r1char}), the character of $\ketw{1,1}$, in particular, is
\be
 \chit[\ketw{1,1}](q)\ =\ \chih_{\D=0}(q)+\big(1-\delta_{p,1}\big)\chih_{\D=(p-1)(p'-1)}(q)
\ee
This clearly resembles the decomposition of the character of the reducible yet indecomposable
Kac representation $(1,1)$
\be
 \chit[(1,1)](q)\ =\ \ch_{\D=0}(q)+\big(1-\delta_{p,1}\big)\ch_{\D=(p-1)(p'-1)}(q)
\ee
As we will see below, the representation $\ketw{1,1}$ is the identity of the ${\cal W}$-extended
fundamental fusion algebra. This is true for all ${\cal WLM}(p,p')$. It is recalled in this regard,
that the outer fusion algebra $\mathrm{Out}[{\cal WLM}(p,p')]$
only has an identity for $p=1$.

We stress that the details of the conjectured embedding pattern (\ref{abemb}) and the ensuing
character expression (\ref{abchar}) have no direct bearing on the fusion rules to be discussed below.
That is, the algebraically motivated fusion algebra $\mathrm{Fund}[{\cal WLM}(p,p')]$ remains
well defined even if the conjecture turns out to be false.

\subsection{Supplementary fusion rules}

We are faced with the task of devising rules for 
$\ketw{a,b}\fus\ketw{\R_{i,j}^{\al,\beta}}$, for all
$\ketw{\R_{i,j}^{\al,\beta}}\in\ketw{\Jc_{p,p'}^{\mathrm{Fund}}}$, compatible with the known 
fusion rules of $\mathrm{Out}[{\cal WLM}(p,p')]$.
In particular, the complete set of fusion rules
of $\mathrm{Fund}[{\cal WLM}(p,p')]$ must yield a commutative and associative fusion algebra.
We also require that fusion in $\mathrm{Fund}[{\cal WLM}(p,p')]$
separates into horizontal and vertical components where for every
$\ketw{\R_{i,j}^{\al,\beta}}\in\ketw{\Jc_{p,p'}^{\mathrm{Fund}}}$ there exist 
$\ketw{\R_{i,1}^{\al,0}}\in\ketw{\Jc_{p,p'}^{\mathrm{Fund}}}$ and 
$\ketw{\R_{1,j}^{0,\beta}}\in\ketw{\Jc_{p,p'}^{\mathrm{Fund}}}$ such that 
\be
 \ketw{\R_{i,j}^{\al,\beta}}\ =\ \ketw{\R_{i,1}^{\al,0}}\fus\ketw{\R_{1,j}^{0,\beta}}
\label{sep}
\ee
This extends the similar separation of fusion in 
$\mathrm{Out}[{\cal WLM}(p,p')]$~\cite{PRR0803,RP0804,Ras0805}.
Furthermore, since we wish to consider $\mathrm{Fund}[{\cal WLM}(p,p')]$ as an {\em extension} of
$\mathrm{Out}[{\cal WLM}(p,p')]$, our proposal for the former should have the latter as an
algebraic ideal with respect to the fusion multiplication $\!\fus\!$. As we will see below,
this is indeed the case. A detailed discussion of how the resulting fusion algebra
$\mathrm{Fund}[{\cal WLM}(p,p')]$ can be viewed as a quotient 
of the fundamental fusion algebra $\mathrm{Fund}[{\cal LM}(p,p')]$ is given as
Proposition~\ref{SecQuo}.1 in Section~\ref{SecFinite}.

Now, our proposal for the fusion rules of $\mathrm{Fund}[{\cal WLM}(p,p')]$ 
is obtained by supplementing the fusion rules of $\mathrm{Out}[{\cal WLM}(p,p')]$, 
determined in~\cite{Ras0805} and listed in 
Appendix~\ref{AppFus}, with fusion rules for $\ketw{a,b}\fus\ketw{a',b'}$ and 
$\ketw{a,b}\fus\ketw{\R_{i,j}^{\al,\beta}}$ for all
$\ketw{\R_{i,j}^{\al,\beta}}\in\ketw{\Jc_{p,p'}^{\mathrm{Out}}}$.
These are given in (\ref{sup0}) through (\ref{sup3}) and are motivated as follows.
Implementing the separation property (\ref{sep}), 
we consider the first type of fusion in (\ref{fus11})
\bea
  \ketw{\kappa p,b}\fus\ketw{\kappa'p,b'}&=&\bigoplus_{\al}^{p-1}\Big\{
  \!\bigoplus_{j=|b-b'|+1,\ \!{\rm by}\ \!2}^{p'-|p'-b-b'|-1}
  \!\!\!\ketw{\R_{(\kappa\cdot\kappa')p,1}^{\al,0}}\fus\ketw{1,j}\nn
  &&\qquad\qquad\qquad\qquad \oplus\!\bigoplus_{\beta}^{b+b'-p'-1}
    \!\ketw{\R_{(\kappa\cdot\kappa') p,1}^{\al,0}}\fus\ketw{\R_{1,p'}^{0,\beta}}
        \Big\}   \nn
 &=&\Big\{\bigoplus_{\al}^{p-1}\ketw{\R_{(\kappa\cdot\kappa') p,1}^{\al,0}}\Big\}\fus
  \Big\{\!\bigoplus_{j=|b-b'|+1,\ \!{\rm by}\ \!2}^{p'-|p'-b-b'|-1}
  \!\!\!\ketw{1,j}\oplus\!\bigoplus_{\beta}^{b+b'-p'-1}\!\ketw{\R_{1,p'}^{0,\beta}}\Big\}
\eea
Since
\bea
 \ketw{\kappa p,s}\fus\ketw{\kappa'p,s'}&=&\big\{\ketw{\kappa p,1}\fus\ketw{\kappa' p,1}\big\}
   \fus\big\{\ketw{1,s}\fus\ketw{1,s'}\big\}\nn
  &=&\Big\{\bigoplus_{\al}^{p-1}\ketw{\R_{(\kappa\cdot\kappa') p,1}^{\al,0}}\Big\}\fus
   \big\{\ketw{1,s}\fus\ketw{1,s'}\big\}
\eea
a naive application of (\ref{sep}) yields
\be
 \ketw{1,b}\fus\ketw{1,b'}\ =\ \bigoplus_{j=|b-b'|+1,\ \!{\rm by}\ \!2}^{p'-|p'-b-b'|-1}
  \!\!\!\ketw{1,j}\oplus\!\bigoplus_{\beta}^{b+b'-p'-1}\!\ketw{\R_{1,p'}^{0,\beta}}
\label{1b1b}
\ee
We likewise find
\be
 \ketw{a,1}\fus\ketw{a',1}\ =\ \bigoplus_{i=|a-a'|+1,\ \!{\rm by}\ \!2}^{p-|p-a-a'|-1}
  \!\!\!\ketw{i,1}\oplus\!\bigoplus_{\al}^{a+a'-p-1}\!\ketw{\R_{p,1}^{\al,0}}
\label{a1a1}
\ee
Similarly using (\ref{fus12}), we consider
\bea
  \ketw{\kappa p,b}\fus\ketw{\R_{r,\kappa'p'}^{0,b'}}
  &=&\bigoplus_{\al}^{r-1}\Big\{
    \!\bigoplus_{\beta=|b-b'|+1,\ \!{\rm by}\ \!2}^{p'-|p'-b-b'|-1}
      \!\!\!\!\ketw{\R_{\kappa p,1}^{\al,0}}\fus\ketw{\R_{1,\kappa'p'}^{0,\beta}}
     \oplus\!\bigoplus_{\beta}^{b-b'-1}\!
      2\ketw{\R_{\kappa p,1}^{\al,0}}\fus\ketw{\R_{1,\kappa'p'}^{0,\beta}}
    \nn
  &&\qquad\qquad\qquad\qquad
     \oplus
     \!\!\!\bigoplus_{\beta}^{b+b'-p'-1}
        \!\!\!2\ketw{\R_{\kappa p,1}^{\al,0}}
      \fus\ketw{\R_{1,(2\cdot\kappa')p'}^{0,\beta}}
        \Big\}   \nn
 &=&\Big\{\bigoplus_{\al}^{r-1}\ketw{\R_{\kappa p,1}^{\al,0}}\Big\}
  \fus\Big\{\bigoplus_{\beta=|b-b'|+1,\ \!{\rm by}\ \!2}^{p'-|p'-b-b'|-1}\ketw{\R_{1,\kappa'p'}^{0,\beta}}\nn
  &&\qquad\qquad
     \oplus\!\bigoplus_{\beta}^{b-b'-1}\!
      2\ketw{\R_{1,\kappa'p'}^{0,\beta}}
     \oplus
     \!\!\!\bigoplus_{\beta}^{b+b'-p'-1}
        \!\!\!2\ketw{\R_{1,(2\cdot\kappa')p'}^{0,\beta}}
        \Big\}
\eea
{}From the separation, 
\be
 \ketw{\kappa p,b}\fus\ketw{\R_{r,\kappa'p'}^{0,b'}}\ =\ 
  \big\{\ketw{\kappa p,1}\fus\ketw{r,1}\big\}\fus\big\{\ketw{1,b}\fus\ketw{\R_{1,\kappa'p'}^{0,b'}}\big\}
\ee
we thus deduce that
\bea
 \ketw{a,1}\fus\ketw{\kappa p,1}&=&\bigoplus_{\al}^{a-1}\ketw{\R_{\kappa p,1}^{\al,0}}\nn
 \ketw{1,b}\fus\ketw{\R_{1,\kappa p'}^{0,b'}}&=&\!\!\!
   \bigoplus_{\beta=|b-b'|+1,\ \!{\rm by}\ \!2}^{p'-|p'-b-b'|-1}\!\!\!\ketw{\R_{1,\kappa p'}^{0,\beta}}
     \oplus\!\bigoplus_{\beta}^{b-b'-1}\!
      2\ketw{\R_{1,\kappa p'}^{0,\beta}}
     \oplus
     \!\!\!\bigoplus_{\beta}^{b+b'-p'-1}
        \!\!\!2\ketw{\R_{1,(2\cdot\kappa)p'}^{0,\beta}}
\label{a1p1}
\eea
We likewise find
\bea
 \ketw{1,b}\fus\ketw{1,\kappa p'}&=&\bigoplus_{\beta}^{b-1}\ketw{\R_{1,\kappa p'}^{0,\beta}}\nn
 \ketw{a,1}\fus\ketw{\R_{\kappa p,1}^{a',0}}&=&\!\!\!
   \bigoplus_{\al=|a-a'|+1,\ \!{\rm by}\ \!2}^{p-|p-a-a'|-1}\!\!\!\ketw{\R_{\kappa p,1}^{\al,0}}
     \oplus\!\bigoplus_{\al}^{a-a'-1}\!
      2\ketw{\R_{\kappa p,1}^{\al,0}}
     \oplus
     \!\!\!\bigoplus_{\al}^{a+a'-p-1}
        \!\!\!2\ketw{\R_{(2\cdot\kappa)p,1}^{\al,0}}
\label{1b1p}
\eea

The horizontal and vertical fusion rules are merged by yet another application
of (\ref{sep}).
In conclusion, we {\em posit the supplementary fusion rules}
\bea
 \ketw{a,b}\fus\ketw{a',b'}&=&
   \bigoplus_{i=|a-a'|+1,\ \!{\rm by}\ \!2}^{p-|p-a-a'|-1}
     \Big\{\bigoplus_{j=|b-b'|+1,\ \!{\rm by}\ \!2}^{p'-|p'-b-b'|-1}
  \!\!\!\ketw{i,j}\Big\}\oplus\!
  \bigoplus_{\al}^{a+a'-p-1}\!
   \Big\{\bigoplus_{j=|b-b'|+1,\ \!{\rm by}\ \!2}^{p'-|p'-b-b'|-1}\ketw{\R_{p,j}^{\al,0}}\Big\}\nn
 &&\oplus\!\bigoplus_{\beta}^{b+b'-p'-1}\!
   \Big\{\bigoplus_{i=|a-a'|+1,\ \!{\rm by}\ \!2}^{p-|p-a-a'|-1}\ketw{\R_{i,p'}^{0,\beta}}\Big\}
  \oplus\bigoplus_{\al}^{a+a'-p-1}
    \Big\{\bigoplus_{\beta}^{b+b'-p'-1}\!\ketw{\R_{p,p'}^{\al,\beta}}\Big\}
\label{sup0}
\eea
and
\bea
 \ketw{a,b}\fus\ketw{\kappa p,b'}&=&\bigoplus_{\al}^{a-1}\Big\{
    \bigoplus_{j=|b-b'|+1,\ \!{\rm by}\ \!2}^{p'-|p'-b-b'|-1}\!\!\ketw{\R_{\kappa p,j}^{\al,0}}
    \oplus  \bigoplus_{\beta}^{b+b'-p'-1}\!\ketw{\R_{\kappa p,p'}^{\al,\beta}}\Big\}\nn
 \ketw{a,b}\fus\ketw{a',\kappa p'}&=& \bigoplus_{\beta}^{b-1}\Big\{
    \bigoplus_{i=|a-a'|+1,\ \!{\rm by}\ \!2}^{p-|p-a-a'|-1}\!\!\ketw{\R_{i,\kappa p'}^{0,\beta}}
    \oplus  \bigoplus_{\al}^{a+a'-p-1}\!\ketw{\R_{\kappa p,p'}^{\al,\beta}}\Big\}  \nn
 \ketw{a,b}\fus\ketw{\kappa p,p'}&=&\bigoplus_{\al}^{a-1}\Big\{
  \bigoplus_{\beta}^{b-1}\ketw{\R_{\kappa p,p'}^{\al,\beta}}\Big\} 
\label{sup1}
\eea
and
\bea
 \ketw{a,b}\fus\ketw{\R_{\kappa p,s}^{a',0}}&=&\bigoplus_{\al=|a-a'|+1,\ \!{\rm by}\ \!2}^{p-|p-a-a'|-1}
   \Big\{\bigoplus_{j=|b-s|+1,\ \!{\rm by}\ \!2}^{p'-|p'-b-s|-1}\ketw{\R_{\kappa p,j}^{\al,0}}
    \oplus \bigoplus_{\beta}^{b+s-p'-1}\ketw{\R_{\kappa p,p'}^{\al,\beta}}\Big\}  \nn
  &&\oplus  \bigoplus_{\al}^{a-a'-1}
   \Big\{\bigoplus_{j=|b-s|+1,\ \!{\rm by}\ \!2}^{p'-|p'-b-s|-1}2\ketw{\R_{\kappa p,j}^{\al,0}}
    \oplus \bigoplus_{\beta}^{b+s-p'-1}2\ketw{\R_{\kappa p,p'}^{\al,\beta}}\Big\}  \nn
        \nn
  &&\oplus  \bigoplus_{\al}^{a+a'-p-1}
   \Big\{\bigoplus_{j=|b-s|+1,\ \!{\rm by}\ \!2}^{p'-|p'-b-s|-1}2\ketw{\R_{(2\cdot\kappa)p,j}^{\al,0}}
    \oplus \bigoplus_{\beta}^{b+s-p'-1}2\ketw{\R_{(2\cdot\kappa)p,p'}^{\al,\beta}}\Big\} 
        \nn
 \ketw{a,b}\fus\ketw{\R_{r,\kappa p'}^{0,b'}}&=&   \bigoplus_{\beta=|b-b'|+1,\ \!{\rm by}\ \!2}^{p'-|p'-b-b'|-1}
   \Big\{\bigoplus_{i=|a-r|+1,\ \!{\rm by}\ \!2}^{p-|p-a-r|-1}\ketw{\R_{i,\kappa p'}^{0,\beta}}
    \oplus \bigoplus_{\al}^{a+r-p-1}\ketw{\R_{\kappa p,p'}^{\al,\beta}}\Big\}  \nn
  &&\oplus  \bigoplus_{\beta}^{b-b'-1}
   \Big\{\bigoplus_{i=|a-r|+1,\ \!{\rm by}\ \!2}^{p-|p-a-r|-1}2\ketw{\R_{i,\kappa p'}^{0,\beta}}
    \oplus \bigoplus_{\al}^{a+r-p-1}2\ketw{\R_{\kappa p,p'}^{\al,\beta}}\Big\}  \nn
        \nn
  &&\oplus  \bigoplus_{\beta}^{b+b'-p'-1}
   \Big\{\bigoplus_{i=|a-r|+1,\ \!{\rm by}\ \!2}^{p-|p-a-r|-1}2\ketw{\R_{i,(2\cdot\kappa)p'}^{0,\beta}}
    \oplus \bigoplus_{\al}^{a+r-p-1}2\ketw{\R_{(2\cdot\kappa)p,p'}^{\al,\beta}}\Big\} 
\label{sup2}
\eea
and finally
\bea
 \ketw{a,b}\fus\ketw{\R_{\kappa p,p'}^{a',b'}}&=&\bigoplus_{\al=|a-a'|+1,\ \!{\rm by}\ \!2}^{p-|p-a-a'|-1}
   \Big\{\bigoplus_{\beta=|b-b'|+1,\ \!{\rm by}\ \!2}^{p'-|p'-b-b'|-1}\!\!\ketw{\R_{\kappa p,p'}^{\al,\beta}}\Big\}
   \nn
 &&\oplus\bigoplus_{\beta}^{b-b'-1}\Big\{
  \bigoplus_{\al=|a-a'|+1,\ \!{\rm by}\ \!2}^{p-|p-a-a'|-1}\!\!2\ketw{\R_{\kappa p,p'}^{\al,\beta}}\Big\}
  \oplus\bigoplus_{\al}^{a-a'-1}\Big\{
   \bigoplus_{\beta=|b-b'|+1,\ \!{\rm by}\ \!2}^{p'-|p'-b-b'|-1}\!\!2\ketw{\R_{\kappa p,p'}^{\al,\beta}}\Big\}
  \nn
 &&\oplus\bigoplus_{\al}^{a-a'-1}\Big\{\bigoplus_{\beta}^{b-b'-1}4\ketw{\R_{\kappa p,p'}^{\al,\beta}}\Big\}
  \oplus
    \bigoplus_{\al}^{a+a'-p-1}\Big\{\bigoplus_{\beta}^{b+b'-p'-1}4\ketw{\R_{\kappa p,p'}^{\al,\beta}}\Big\}
  \nn
&&\oplus\bigoplus_{\al}^{a+a'-p-1}\Big\{
  \bigoplus_{\beta=|b-b'|+1,\ \!{\rm by}\ \!2}^{p'-|p'-b-b'|-1}\!\!2\ketw{\R_{(2\cdot\kappa)p,p'}^{\al,\beta}}
    \oplus\bigoplus_{\beta}^{b-b'-1}4\ketw{\R_{(2\cdot\kappa)p,p'}^{\al,\beta}}\Big\}
      \nn
 &&\oplus\bigoplus_{\beta}^{b+b'-p'-1}\Big\{
  \bigoplus_{\al=|a-a'|+1,\ \!{\rm by}\ \!2}^{p-|p-a-a'|-1}\!\!2\ketw{\R_{(2\cdot\kappa)p,p'}^{\al,\beta}}
    \oplus\bigoplus_{\al}^{a-a'-1}4\ketw{\R_{(2\cdot\kappa)p,p'}^{\al,\beta}}\Big\}
\label{sup3}
\eea

\subsubsection{${\cal W}$-fundamental representations}
\label{SecWfunRep}

Using the complete set of fusion rules underlying $\mathrm{Fund}[{\cal WLM}(p,p')]$, it is not hard to
see that this ${\cal W}$-extended fundamental fusion algebra is built from repeated fusions
of the two ${\cal W}$-indecomposable representations $\ketw{2,1}$ and $\ketw{1,2}$. We thus have
(\ref{Fund2112}) and naturally refer to $\ketw{2,1}$ and $\ketw{1,2}$ as 
${\cal W}$-{\em fundamental} representations.

Strictly speaking, though, for $p>1$, we are implicitly assuming that the identity $\ketw{1,1}$,
say, is also a generator of the fusion algebra. From the fusion 
$\ketw{1,2}\fus\ketw{1,2}=\ketw{1,1}\oplus\ketw{1,3}$, we can otherwise not identify $\ketw{1,1}$
and $\ketw{1,3}$ as {\em separate} representations. 
A similar and likewise implicit assumption is made when writing $\langle(2,1),(1,2)\rangle_{p,p'}$
for the fundamental fusion algebra in the Virasoro picture (\ref{A2112}).
In that case, on the other hand, lattice considerations allow us to disentangle the fusion
$(1,2)\otimes(1,2)=(1,1)\oplus(1,3)$. At present, the ${\cal W}$-indecomposable representations
$\{\ketw{a,b}\}$ are introduced algebraically only, so there is no disentangling
procedure to rely on. In any case, $\langle\ketw{2,1},\ketw{1,2}\rangle_{p,p'}$
is merely a suggestive (but in the sense just described, also a bit misleading) shorthand
for $\mathrm{Fund}[{\cal WLM}(p,p')]=\langle\ketw{1,1},\ketw{2,1},\ketw{1,2}\rangle_{p,p'}$ 
(\ref{WfusFund}).

We note that $\ketw{1,2}\in\ketw{\Jc_{p,p'}^{\mathrm{Out}}}$ iff $p'=2$, while
$\ketw{2,1}\in\ketw{\Jc_{p,p'}^{\mathrm{Out}}}$ iff $p=1,2$, and now list
all fusions involving these ${\cal W}$-fundamental representations.
For $p=1$, we have
\bea
 \ketw{2,1}\fus\ketw{\kappa,s}&=&\ketw{2\cdot\kappa,s}\nn
 \ketw{2,1}\fus\ketw{\R_{1,\kappa p'}^{0,b}}&=&\ketw{\R_{1,(2\cdot\kappa)p'}^{0,b}}
\label{21a}
\eea
while for $p>1$, we have
\bea
 \ketw{2,1}\fus\ketw{a,b}&=&\big(1-\delta_{a,1}\big)\ketw{a-1,b}\oplus\ketw{a+1,b}\nn
 \ketw{2,1}\fus\ketw{\kappa p,s}&=&\ketw{\R_{\kappa p,s}^{1,0}}\nn
 \ketw{2,1}\fus\ketw{a,\kappa p'}&=&\big(1-\delta_{a,1}\big)\ketw{a-1,\kappa p'}\oplus\ketw{a+1,\kappa p'}
    \nn
 \ketw{2,1}\fus\ketw{\R_{\kappa p,s}^{a,0}}&=&2\delta_{a,1}\ketw{\kappa p,s}
   \oplus2\delta_{a,p-1}\ketw{(2\cdot\kappa)p,s} \nn
  &&\oplus \big(1-\delta_{a,1}\big)\ketw{\R_{\kappa p,s}^{a-1,0}}    
    \oplus\big(1-\delta_{a,p-1}\big)\ketw{\R_{\kappa p,s}^{a+1,0}}
    \nn
 \ketw{2,1}\fus\ketw{\R_{r,\kappa p'}^{0,b}}&=&\delta_{r,1}\ketw{\R_{2,\kappa p'}^{0,b}}
   \oplus\delta_{r,p}\ketw{\R_{\kappa p,p'}^{1,b}}\nn
 &&\oplus\big(1-\delta_{r,1}\big)\big(1-\delta_{r,p}\big)\big(\ketw{\R_{r-1,\kappa p'}^{0,b}}
    \oplus\ketw{\R_{r+1,\kappa p'}^{0,b}}\big)\nn
 \ketw{2,1}\fus\ketw{\R_{\kappa p,p'}^{a,b}}&=&2\delta_{a,1}\ketw{\R_{p,\kappa p'}^{0,b}}
  \oplus2\delta_{a,p-1}\ketw{\R_{p,(2\cdot\kappa)p'}^{0,b}}\nn
 &&
  \oplus \big(1-\delta_{a,1}\big)\ketw{\R_{\kappa p,p'}^{a-1,b}}
   \oplus\big(1-\delta_{a,p-1}\big)\ketw{\R_{\kappa p,p'}^{a+1,b}}
\label{21}
\eea
Since $p'>p\geq1$, we have
\bea
 \ketw{1,2}\fus\ketw{a,b}&=&\big(1-\delta_{b,1}\big)\ketw{a,b-1}\oplus\ketw{a,b+1}\nn
 \ketw{1,2}\fus\ketw{\kappa p,b}&=&\big(1-\delta_{b,1}\big)\ketw{\kappa p,b-1}\oplus\ketw{\kappa p,b+1}
   \nn
 \ketw{1,2}\fus\ketw{r,\kappa p'}&=&\ketw{\R_{r,\kappa p'}^{0,1}}\nn
 \ketw{1,2}\fus\ketw{\R_{\kappa p,s}^{a,0}}&=&\delta_{s,1}\ketw{\R_{\kappa p,2}^{a,0}}
   \oplus\delta_{s,p'}\ketw{\R_{\kappa p,p'}^{a,1}}\nn
  &&\oplus\big(1-\delta_{s,1}\big)\big(1-\delta_{s,p'}\big)\big(\ketw{\R_{\kappa p,s-1}^{a,0}}
    \oplus\ketw{\R_{\kappa p,s+1}^{a,0}}\big)\nn
 \ketw{1,2}\fus\ketw{\R_{r,\kappa p'}^{0,b}}&=&2\delta_{b,1}\ketw{r,\kappa p'}
   \oplus2\delta_{b,p'-1}\ketw{r,(2\cdot\kappa)p'}\nn
  &&\oplus\big(1-\delta_{b,1}\big)\ketw{\R_{r,\kappa p'}^{0,b-1}}\oplus
   \big(1-\delta_{b,p'-1}\big)\ketw{\R_{r,\kappa p'}^{0,b+1}}\nn
 \ketw{1,2}\fus\ketw{\R_{\kappa p,p'}^{a,b}}&=&2\delta_{b,1}\ketw{\R_{\kappa p,p'}^{a,0}}
   \oplus2\delta_{b,p'-1}\ketw{\R_{(2\cdot\kappa)p,p'}^{a,0}}\nn
  &&\oplus\big(1-\delta_{b,1}\big)\ketw{\R_{\kappa p,p'}^{a,b-1}}\oplus\big(1-\delta_{b,p'-1}\big)
    \ketw{\R_{\kappa p,p'}^{a,b+1}}
\label{12}
\eea
for all $p\in\mathbb{N}$.

\subsection{Fusion matrices and quotient polynomial fusion rings}

For every $n\in\mathbb{N}$, we introduce the polynomial
\be
 P_n(x)\ =\ U_{3n-1}\big(\frac{x}{2}\big)-3U_{n-1}\big(\frac{x}{2}\big)\ =\ 
  2\Big(T_{2n}\big(\frac{x}{2}\big)-1\Big)U_{n-1}\big(\frac{x}{2}\big)\ =\ 
  (x^2-4)U_{n-1}^3\big(\frac{x}{2}\big)
\label{Pn}
\ee
where the rewritings are due to (\ref{2TU}) and (\ref{T2n}).
These polynomials play an important role in the following description of 
$\mathrm{Fund}[{\cal WLM}(p,p')]$ in terms of fusion matrices
and in the description of the associated quotient polynomial fusion ring.
The fusion matrix associated to $\ketw{\R}\in\ketw{\Jc_{p,p'}^{\mathrm{Fund}}}$ is
denoted by $N_{\ketw{\R}}$. For convenience of notation, we also introduce the 
fusion matrices $X$ and $Y$ by
\be
 X\ =\ \big(1+\delta_{p,1}\big)N_{\ketw{2,1}},\qquad\quad Y\ =\ N_{\ketw{1,2}}
\label{XYexpli}
\ee
The normalization of $X$ is to ensure universality of notation.
\\[.2cm]
{\bf Proposition \ref{SecFundWLM}.1}\ \ \ Modulo the polynomials $P_{p}(X)$, $P_{p'}(Y)$ and 
$P_{p,p'}(X,Y)$ defined in (\ref{Pn}) and (\ref{Pnn}), the matrices
\be
 N_{\ketw{a,b}}(X,Y)\ =\ U_{a-1}\big(\frac{X}{2}\big)U_{b-1}\big(\frac{Y}{2}\big)
\label{NfundWLM}
\ee
and
\bea
 N_{\ketw{\kappa p,b}}(X,Y)\!\!&=&\!\!
    \frac{1}{\kappa}U_{\kappa p-1}\big(\frac{X}{2}\big)U_{b-1}\big(\frac{Y}{2}\big)\nn
 N_{\ketw{a,\kappa p'}}(X,Y)\!\!&=&\!\!
    \frac{1}{\kappa}U_{a-1}\big(\frac{X}{2}\big)U_{\kappa p'-1}\big(\frac{Y}{2}\big)\nn
 N_{\ketw{\kappa p,p'}}(X,Y)\!\!&=&\!\!
    \frac{1}{\kappa}U_{\kappa p-1}\big(\frac{X}{2}\big)U_{p'-1}\big(\frac{Y}{2}\big)
\label{NWLM}
\eea
and
\bea
 N_{\ketw{\R_{\kappa p,s}^{a,0}}}(X,Y)\!\!&=&\!\!
   2T_{a}\big(\frac{X}{2}\big)N_{\ketw{\kappa p,s}}(X,Y)\nn
 N_{\ketw{\R_{r,\kappa p'}^{0,b}}}(X,Y)\!\!&=&\!\! 
   2N_{\ketw{r,\kappa p'}}(X,Y)T_{b}\big(\frac{Y}{2}\big)\nn
 N_{\ketw{\R_{\kappa p,p'}^{a,b}}}(X,Y)\!\!&=&\!\! 
   4T_{a}\big(\frac{X}{2}\big)N_{\ketw{\kappa p,p'}}(X,Y)T_{b}\big(\frac{Y}{2}\big)
\label{NWR}
\eea
constitute a fusion-matrix realization of the ${\cal W}$-extended fundamental fusion algebra
$\mathrm{Fund}[{\cal WLM}(p,p')]$
with the fusion multiplication $\!\fus\!$ and direct summation
$\oplus$ replaced by matrix multiplication and addition, respectively.
\\[.2cm]
{\bf Proof}\ \ \ First, we address how the separation property (\ref{sep}) is respected by the
proposed fusion matrices. In particular, every fusion matrix can indeed be factored as in the
example
\bea
 N_{\ketw{\R_{\kappa p,p'}^{a,b}}}(X,Y)\!\!&=&\!\! 4T_a\big(\frac{X}{2}\big)\Big(\frac{1}{\kappa}
  U_{\kappa p-1}\big(\frac{X}{2}\big)U_{p'-1}\big(\frac{Y}{2}\big)\Big)T_b\big(\frac{Y}{2}\big)\nn
 &=&\!\! \Big(2T_a\big(\frac{X}{2}\big)N_{\ketw{\kappa p,1}}(X,Y)\Big)
   \Big(2N_{\ketw{1,p'}}(X,Y)T_b\big(\frac{Y}{2}\big)\Big)\nn
 &=&\!\!N_{\ketw{\R_{\kappa p,1}^{a,0}}}(X,Y)N_{\ketw{\R_{1,p'}^{0,b}}}(X,Y)
\eea
In addition, the separation property implicitly assumes the existence of the fusion matrices 
\bea
 N_{\ketw{\R_{\kappa p,\kappa'p'}^{\al,\beta}}}\!\!&=&\!\!
  N_{\ketw{\R_{\kappa p,1}^{\al,0}}}N_{\ketw{\R_{1,\kappa'p'}^{0,\beta}}}\nn
 &=&\!\! \Big(
  \frac{2-\delta_{\al,0}}{\kappa}T_{\al}\big(\frac{X}{2}\big)U_{\kappa p-1}\big(\frac{X}{2}\big)\Big)
  \Big(\frac{2-\delta_{\beta,0}}{\kappa'}T_{\beta}\big(\frac{Y}{2}\big)U_{\kappa' p'-1}\big(\frac{Y}{2}
   \big)\Big)
\eea
Just as the identifications (\ref{identi}) apply to certain representations, the corresponding 
fusion matrices are equivalent to each other modulo the polynomials 
$P_p(X)$, $P_{p'}(Y)$ and $P_{p,p'}(X,Y)$ 
\be
 N_{\ketw{\R_{(\kappa\cdot\kappa')p,p'}^{\al,\beta}}}\ \equiv\ 
   N_{\ketw{\R_{\kappa p,\kappa'p'}^{\al,\beta}}}\ \equiv\ 
   N_{\ketw{\R_{p,(\kappa\cdot\kappa')p'}^{\al,\beta}}}
\label{NNN}
\ee
These congruences are trivial for $\kappa=\kappa'=1$, while they are simple consequences of
$U_{2p-1}\big(\frac{X}{2}\big)U_{p'-1}\big(\frac{Y}{2}\big)
=U_{p-1}\big(\frac{X}{2}\big)U_{2p'-1}\big(\frac{Y}{2}\big)$ for $\kappa\neq\kappa'$.
For $\kappa=\kappa'=2$, they follow from
\bea
 \frac{1}{4}U_{2p-1}\big(\frac{X}{2}\big)U_{2p'-1}\big(\frac{Y}{2}\big)
 &=&\!\!\frac{1}{2}T_p\big(\frac{X}{2}\big)U_{p-1}\big(\frac{X}{2}\big)U_{2p'-1}\big(\frac{Y}{2}\big)
  \ \equiv\ \frac{1}{2}T_p\big(\frac{X}{2}\big)U_{2p-1}\big(\frac{X}{2}\big)U_{p'-1}\big(\frac{Y}{2}\big)\nn
  &=&\!\!\frac{1}{4}\Big(U_{3p-1}\big(\frac{X}{2}\big)+U_{p-1}\big(\frac{X}{2}\big)\Big)
   U_{p'-1}\big(\frac{Y}{2}\big)
  \ \equiv\ U_{p-1}\big(\frac{X}{2}\big)U_{p'-1}\big(\frac{Y}{2}\big)
\label{k2k2}
\eea 
where the second congruence is modulo $P_p(X)$.
Completing the proof now amounts to establishing the proposition for 
the horizontal (described by $X$) and vertical (described by $Y$) components separately. 
Since the ${\cal W}$-fundamental fusion algebra is generated from repeated
fusions of $\ketw{2,1}$ and $\ketw{1,2}$, every fusion can be realized in terms of multiple
fusions of $\ketw{2,1}$ and $\ketw{1,2}$. Aside from performing linear combinations, every step
in such a multiple fusion corresponds to a single fusion by $\ketw{2,1}$ or $\ketw{1,2}$.
Combined with the separation observations, and since $p<p'$, 
it thus suffices to verify the fusions involving $\ketw{2,1}$ in terms of the fusion matrices.
When considering $\ketw{a,1}$ or $\ketw{\R_{\kappa p,1}^{a,0}}$, we are implicitly assuming
that $a\in\mathbb{Z}_{1,p-1}$ and hence $p>1$. We can therefore ignore the normalization in 
(\ref{XYexpli}) in those cases. We now have
\bea
 \ketw{2,1}\fus\ketw{a,1}&\leftrightarrow&XU_{a-1}\big(\frac{X}{2}\big)\ =\ 
  \big(1-\delta_{a,1}\big)U_{a-2}\big(\frac{X}{2}\big)+U_a\big(\frac{X}{2}\big)\nn
 &\leftrightarrow&\big(1-\delta_{a,1}\big)\ketw{a-1,1}\oplus\ketw{a+1,1}
\eea
and
\bea
 \ketw{2,1}\fus\ketw{\kappa p,1}&\leftrightarrow&
   \frac{1}{1+\delta_{p,1}}X\frac{1}{\kappa}U_{\kappa p-1}\big(\frac{X}{2}\big)\nn
   &=&\delta_{p,1}\Big(\frac{1}{2\cdot\kappa}U_{(2\cdot\kappa)p-1}\big(\frac{X}{2}\big)\Big)+
     (1-\delta_{p,1}\big)\frac{2}{\kappa}T_1\big(\frac{X}{2}\big)U_{\kappa p-1}\big(\frac{X}{2}\big)\nn
  &\leftrightarrow&\delta_{p,1}\ketw{2\cdot\kappa,1}
   \oplus\big(1-\delta_{p,1}\big)\ketw{\R_{\kappa p,1}^{1,0}}
\eea
both in accordance with (\ref{21}).
The final fusion to examine is
\be
 \ketw{2,1}\fus\ketw{\R_{\kappa p,1}^{a,0}}\ \leftrightarrow\ 
   X\frac{2}{\kappa}T_a\big(\frac{X}{2}\big)U_{\kappa p-1}\big(\frac{X}{2}\big)
  \ =\ \frac{2}{\kappa}\Big(T_{a-1}\big(\frac{X}{2}\big)+T_{a+1}\big(\frac{X}{2}\big)\Big)
   U_{\kappa p-1}\big(\frac{X}{2}\big)
\ee
For $a+1<p$, this immediately yields
\be
 2\delta_{a,1}\ketw{\kappa p,1}\oplus\big(1-\delta_{a,1}\big)\ketw{\R_{\kappa p,1}^{a-1,0}}
    \oplus\ketw{\R_{\kappa p,1}^{a+1,0}}
\ee
while for $a+1=p$, we have
\bea
&& \frac{2}{\kappa}\Big(T_{a-1}\big(\frac{X}{2}\big)+T_{a+1}\big(\frac{X}{2}\big)\Big)
   U_{\kappa p-1}\big(\frac{X}{2}\big)\nn
 &=&\frac{2}{\kappa}T_{a-1}\big(\frac{X}{2}\big)U_{\kappa p-1}\big(\frac{X}{2}\big)
  +\delta_{\kappa,1}U_{2p-1}\big(\frac{X}{2}\big)+\delta_{\kappa,2}\frac{1}{2}\Big(
   U_{3p-1}\big(\frac{X}{2}\big)+U_{p-1}\big(\frac{X}{2}\big)\Big)\nn
  &\equiv&\frac{2}{\kappa}T_{a-1}\big(\frac{X}{2}\big)U_{\kappa p-1}\big(\frac{X}{2}\big)
    +\delta_{\kappa,1}U_{2p-1}\big(\frac{X}{2}\big)+2\delta_{\kappa,2}U_{p-1}\big(\frac{X}{2}\big)\nn
  &\leftrightarrow&\delta_{a,1}\ketw{\kappa p,1}
    \oplus\big(1-\delta_{a,1}\big)\ketw{\R_{\kappa p,1}^{a-1,0}} \oplus2\ketw{(2\cdot\kappa)p,1}
\label{2TTU}
\eea
Both of these results are in accordance with (\ref{21}).
\\
$\Box$
\\[.2cm]
It is noted that $P_p(X)$ and $P_{p'}(Y)$ are the minimal polynomials of the 
fusion matrices $X$ and $Y$, respectively, while $P_{p,p'}(X,Y)$ governs the merge
of the $X$ and $Y$ components. How this last observation is 
related to the identities (\ref{identi}) is detailed in the discussion of (\ref{NNN}) and (\ref{k2k2}).
As an almost immediate consequence of Proposition \ref{SecFundWLM}.1, 
we have the following proposition.
\\[.2cm]
{\bf Proposition \ref{SecFundWLM}.2}\ \ \ 
The ${\cal W}$-extended fundamental fusion algebra $\mathrm{Fund}[{\cal WLM}(p,p')]$ 
is isomorphic to the polynomial
ring generated by $X$ and $Y$ modulo the ideal $(P_{p}(X),P_{p'}(Y),P_{p,p'}(X,Y))$, that is,
\be
 \big\langle \ketw{1,1},\ketw{2,1},\ketw{1,2}\big\rangle_{p,p'}
   \ \simeq\ \mathbb{C}[X,Y]/\big(P_{p}(X),P_{p'}(Y),P_{p,p'}(X,Y)\big)
\ee
The isomorphism reads
\bea
 \ketw{\R_{a,b}^{0,0}}&\leftrightarrow&\mathrm{pol}_{\ketw{\R_{a,b}^{0,0}}}(X,Y)\ =\ 
   U_{a-1}\big(\frac{X}{2}\big)U_{b-1}\big(\frac{Y}{2}\big)\nn
 \ketw{\R_{\kappa p,b}^{\al,0}}&\leftrightarrow&\mathrm{pol}_{\ketw{\R_{\kappa p,b}^{\al,0}}}(X,Y)\ =\ 
   \frac{2-\delta_{\al,0}}{\kappa}T_\al\big(\frac{X}{2}\big)
    U_{\kappa p-1}\big(\frac{X}{2}\big)U_{b-1}\big(\frac{Y}{2}\big)\nn
 \!\!\!\ketw{\R_{a,\kappa'p'}^{0,\beta}}&\leftrightarrow&
    \mathrm{pol}_{\ketw{\R_{a,\kappa'p'}^{0,\beta}}}(X,Y)\ =\ 
   U_{a-1}\big(\frac{X}{2}\big) 
   \frac{2-\delta_{\beta,0}}{\kappa'}T_\beta\big(\frac{Y}{2}\big)U_{\kappa'p'-1}\big(\frac{Y}{2}\big)\nn
 \ketw{\R_{\kappa p,p'}^{\al,\beta}}&\leftrightarrow&
  \mathrm{pol}_{\ketw{\R_{\kappa p,p'}^{\al,\beta}}}(X,Y)\ =\
  \frac{2-\delta_{\al,0}}{\kappa}T_\al\big(\frac{X}{2}\big)
  U_{\kappa p-1}\big(\frac{X}{2}\big)\big(2-\delta_{\beta,0}\big)
  T_\beta\big(\frac{Y}{2}\big)U_{p'-1}\big(\frac{Y}{2}\big)
\label{abR4}
\eea

\subsection{Fusion matrices and fusion ring of $\mathrm{Out}[{\cal WLM}(p,p')]$}

The fusion-matrix realization and its associated quotient polynomial ring of
$\mathrm{Out}[{\cal WLM}(p,p')]$ follow from those of $\mathrm{Fund}[{\cal WLM}(p,p')]$
in very much the same way as the similar constructions for
$\mathrm{Out}[{\cal LM}(p,p')]$ follow from those for $\mathrm{Fund}[{\cal LM}(p,p')]$,
see Section \ref{FusMat}.
Thus, the ${\cal W}$-extended analogues of Proposition \ref{SecLM}.3 and Proposition \ref{SecLM}.4
are here given as Proposition \ref{SecFundWLM}.3 and Proposition \ref{SecFundWLM}.4.
Their proofs are simple adaptations of the proofs of the two propositions of Section \ref{FusMat}.
\\[.2cm]
{\bf Proposition \ref{SecFundWLM}.3} \ \ \ Let 
$\{\tilde{N}_{\ketw{\R}};\ \ketw{\R}\in\ketw{\Jc_{p,p'}^{\mathrm{Fund}}}\}$ 
be a fusion-matrix
realization of the ${\cal W}$-extended fundamental fusion algebra
$\mathrm{Fund}[{\cal WLM}(p,p')]$ in some basis, that is, some ordering of the elements 
$\ketw{\Jc_{p,p'}^{\mathrm{Fund}}}$. 
For every $\ketw{\R}\in\ketw{\Jc_{p,p'}^{\mathrm{Out}}}$, the matrix 
$N_{\ketw{\R}}$ 
is constructed from the fusion matrix
$\tilde{N}_{\ketw{\R}}\in
\{\tilde{N}_{\ketw{\R}};\ \ketw{\R}\in\ketw{\Jc_{p,p'}^{\mathrm{Out}}}\}
\subseteq\{\tilde{N}_{\ketw{\R}};\ \ketw{\R}\in\ketw{\Jc_{p,p'}^{\mathrm{Fund}}}\}$
by deleting the rows and columns corresponding to
the $(p-1)(p'-1)$ ${\cal W}$-reducible yet ${\cal W}$-indecomposable representations $\ketw{a,b}$.
The set $\{N_{\ketw{\R}};\ \ketw{\R}\in\ketw{\Jc_{p,p'}^{\mathrm{Out}}}\}$
constitutes a fusion-matrix realization of the ${\cal W}$-extended outer fusion algebra
$\mathrm{Out}[{\cal WLM}(p,p')]$.
\\[.2cm]
{\bf Proposition \ref{SecFundWLM}.4} \ \ \ Corresponding to the elements
$\ketw{\R_{i,j}^{\al,\beta}}\in\ketw{\Jc_{p,p'}^{\mathrm{Out}}}$, the polynomials
\bea
 \mathrm{pol}_{\ketw{\R_{\kappa p,b}^{\al,0}}}(X,Y)&=&
   \frac{2-\delta_{\al,0}}{\kappa}T_\al\big(\frac{X}{2}\big)
    U_{\kappa p-1}\big(\frac{X}{2}\big)U_{b-1}\big(\frac{Y}{2}\big)\nn
 \mathrm{pol}_{\ketw{\R_{a,\kappa'p'}^{0,\beta}}}(X,Y)&=&
   U_{a-1}\big(\frac{X}{2}\big) 
   \frac{2-\delta_{\beta,0}}{\kappa'}T_\beta\big(\frac{Y}{2}\big)U_{\kappa'p'-1}\big(\frac{Y}{2}\big)\nn
\mathrm{pol}_{\ketw{\R_{\kappa p,p'}^{\al,\beta}}}(X,Y)&=&
  \frac{2-\delta_{\al,0}}{\kappa}T_\al\big(\frac{X}{2}\big)
  U_{\kappa p-1}\big(\frac{X}{2}\big)\big(2-\delta_{\beta,0}\big)
  T_\beta\big(\frac{Y}{2}\big)U_{p'-1}\big(\frac{Y}{2}\big)
\eea
generate an ideal of the quotient polynomial ring $\mathbb{C}[X,Y]/(P_p(X),P_{p'}(Y),P_{p,p'}(X,Y))$. 
The ${\cal W}$-extended outer fusion algebra $\mathrm{Out}[{\cal WLM}(p,p')]$ 
is isomorphic to this ideal.

\section{Enlarged system}
\label{SecEnlarged}

According to Proposition \ref{SecFundWLM}.1, explicit expressions for $X$ and $Y$ of
$\mathrm{Fund}[{\cal WLM}(p,p')]$ yield the explicit form of all the fusion matrices
by polynomial constructions.
However, even with the fusion rules for the 
${\cal W}$-fundamental representations $\ketw{2,1}$ and $\ketw{1,2}$ at hand,
(\ref{21a}) through (\ref{12}), all natural choices of ordering of the elements of
$\ketw{\Jc_{p,p'}^{\mathrm{Fund}}}$ seem to leave the matrix realization of $X$ or $Y$ rather messy.
We propose to partly circumnavigate this technical problem by enlarging the basis.
Keeping in mind that we are generally dealing with a triplet of interrelated entities
(representations, fusion matrices and ring elements), this corresponds to
the fusion-matrix perspective of an enlarged system. We start by describing this system from
the representation viewpoint.

It is primarily the identities (\ref{identi}) which present a stumbling-block when choosing a suitable basis 
(ordering of the elements of $\ketw{\Jc_{p,p'}^{\mathrm{Fund}}}$) for the fusion-matrix realization of 
$\mathrm{Fund}[{\cal WLM}(p,p')]$.
It is therefore natural to attempt to treat all of the involved entities
as {\em independent} `representations' in the intermediate steps of evaluation. 
The set of these generalized representations is written
\be
 \ketw{\Jc_{p,p'}^{\mathrm{Enl}}}\ =\ \big\{\ketw{\tilde{\R}_{a,b}^{0,0}},\ketw{\tilde{\R}_{\kappa p,b}^{\al,0}},
  \ketw{\tilde{\R}_{a,\kappa'p'}^{0,\beta}},\ketw{\tilde{\R}_{\kappa p,\kappa'p'}^{\al,\beta}}\big\}
\label{JcEnl}
\ee
and has cardinality
\be
 |\ketw{\Jc_{p,p'}^{\mathrm{Enl}}}|\ =\ 
  |\ketw{\Jc_{p,p'}^{\mathrm{Fund}}}|+2pp'\ =\ (3p-1)(3p'-1)\ =\ 9pp'-3p-3p'+1
\ee
We denote a generic element in (\ref{JcEnl}) by $\ketw{\tilde{\R}}$.
The ensuing {\em enlarged fusion algebra} $\mathrm{Enl}[{\cal WLM}(p,p')]$
of the elements of $\ketw{\Jc_{p,p'}^{\mathrm{Enl}}}$
is obtained by applying the separation property (\ref{sep}) and the fusion rules for the horizontal
and vertical components, but without employing the
identities (\ref{identi}). This means that the enlarged fusion algebra is isomorphic to the
tensor product of the horizontal and vertical subalgebras of the fundamental fusion algebra
\be
 \mathrm{Enl}[{\cal WLM}(p,p')]\ \simeq\ 
  \big\langle\ketw{1,1},\ketw{2,1}\big\rangle_{p,p'}\times \big\langle\ketw{1,1},\ketw{1,2}\big\rangle_{p,p'}
\label{112112}
\ee
The horizontal fusion algebra, for example, is actually independent of $p'$ and could therefore
be denoted simply by $\langle\ketw{1,1},\ketw{2,1}\rangle_p$.  
The fusion of two elements of (\ref{JcEnl}) is evaluated as in the formal example
\bea
 \ketw{\tilde{\R}_{i,j}^{\al,\beta}}\tilde{\otimes}\ketw{\tilde{\R}_{i',j'}^{\al',\beta'}}&=&
   \Big(\ketw{\R_{i,1}^{\al,0}}\fus\ketw{\R_{i',1}^{\al',0}}\Big)\fus\Big(\ketw{\R_{1,j}^{0,\beta}}\fus
   \ketw{\R_{1,j'}^{0,\beta'}}\Big)\nn
 &=&\Big(\bigoplus_{i'',\al''}\ketw{\R_{i'',1}^{\al'',0}}\Big)\fus\Big(\bigoplus_{j'',\beta''}
   \ketw{\R_{1,j''}^{0,\beta''}}\Big)\nn
 &=&\bigoplus_{i'',j'',\al'',\beta''}\ketw{\tilde{\R}_{i'',j''}^{\al'',\beta''}}
\label{enlfus}
\eea
where we have introduced $\tilde{\otimes}$ as the fusion multiplication in 
$\mathrm{Enl}[{\cal WLM}(p,p')]$.

{}From the polynomial-ring perspective, this enlargement corresponds to working with 
the tensor-product structure
\be
 \mathbb{C}[X,Y]/\big(P_{p}(X),P_{p'}(Y)\big)\ \simeq\ 
 \Big(\mathbb{C}[X]/\big(P_{p}(X)\big)\Big)\times \Big(\mathbb{C}[Y]/\big(P_{p'}(Y)\big)\Big)
\ee
clearly reminiscent of (\ref{112112}). 
This quotient polynomial ring indeed has
\be
 \mathrm{deg}(P_p)\mathrm{deg}(P_{p'})\ =\ (3p-1)(3p'-1)\ =\ |\ketw{\Jc_{p,p'}^{\mathrm{Enl}}}|
\label{3p3p}
\ee
linearly independent generators, see Appendix \ref{AppQuo}, and corresponds to 
a finite lift of the quotient polynomial
ring $\mathbb{C}[X,Y]/(P_{p}(X),P_{p'}(Y),P_{p,p'}(X,Y))$. 
The isomorphism between 
$\mathrm{Enl}[{\cal WLM}(p,p')]$ and $\mathbb{C}[X,Y]/(P_{p}(X),P_{p'}(Y))$ reads
\bea
 \ketw{\tilde{\R}_{a,b}^{0,0}}&\leftrightarrow&\mathrm{pol}_{\ketw{\tilde{\R}_{a,b}^{0,0}}}(X,Y)\ =\ 
   U_{a-1}\big(\frac{X}{2}\big)U_{b-1}\big(\frac{Y}{2}\big)\nn
 \ketw{\tilde{\R}_{\kappa p,b}^{\al,0}}&\leftrightarrow&
   \mathrm{pol}_{\ketw{\tilde{\R}_{\kappa p,b}^{\al,0}}}(X,Y)\ =\ 
   \frac{2-\delta_{\al,0}}{\kappa}T_\al\big(\frac{X}{2}\big)
    U_{\kappa p-1}\big(\frac{X}{2}\big)U_{b-1}\big(\frac{Y}{2}\big)\nn
 \ketw{\tilde{\R}_{a,\kappa'p'}^{0,\beta}}&\leftrightarrow&
    \mathrm{pol}_{\ketw{\tilde{\R}_{a,\kappa'p'}^{0,\beta}}}(X,Y)\ =\ U_{a-1}\big(\frac{X}{2}\big) 
   \frac{2-\delta_{\beta,0}}{\kappa'}T_\beta\big(\frac{Y}{2}\big)U_{\kappa'p'-1}\big(\frac{Y}{2}\big)\nn
 \!\!\!\!  \ketw{\tilde{\R}_{\kappa p,\kappa'p'}^{\al,\beta}}&\leftrightarrow&
   \mathrm{pol}_{\ketw{\tilde{\R}_{\kappa p,\kappa'p'}^{\al,\beta}}}(X,Y)\ =\ 
  \frac{2-\delta_{\al,0}}{\kappa}T_\al\big(\frac{X}{2}\big)
  U_{\kappa p-1}\big(\frac{X}{2}\big)\frac{2-\delta_{\beta,0}}{\kappa'}
  T_\beta\big(\frac{Y}{2}\big)U_{\kappa'p'-1}\big(\frac{Y}{2}\big)
\label{abR}
\eea
The fusion rules underlying the enlarged fusion algebra $\mathrm{Enl}[{\cal WLM}(p,p')]$
thus also follow from the multiplication rules of
$\mathbb{C}[X,Y]/(P_{p}(X),P_{p'}(Y))$. 

Every ordering of the set (\ref{JcEnl}) provides a basis in which we can realize
the enlarged fusion algebra $\mathrm{Enl}[{\cal WLM}(p,p')]$ in terms of enlarged fusion matrices,
here denoted by $E_{\ketw{\tilde{R}}}$ where $\ketw{\tilde{\R}}\in\ketw{\Jc_{p,p'}^{\mathrm{Enl}}}$.
It follows that
\be
 E_{\ketw{\tilde{R}}}(\Xt,\Yt)\ =\ \mathrm{pol}_{\ketw{\tilde{R}}}(\Xt,\Yt)
\label{EEEE}
\ee
where the fundamental, enlarged fusion matrices $\Xt$ and $\Yt$ are given by  
\be
 \Xt\ =\ \big(1+\delta_{p,1}\big)E_{\ketw{\tilde{\R}_{2,1}^{0,0}}},\qquad\quad 
   \Yt\ =\ E_{\ketw{\tilde{\R}_{1,2}^{0,0}}}
\label{XYenl}
\ee
We stress that $X$ and $Y$ are formal entities in ring considerations such as (\ref{abR}),
while $\Xt$ and $\Yt$ are enlarged fusion matrices in (\ref{EEEE}) and (\ref{XYenl}).

It is recalled that the identities (\ref{identi}), on one hand, and the polynomial $P_{p,p'}(X,Y)$,
on the other, govern the merge of the horizontal and vertical components resulting in
$\mathrm{Fund}[{\cal WLM}(p,p')]$ and 
$\mathbb{C}[X,Y]/(P_{p}(X),P_{p'}(Y),P_{p,p'}(X,Y))$, respectively.

As a side remark, we note that the number of linearly independent generators of the quotient
polynomial ring $\mathbb{C}[X,Y]/(P_{p}(X),P_{p'}(Y))$ given in (\ref{3p3p}) is exactly twice the
dimension of the maximal representation of the modular group discussed in~\cite{FGST0606b}.

\subsection{Fusion matrices revisited}
\label{SecFusMatRev}

To utilize the enlarged basis when discussing the
fusion-matrix realization of $\mathrm{Fund}[{\cal WLM}(p,p')]$, we need to construct
two rectangular matrices: in the language of the associated ring structures, one lifts from
$\mathbb{C}[X,Y]/(P_{p}(X),P_{p'}(Y),P_{p,p'}(X,Y))$ to
$\mathbb{C}[X,Y]/(P_{p}(X),P_{p'}(Y))$; the other projects back down onto
$\mathbb{C}[X,Y]/(P_{p}(X),P_{p'}(Y),P_{p,p'}(X,Y))$.
That is, we wish to devise a pair of matrices $R,L$ intertwining between the matrix realizations
$N_{\ketw{\R}}$ and $E_{\ketw{\tilde{\R}}}$
\be
 N_{\ketw{\R}}\ =\ LE_{\ketw{\tilde{\R}}}R
\label{NLER}
\ee
where $\ketw{\tilde{\R}}\in\ketw{\Jc_{p,p'}^{\mathrm{Enl}}}$ is a lift of (in the class of) 
$\ketw{\R}\in\ketw{\Jc_{p,p'}^{\mathrm{Fund}}}$. Before describing $L$ and $R$, 
let us introduce the projection $\rho$ which maps 
$\ketw{\tilde{\R}}\in\ketw{\Jc_{p,p'}^{\mathrm{Enl}}}$ to 
$\ketw{\R}\in\ketw{\Jc_{p,p'}^{\mathrm{Fund}}}$ if $\ketw{\tilde{\R}}$ is a lift of 
$\ketw{\R}$, that is,
\bea
 \rho\big(\tilde{\R}_{p,p'}^{\al,\beta}\big)\ =\ 
   \rho\big(\tilde{\R}_{2p,2p'}^{\al,\beta}\big)&=&\ketw{\R_{p,p'}^{\al,\beta}}\nn
 \rho\big(\tilde{\R}_{2p,p'}^{\al,\beta}\big)\ =\ 
   \rho\big(\tilde{\R}_{p,2p'}^{\al,\beta}\big)&=&\ketw{\R_{2p,p'}^{\al,\beta}}\nn
 \rho\big(\ketw{\tilde{\R}_{i,j}^{\al,\beta}}\big)&=&\ketw{\R_{i,j}^{\al,\beta}},\qquad\quad
  \ketw{\tilde{\R}_{i,j}^{\al,\beta}}\in\ketw{\Jc_{p,p'}^{\mathrm{Enl}}}
   \setminus\{\ketw{\tilde{\R}_{\kappa p,\kappa'p'}^{\al,\beta}}\}
\label{rho}
\eea  
The map $\rho$ is an epimorphism (surjective homomorphism) from
$\mathrm{Enl}[{\cal WLM}(p,p')]$ to $\mathrm{Fund}[{\cal WLM}(p,p')]$ since 
\be
 \rho\big(\ketw{\tilde{\R}}\tilde{\otimes}\ketw{\tilde{\R}'}\big)\ =\ \rho\big(\ketw{\tilde{\R}}\big)
   \fus\rho\big(\ketw{\tilde{\R}'}\big)
\ee
We also introduce the involution $\g$ whose action on the elements in 
$\ketw{\Jc_{p,p'}^{\mathrm{Enl}}}$ is given by
\bea
 \g\big(\ketw{\tilde{\R}_{\kappa p,\kappa'p'}^{\al,\beta}}\big)&=&
   \ketw{\tilde{\R}_{(2\cdot\kappa)p,(2\cdot\kappa')p'}^{\al,\beta}}\nn
 \g\big(\ketw{\tilde{\R}}\big)&=& \ketw{\tilde{\R}},\qquad\quad \ketw{\tilde{\R}}\in
   \ketw{\Jc_{p,p'}^{\mathrm{Enl}}}\setminus\big\{\ketw{\tilde{\R}_{\kappa p,\kappa'p'}^{\al,\beta}}\big\}
\eea

Selecting bases for the fusion-matrix realizations of 
$\mathrm{Fund}[{\cal WLM}(p,p')]$ and $\mathrm{Enl}[{\cal WLM}(p,p')]$,
corresponds to choosing among the many obvious bijections between
$\ketw{\Jc_{p,p'}^{\mathrm{Fund}}}$ and $\mathbb{Z}_{1,7pp'-3p-3p'+1}$
and between $\ketw{\Jc_{p,p'}^{\mathrm{Enl}}}$ and $\mathbb{Z}_{1,9pp'-3p-3p'+1}$.
These simply describe separate permutations of the elements of $\ketw{\Jc_{p,p'}^{\mathrm{Fund}}}$
and $\ketw{\Jc_{p,p'}^{\mathrm{Enl}}}$. Let $\la,\mu,\nu\in\mathbb{Z}_{1,7pp'-3p-3p'+1}$ and 
$\tilde{\la},\tilde{\mu},\tilde{\nu}\in\mathbb{Z}_{1,9pp'-3p-3p'+1}$
such that $\mu$ and $\tilde{\mu}$, for example, correspond to 
$\ketw{\R}\in\ketw{\Jc_{p,p'}^{\mathrm{Fund}}}$ and a lift 
$\ketw{\tilde{\R}}\in\ketw{\Jc_{p,p'}^{\mathrm{Enl}}}$ thereof, respectively.
We can then express the fusion matrix $N_\la$ in terms of the enlarged fusion matrix $E_{\tilde{\la}}$
as
\be
 {N_{\la,\mu}}^\nu\ =\ \sum_{\tilde{\nu};\ \! \rho(\tilde(\nu)=\nu}
   \Big(c_\mu {E_{\tilde{\la},\tilde{\mu}}}^{\tilde{\nu}}
   +(1-c_\mu){E_{\tilde{\la},\g(\tilde{\mu})}}^{\tilde{\nu}}\Big)
\label{NE}
\ee
where $c_\mu$ is a free parameter for each row index $\mu$. Here, we have extended $\g$ and $\rho$,
by composition with the selected bijection above, to an involution on
the set $\mathbb{Z}_{1,9pp'-3p-3p'+1}$ and to a projection from
$\mathbb{Z}_{1,9pp'-3p-3p'+1}$ to $\mathbb{Z}_{1,7pp'-3p-3p'+1}$.
Without loss of generality, we choose $\tilde{\mu}\leq\g(\tilde{\mu})$.
To write (\ref{NE}) in the matrix form (\ref{NLER}), we introduce the
$|\ketw{\Jc^{\mathrm{Fund}}_{p,p'}}|\times|\ketw{\Jc^{\mathrm{Enl}}_{p,p'}}|$-matrix $L$ with entries
\be
 {L_\mu}^{\tilde{\nu}}\ =\ \begin{cases}  1,\qquad &\mathrm{if}\ \rho(\tilde{\nu})=\mu\ \mathrm{and}
     \ \g(\tilde{\nu})=\tilde{\nu}            \\
    c_\mu,\qquad &\mathrm{if} \    \rho(\tilde{\nu})=\mu\ \mathrm{and}
         \ \g(\tilde{\nu})>\tilde{\nu}        \\
   1-c_\mu, \qquad &\mathrm{if}  \    \rho(\tilde{\nu})=\mu\ \mathrm{and}
         \ \g(\tilde{\nu})<\tilde{\nu}        \\
   0,   &\mathrm{otherwise}     \end{cases}
\label{Lc}
\ee
and the $|\ketw{\Jc^{\mathrm{Enl}}_{p,p'}}|\times|\ketw{\Jc^{\mathrm{Fund}}_{p,p'}}|$-matrix 
$R$ with entries
\be
 {R_{\tilde{\mu}}}^{\nu}\ =\ \begin{cases} 1,\qquad&\mathrm{if}\ \rho(\tilde{\mu})=\nu\\
  0,&\mathrm{otherwise}    \end{cases}
\label{R}
\ee
The matrix $R$ is a sparse binary matrix of full column rank.
We stress that (\ref{NLER}) is valid for every lift $\tilde{\la}$ of $\la$ and recall that there is
either one (if $\g(\tilde{\la})=\tilde{\la}$) or two (if $\g(\tilde{\la})\neq\tilde{\la}$) distinct lifts of $\la$.

\subsubsection{Intertwining by pseudoinverses}

There is a canonical choice for $L$, namely the one where $c_\mu=\frac{1}{2}$ for all $\mu$
in (\ref{Lc}), that is,
\be
 {L_\mu}^{\tilde{\nu}}\ =\ \begin{cases}  1,\qquad &\mathrm{if} \ \rho(\tilde{\nu})=\mu\ \mathrm{and}
     \ \g(\tilde{\nu})=\tilde{\nu}                  \\
   \frac{1}{2}, \qquad &\mathrm{if} \ \rho(\tilde{\nu})=\mu\ \mathrm{and}
     \ \g(\tilde{\nu})\neq\tilde{\nu}             \\
   0,   &\mathrm{otherwise}     \end{cases}
\label{L}
\ee 
This matrix $L$ has full row rank.
The reason for singling out this matrix is explained by Proposition~\ref{SecEnlarged}.1 below.
Before getting to this, 
we recall that for every $n\times m$ matrix $A$, there is
a unique matrix $A^\dagger$ satisfying the four Penrose equations (see~\cite{Penrose}, for example)
\be
 AA^\dagger A\ =\ A,\qquad A^\dagger AA^\dagger\ =\ A^\dagger,\qquad
  (AA^\dagger)^\ast\ = \ AA^\dagger,\qquad (A^\dagger A)^\ast\ =\ A^\dagger A
\ee
where $A^\ast$ denotes the conjugate transpose of $A$.
The matrix $A^\dagger$ is called the Moore-Penrose inverse, or pseudoinverse for short, of $A$.
Clearly, $A^\dagger$ is an $m\times n$ matrix, and if $A$ is nonsingular, then $A^\dagger=A^{-1}$.
It also follows readily that $AA^\dagger$ and $A^\dagger A$ are projection matrices.
Furthermore, if $A$ has full column rank, then $A^\ast A$ is invertible and 
$A^\dagger=(A^\ast A)^{-1}A^\ast$ implying, in particular, that $A^\dagger A=I$.
\\[.2cm]
{\bf Proposition \ref{SecEnlarged}.1}\ \ \ 
The $|\ketw{\Jc^{\mathrm{Fund}}_{p,p'}}|\times|\ketw{\Jc^{\mathrm{Enl}}_{p,p'}}|$-matrix $L$, defined in
(\ref{L}), is the Moore-Penrose inverse of the 
$|\ketw{\Jc^{\mathrm{Enl}}_{p,p'}}|\times|\ketw{\Jc^{\mathrm{Fund}}_{p,p'}}|$-matrix $R$,
defined in (\ref{R}), and is given in terms of $R$ by
\be
 L\ =\ (R^TR)^{-1}R^T
\label{LR}
\ee  
{\bf Proof}\ \ \ 
Since $R$ is real and has full column rank, we have $R^\dagger=(R^TR)^{-1}R^T$. 
A straightforward evaluation of this expression then verifies (\ref{LR}).
\\
$\Box$
\\[.2cm]
That $L$ is the Moore-Penrose inverse of $R$ also follows immediately from the 
observations that
\be
 LR\ =\ I
\ee
where $I$ is the $(7pp'-3p-3p'+1)$-dimensional identity matrix, and that the entries of $RL$ are given by
\be
 {(RL)_{\tilde{\mu}}}^{\tilde{\nu}}\ =\ \begin{cases} 1,\qquad &\mathrm{if}\ \tilde{\nu}=\tilde{\mu}\
      \mathrm{and}\ \g(\tilde{\nu})=\tilde{\nu}\\
  \frac{1}{2}, &\mathrm{if}\ \tilde{\nu}=\tilde{\mu}\
      \mathrm{and}\ \g(\tilde{\nu})\neq\tilde{\nu}\\
  \frac{1}{2}, &\mathrm{if}\ \tilde{\nu}\neq\tilde{\mu}\
      \mathrm{and}\ \g(\tilde{\nu})=\tilde{\mu}\\
 0, &\mathrm{otherwise}    \end{cases}
\label{RLmu}
\ee
This clearly yields a symmetric matrix, as required by one of the Penrose equations, 
since the idempotency of $\g$ ensures that the
condition $\g(\tilde{\nu})=\tilde{\mu}$ is equivalent to $\g(\tilde{\mu})=\tilde{\nu}$.

We can use (\ref{RLmu}) to determine the rank of $RL$, and we find
\be
 rank(RL)\ =\ rank(L)\ =\ rank(R)\ =\ |\ketw{\Jc^{\mathrm{Fund}}_{p,p'}}|
\ee
Since $L$ has full row rank and $R$ has full column rank, this implies that 
$RL$ is written as a full-rank factorization.
The matrix $RL$ also encapsulates rather explicitly the action of $\g$ since
\be
 \big(2RL-I\big)v\ =\ \g(v)
\ee
where $\g$ acts componentwise on the vector $v$ and has been extended by linearity.

\subsection{Explicit fusion matrices}

Recalling the sets $\{\mathrm{pol}_{\ketw{\R}}(X,Y);\ \ketw{\R}\in\ketw{\Jc^{\mathrm{Fund}}_{p,p'}}\}$,
defined in (\ref{abR4}), and 
$\{\mathrm{pol}_{\ketw{\tilde{\R}}}(X,Y);\ \ketw{\tilde{\R}}\in\ketw{\Jc^{\mathrm{Enl}}_{p,p'}}\}$,
defined in (\ref{abR}), we note that
\be
 \big\{\mathrm{pol}_{\ketw{\R}}(X,Y);\ \ketw{\R}\in\ketw{\Jc^{\mathrm{Fund}}_{p,p'}}\big\}\subsetneq
 \big\{\mathrm{pol}_{\ketw{\tilde{\R}}}(X,Y);\ \ketw{\tilde{\R}}\in\ketw{\Jc^{\mathrm{Enl}}_{p,p'}}\big\}
\ee
and that 
\be
 \mathrm{pol}_{\ketw{\tilde{\R}}}(X,Y)\ \equiv\ \mathrm{pol}_{\g(\ketw{\tilde{\R}})}(X,Y),\qquad
 \mathrm{pol}_{\ketw{\tilde{\R}}}(X,Y)\ \equiv\ \mathrm{pol}_{\rho(\ketw{\tilde{\R}})}(X,Y)
\ee
modulo $P_{p,p'}(X,Y)$.
We also observe that the fundamental fusion matrices $X$ and $Y$ are given by
\be
 X\ =\ L\Xt R,\qquad Y\ =\ L\Yt R
\ee
\\[.2cm]
{\bf Lemma \ref{SecEnlarged}.2}\ \ \ For every $\ketw{\tilde{\R}}\in\ketw{\Jc^{\mathrm{Enl}}_{p,p'}}$
and modulo $P_{p,p'}(L\Xt R,L\Yt R)$, we have
\be
 \mathrm{pol}_{\ketw{\tilde{\R}}}(L\Xt R,L\Yt R)\ \equiv\ L\mathrm{pol}_{\ketw{\tilde{\R}}}(\Xt,\Yt)R
\ee
where $L$, $R$, $\Xt$ and $\Yt$ are defined in (\ref{L}), (\ref{R}) and (\ref{XYenl}).
\\[.2cm]
{\bf Proof}\ \ \ 
We not only have
\be
 N_{\rho(\ketw{\tilde{\R}})}\ =\ LE_{\ketw{\tilde{\R}}}R\ =\ L\mathrm{pol}_{\ketw{\tilde{\R}}}(\Xt,\Yt)R
\ee
but also
\be
 N_{\rho(\ketw{\tilde{\R}})}\ =\ \mathrm{pol}_{\rho(\ketw{\tilde{\R}})}(X,Y)\ =\ 
    \mathrm{pol}_{\rho(\ketw{\tilde{\R}})}(L\Xt R,L\Yt R)\ \equiv\ 
    \mathrm{pol}_{\ketw{\tilde{\R}}}(L\Xt R,L\Yt R)
\ee
modulo $P_{p,p'}(L\Xt R,L\Yt R)$.
\\
$\Box$
\\[.2cm]
Due to the tensor-product structure of (\ref{112112}), it is easier to construct the generalized
fusion matrices $\Xt$ and $\Yt$ explicitly than the fusion matrices $X$ and $Y$.
The somewhat intricate combinatorics underlying the construction of
$\Xt$ and $\Yt$ for general $p,p'$ will be discussed elsewhere.
The fusion matrices $N_{\ketw{\R}}$ are subsequently obtained in one of two 
straightforward ways: 
(i) by applying Proposition~\ref{SecFundWLM}.1 to $X=L\Xt L$ and $Y=L\Yt R$, 
or (ii) by sandwiching between $L$ and $R$ the result of applying (\ref{EEEE}) to $\Xt$ 
and $\Yt$ directly.
The matrices $X$ and $Y$ can, of course, also be read off directly from the fusion
rules in Section~\ref{SecWfunRep}.

\section{Fusion-algebra epimorphisms and quotient polynomial rings}
\label{SecQuo}

We have already encountered an example of a commutative diagram like
\be
 \mbox{
 \begin{picture}(100,145)(10,-10)
    \unitlength=0.9cm
  \thinlines
\put(0,4.5){\vector(1,0){4}}
\put(0,0.5){\vector(1,0){4}}
\put(-0.3,4.2){\vector(0,-1){3.4}}
\put(4.3,4.2){\vector(0,-1){3.4}}
\put(1.85,5){$\varphi$}
\put(1.85,-0.2){$\phi$}
\put(-1.01,2.45){$\psi$}
\put(4.7,2.45){$\pi$}
\put(-0.7,4.45){$A$}
\put(4.2,4.45){$Q$}
\put(-1,0.25){$A/\!\!\sim$}
\put(4.2,0.25){$Q/\!\!\equiv$}
 \end{picture}
}
\label{inter}
\ee
with morphisms between fusion algebras and polynomial rings.
Here, $A$ is a fusion algebra, $Q$ is a (quotient polynomial) fusion ring, $\varphi$ and $\phi$
are isomorphisms, $\psi$ is an algebra epimorphism whose kernel is the equivalence relation
$\sim$, while $\pi$ is the projection whose kernel is the equivalence $\equiv$.
Indeed, the epimorphism $\rho$, introduced in Section~\ref{SecFusMatRev}, 
defines the equivalence relation $\sim$
on $\ketw{\Jc^{\mathrm{Enl}}_{p,p'}}$ by $\ketw{\tilde{\R}}\sim\ketw{\tilde{\R}'}$ if
$\rho(\ketw{\tilde{\R}})=\rho(\ketw{\tilde{\R}'})$. This corresponds to imposing the
identities (\ref{identi}) and hence to quotienting the (quotient) polynomial ring
$\mathbb{C}[X,Y]/(P_{p}(X),P_{p'}(Y))$ by the ideal generated by $P_{p,p'}(X,Y)$.
The projection $\pi$ then maps the elements of $\mathbb{C}[X,Y]/(P_{p}(X),P_{p'}(Y))$ to
their respective classes in $\mathbb{C}[X,Y]/(P_{p}(X),P_{p'}(Y),P_{p,p'}(X,Y))$.

The homomorphic property of $\psi$ in (\ref{inter}) is required to ensure 
the intuitive notion of `compatibility' of the two fusion algebras, while the surjectivity 
of $\psi$ merely reflects
that we are dealing with a partitioning into equivalence classes.
If the right (lower) edge of a commutative diagram $D_1$ matches the left (upper) edge of another
commutative diagram $D_2$, the two diagrams can be concatenated horizontally (vertically)
by identifying the matching edges. The horizontal (vertical) 
maps extending across a pair of identified vertical 
(horizontal) edges are obtained by simple compositions: $\varphi_{12}=\varphi_2\circ\varphi_1$
and $\phi_{12}=\phi_2\circ\phi_1$ ($\psi_{12}=\psi_2\circ\psi_1$ and $\pi_{12}=\pi_2\circ\pi_1$).

It is beyond the scope of the present work to examine general commutative diagrams 
of the form (\ref{inter}) related to logarithmic minimal models.
We will not, for example, attempt to classify which projections $\pi$ arise
from taking a quotient with respect to a set of polynomials $\{f_1,\ldots,f_n\}$ in the
elements of $Q$, that is, where the equivalence relation $\equiv$ corresponds
to algebraic congruence modulo the polynomials $f_1,\ldots,f_n$
\be
 \pi:\  Q\ \to\ Q/\big(f_{1},\ldots,f_{n}\big) 
\ee
This section concerns the description of certain commutative diagrams only. 
Of particular interest to us are the situations where both isomorphisms 
and either $\psi$ or $\pi$ are known, in which case the a priori unknown map follows from
\be
 \psi\ =\ \phi^{-1}\circ\pi\circ\varphi,\qquad\quad \pi\ =\ \phi\circ\psi\circ\varphi^{-1}
\label{psipi}
\ee

We conclude this introduction by stating the existence of a `maximal lift' 
of Proposition \ref{SecLM}.2 and Proposition \ref{SecFundWLM}.2 to
\be
 \varphi:\ \big(\mathrm{sl}(2)\times\mathrm{sl}(2)\big)_{p,p'}\ \to\ \mathbb{C}[X,Y]
\label{maxlift}
\ee
whose details and dependence on the labels $p$ and $p'$
will be discussed elsewhere in connection with the combinatorics alluded to 
at the end of Section~\ref{SecEnlarged}. This gives rise to a partial ordering
of the set of pairs $A,Q$ appearing in the commutative diagrams (\ref{inter}).
{}From lattice considerations, however, we believe there exists an extension of the fundamental
fusion algebra, called the ``full fusion algebra" of ${\cal LM}(p,p')$ in~\cite{RP0706,RP0707},
{\em not} obtainable from the maximal lift by an algebra epimorphism. We conjecture that more
variables than $X$ and $Y$ are required in their fusion-ring descriptions, and
hope to address this again elsewhere.

\subsection{$\mathrm{Fund}[{\cal WLM}(p,p')]$ as a quotient of $\mathrm{Fund}[{\cal LM}(p,p')]$}
\label{SecFinite}

As a consequence of Proposition \ref{SecLM}.2 and
Proposition \ref{SecFundWLM}.2, the ${\cal W}$-extended
fundamental fusion algebra can be viewed as a `finitization' or `finite version'
of the fundamental fusion algebra
in the Virasoro picture. The details of this allegory are the content of Proposition~\ref{SecQuo}.1 below.
\\[.2cm]
{\bf Proposition \ref{SecQuo}.1}\ \ \ Let $\varphi$ be the isomorphism
\be 
 \varphi:\ \mathrm{Fund}[{\cal LM}(p,p')]\ \to\ \mathbb{C}[X,Y]/\big(P_{p,p'}(X,Y)\big)
\ee
in Proposition \ref{SecLM}.2, and let $\phi$ be the isomorphism 
\be
 \phi:\ \mathrm{Fund}[{\cal WLM}(p,p')]\ \to\ 
    \mathbb{C}[X,Y]/\big(P_p(X),P_{p'}(Y),P_{p,p'}(X,Y)\big)
\ee
in Proposition \ref{SecFundWLM}.2. With $\pi$ being the projection
\be
 \pi:\  \mathbb{C}[X,Y]/\big(P_{p,p'}(X,Y)\big)\ \to\ 
    \Big(\mathbb{C}[X,Y]/\big(P_{p,p'}(X,Y)\big)\Big)/\big(P_p(X),P_{p'}(Y)\big)
\ee
and $\psi$ the map from $\mathrm{Fund}[{\cal LM}(p,p')]$ to $\mathrm{Fund}[{\cal WLM}(p,p')]$
defined by 
\bea
 \!\!\!\!\psi\big((a,b)\big)\ =\ \ketw{a,b},\quad\!\!\!
&&\psi\big(\R_{(2k-1)p,s}^{\al,\beta}\big)\ =\ (2k-1)\ketw{\R_{p,s}^{\al,\beta}},\qquad
  \psi\big(\R_{2kp,s}^{\al,\beta}\big)\ =\ 2k\ketw{\R_{2p,s}^{\al,\beta}}\nn
&&
  \psi\big(\R_{r,(2k-1)p'}^{\al,\beta}\big)\ =\ (2k-1)\ketw{\R_{r,p'}^{\al,\beta}},\qquad
  \!\psi\big(\R_{r,2kp'}^{\al,\beta}\big)\ =\ 2k\ketw{\R_{r,2p'}^{\al,\beta}}
\label{psiprop}
\eea
we have the commutative diagram 
\be
 \mbox{
 \begin{picture}(100,150)(30,-10)
    \unitlength=0.9cm
  \thinlines
\put(0,4.5){\vector(1,0){4}}
\put(0,0.5){\vector(1,0){4}}
\put(-0.3,4.2){\vector(0,-1){3.4}}
\put(4.3,4.2){\vector(0,-1){3.4}}
\put(1.85,5){$\varphi$}
\put(1.85,-0.2){$\phi$}
\put(-1.01,2.45){$\psi$}
\put(4.7,2.45){$\pi$}
\put(-3.2,4.45){$\mathrm{Fund}[{\cal LM}(p,p')]$}
\put(4.2,4.45){$\mathbb{C}[X,Y]/\big(P_{p,p'}(X,Y)\big)$}
\put(-3.65,0.25){$\mathrm{Fund}[{\cal WLM}(p,p')]$}
\put(4.2,0.25){$\mathbb{C}[X,Y]/\big(P_p(X),P_{p'}(Y),P_{p,p'}(X,Y)\big)$}
 \end{picture}
}
\label{LMinter}
\ee
\\[.2cm]
{\bf Proof}\ \ \ Since the form of the lower-right corner is an immediate consequence of
\be
 \Big(\mathbb{C}[X,Y]/\big(P_{p,p'}(X,Y)\big)\Big)/\big(P_p(X),P_{p'}(Y)\big)\ \simeq\ 
 \mathbb{C}[X,Y]/\big(P_p(X),P_{p'}(Y),P_{p,p'}(X,Y)\big)
\ee
completing the proof amounts to demonstrating that the map $\psi$ in (\ref{psiprop})
is the algebra epimorphism from $\mathrm{Fund}[{\cal LM}(p,p')]$ to 
$\mathrm{Fund}[{\cal WLM}(p,p')]$ satisfying (\ref{psipi}).
We straightforwardly have
\be
 \phi^{-1}\circ\pi\circ\varphi\big((a,b)\big)\ =\ \phi^{-1}\circ\pi\Big(U_{a-1}\big(\frac{X}{2}\big)
   U_{b-1}\big(\frac{Y}{2}\big)\Big)\ =\ \phi^{-1}\Big(U_{a-1}\big(\frac{X}{2}\big)
   U_{b-1}\big(\frac{Y}{2}\big)\Big)\ =\ \ketw{a,b}
\ee
in accordance with the first equality of (\ref{psiprop}). Using Lemma~\ref{AppCheb}.1,
we also have
\bea
 \phi^{-1}\circ\pi\circ\varphi\big(\R_{(2k-1)p,s}^{\al,\beta}\big)
 &=&\phi^{-1}\circ\pi\Big(\big(2-\delta_{\al,0}\big)T_\al\big(\frac{X}{2}\big)U_{(2k-1)p-1}\big(\frac{X}{2}\big)
   \big(2-\delta_{\beta,0}\big)T_\beta\big(\frac{Y}{2}\big)U_{s-1}\big(\frac{Y}{2}\big)\Big)\nn
 &\equiv&\phi^{-1}\circ\pi\Big(\big(2-\delta_{\al,0}\big)T_\al\big(\frac{X}{2}\big)
   (2k-1)U_{p-1}\big(\frac{X}{2}\big)
   \big(2-\delta_{\beta,0}\big)T_\beta\big(\frac{Y}{2}\big)U_{s-1}\big(\frac{Y}{2}\big)\Big)\nn
 &=&(2k-1)\phi^{-1}\Big(\big(2-\delta_{\al,0}\big)T_\al\big(\frac{X}{2}\big)U_{p-1}\big(\frac{X}{2}\big)
   \big(2-\delta_{\beta,0}\big)T_\beta\big(\frac{Y}{2}\big)U_{s-1}\big(\frac{Y}{2}\big)\Big)\nn
 &=&(2k-1)\ketw{\R_{p,s}^{\al,\beta}}
\eea
and
\bea
 \phi^{-1}\circ\pi\circ\varphi\big(\R_{2kp,s}^{\al,\beta}\big)
 &=&\phi^{-1}\circ\pi\Big(\big(2-\delta_{\al,0}\big)T_\al\big(\frac{X}{2}\big)U_{2kp-1}\big(\frac{X}{2}\big)
   \big(2-\delta_{\beta,0}\big)T_\beta\big(\frac{Y}{2}\big)U_{s-1}\big(\frac{Y}{2}\big)\Big)\nn
 &\equiv&\phi^{-1}\circ\pi\Big(\big(2-\delta_{\al,0}\big)T_\al\big(\frac{X}{2}\big)
   kU_{2p-1}\big(\frac{X}{2}\big)
   \big(2-\delta_{\beta,0}\big)T_\beta\big(\frac{Y}{2}\big)U_{s-1}\big(\frac{Y}{2}\big)\Big)\nn
 &=&k\phi^{-1}\Big(\big(2-\delta_{\al,0}\big)T_\al\big(\frac{X}{2}\big)U_{2p-1}\big(\frac{X}{2}\big)
   \big(2-\delta_{\beta,0}\big)T_\beta\big(\frac{Y}{2}\big)U_{s-1}\big(\frac{Y}{2}\big)\Big)\nn
 &=&2k\ketw{\R_{2p,s}^{\al,\beta}}
\eea
in accordance with the second and third equalities of (\ref{psiprop}). The remaining two equalities
of (\ref{psiprop}) follow similarly. 
\\
$\Box$

\subsection{A quotient of $\mathrm{Fund}[{\cal WLM}(p,p')]$}

In this section, we let $A=\mathrm{Fund}[{\cal WLM}(p,p')]$, 
$Q=\mathbb{C}[X,Y]/(P_p(X),P_{p'}(Y),P_{p,p'}(X,Y))$ 
and the isomorphism $\varphi$ in (\ref{inter}) be given as in Proposition~\ref{SecFundWLM}.2.
To each partitioning of $\ketw{\Jc^{\mathrm{Fund}}_{p,p'}}$, there is an equivalence relation 
on $A$ and an associated set of equivalence classes.
Many of these do not seem to be of interest from a fusion-algebraic point of view, but some are.
For simplicity, we here restrict ourselves to situations where
$\pi$ in (\ref{inter}) is the projection corresponding to a triplet of polynomials
$f_1(X)$, $f_2(Y)$ and $f_3(X,Y)$ which are divisors of $P_p(X)$, $P_{p'}(Y)$ and 
$P_{p,p'}(X,Y)$, respectively,
\be
 f_1(X)|P_p(X),\qquad f_2(Y)|P_{p'}(Y),\qquad f_3(X,Y)|P_{p,p'}(X,Y)
\label{fP}
\ee
It then follows that
\be
 Q/\big(f_1(X),f_2(Y),f_3(X,Y)\big) \ \simeq\ \mathbb{C}[X,Y]/\big(f_1(X),f_2(Y),f_3(X,Y)\big)
\label{Qfff}
\ee
and a natural objective would be to determine an admissible triplet $\psi$, $A/\!\!\sim$ and $\phi$
completing the commutative diagram (\ref{inter}). Such a triplet is in general not unique.

The example to be discussed here could be introduced without motivation. 
According to~\cite{PR0812}, though, it does
play an important role as the algebra of the fusion matrices obtained from a Verlinde formula
applied to the modular $S$ matrix of the set of so-called projective characters 
in ${\cal WLM}(p,p')$~\cite{FGST0606b,Ras0805}.
This algebra is also related to the fusion algebra arising from lattice considerations 
when omitting the so-called disentangling procedure~\cite{Ras0805}, see Section~\ref{SecDis} below.

Here, we leave this motivation aside and simply introduce the $(p+1)(p'+1)$ generators
\be
 \Hc_{i,j},\qquad i\in\mathbb{Z}_{0,p},\quad j\in\mathbb{Z}_{0,p'}
\label{Hij}
\ee
subject to
\be
 \Hc_{i,j}\ =\ \Hc_{p-i,p'-j}
\label{Hcond}
\ee
This yields a set of $\frac{1}{2}(p+1)(p'+1)$ linearly independent generators of the algebra
\be
 \big\langle \Hc_{i,j} \big\rangle_{p,p'}
\label{Halg}
\ee
whose multiplication is denoted by $\ast$. In due time, this is the algebra to be identified with
$A/\!\!\sim$ in the lower-left corner of the commutative diagram (\ref{inter}).
Two of the generators of (\ref{Halg}) are special and are also denoted by
\be
 I\ =\ \Hc_{0,0}\ =\ \Hc_{p,p'},\qquad C\ =\ \Hc_{p,0}\ =\ \Hc_{0,p'}
\ee
The associated multiplication rules are
\be
 I\ast \Hc_{i,j}\ =\ \Hc_{i,j},\qquad C\ast \Hc_{i,j}\ =\ \Hc_{p-i,j}\ =\ \Hc_{i,p'-j}
\ee
in particular,
\be
 C\ast C\ =\ I
\ee
and we recognize $I$ as the identity of the $\Hc$-algebra while $C$ is seen to act as a conjugation
operator.
The horizontal and vertical multiplication rules (of the $p-1$ and $p'-1$ 
generators $\Hc_{a,0}$ and $\Hc_{0,b}$, respectively) are given by
\bea
 \Hc_{a,0}\ast \Hc_{a',0}&=&\big(1+\delta_{a,a'}\big)\Hc_{|a-a'|,0}
   +\big(1+\delta_{a+a',p}\big)H_{p-|p-a-a'|,0}\nn
 \Hc_{0,b}\ast \Hc_{0,b'}&=&\big(1+\delta_{b,b'}\big)\Hc_{0,|b-b'|}
   +\big(1+\delta_{b+b',p'}\big)\Hc_{0,p'-|p'-b-b'|}
\eea
The two sectors are merged according to
\be
 \Hc_{i,0}\ast \Hc_{0,j}\ =\ \Hc_{i,j}
\label{HHH}
\ee
thus resulting in the multiplication rules (for the $\frac{1}{2}(p-1)(p'-1)$ generators $\Hc_{a,b}$)
\bea
 \Hc_{a,b}\ast \Hc_{a',b'}&=&\big(1+\delta_{a,a'}\big)\big(1+\delta_{b,b'}\big)\Hc_{|a-a'|,|b-b'|}
   +\big(1+\delta_{a+a',p}\big)\big(1+\delta_{b+b',p'}\big)\Hc_{p-|p-a-a'|,p'-|p'-b-b'|}\nn
  &+&\big(1+\delta_{a,a'}\big)\big(1+\delta_{b+b',p'}\big)\Hc_{|a-a'|,p'-|p'-b-b'|}
    +\big(1+\delta_{a+a',p}\big)\big(1+\delta_{b,b'}\big)\Hc_{p-|p-a-a'|,|b-b'|}\nn
\eea
It follows that this $\Hc$-algebra is generated from repeated multiplications of the two fundamental
generators $\Hc_{1,0}$ and $\Hc_{0,1}$ in addition to $I$ or equivalently $C$
\be
  \big\langle \Hc_{i,j} \big\rangle_{p,p'}\ =\  \big\langle I, \Hc_{1,0}, \Hc_{0,1} \big\rangle_{p,p'}
  \ =\ \big\langle C, \Hc_{1,0}, \Hc_{0,1} \big\rangle_{p,p'}
\ee

Turning to the ring side of affairs, we introduce the polynomials
\be
 H_n(x)\ =\ T_{n+1}\big(\frac{x}{2}\big)-T_{n-1}\big(\frac{x}{2}\big)
  \ =\ \Big(T_2\big(\frac{x}{2}\big)-1\Big)U_{n-1}\big(\frac{x}{2}\big)\ =\ 
   \frac{1}{2}(x^2-4)U_{n-1}\big(\frac{x}{2}\big)
\label{Hn}
\ee
and 
\be
 H_{n,n'}(x,y)\ =\ T_n\big(\frac{x}{2}\big)-T_{n'}\big(\frac{y}{2}\big)\ =\ 
  \frac{1}{2}\Big(U_n\big(\frac{x}{2}\big)-U_{n-2}\big(\frac{x}{2}\big)\Big)
   -\frac{1}{2}\Big(U_{n'}\big(\frac{y}{2}\big)-U_{n'-2}\big(\frac{y}{2}\big)\Big)
\label{Hnn}
\ee
We note that they divide the polynomials $P_n(x)$ and $P_{n,n'}(x,y)$, respectively, since it follows
from (\ref{Pnn}) and (\ref{Pn}) that
\be
 P_n(x)\ =\ 2U_{n-1}^2\big(\frac{x}{2}\big)H_n(x),\qquad\quad
 P_{n,n'}(x,y)\ =\ U_{n-1}\big(\frac{x}{2}\big)U_{n'-1}\big(\frac{y}{2}\big)H_{n,n'}(x,y)
\label{PHPH}
\ee
\\[.2cm]
{\bf Proposition \ref{SecQuo}.2}\ \ \ The $\Hc$-algebra is isomorphic to the polynomial ring 
generated by $X$ and $Y$ modulo the ideal $(H_p(X),H_{p'}(Y),H_{p,p'}(X,Y))$, that is, 
\be
 \big\langle I, \Hc_{1,0}, \Hc_{0,1} \big\rangle_{p,p'}\ \simeq\ 
   \mathbb{C}[X,Y]/\big(H_p(X),H_{p'}(Y),H_{p,p'}(X,Y)\big)
\ee
The isomorphism reads
\be 
 \Hc_{i,j}\ \leftrightarrow\ \mathrm{pol}_{\Hc_{i,j}}(X,Y)\ =\ 
   \big(2-\delta_{i,0}-\delta_{i,p}\big)T_i\big(\frac{X}{2}\big)
   \big(2-\delta_{j,0}-\delta_{j,p'}\big)T_j\big(\frac{Y}{2}\big)
\label{HcTT}
\ee
{\bf Proof}\ \ \ The map (\ref{HcTT}) provides an obvious bijection between the tensor-product structure
\be
 \mathbb{C}[X,Y]/\big(H_p(X),H_{p'}(Y)\big)\ \simeq\ \Big(\mathbb{C}[X]/\big(H_p(X)\big)\Big)\times
   \Big(\mathbb{C}[Y]/\big(H_{p'}(Y)\big)\Big)
\ee
and the lift of the $\Hc$-algebra defined by {\em not} imposing the condition (\ref{Hcond}).
Using Lemma~\ref{AppCheb}.2, we see that imposing (\ref{Hcond}) corresponds to quotienting
by the ideal $(H_{p,p'}(X,Y))$ since
\bea
 \mathrm{pol}_{\Hc_{p-i,p'-j}}(X,Y)&=&\big(2-\delta_{i,0}-\delta_{i,p}\big)T_{p-i}\big(\frac{X}{2}\big)
   \big(2-\delta_{j,0}-\delta_{j,p'}\big)T_{p'-j}\big(\frac{Y}{2}\big)\nn
 &\equiv&\big(2-\delta_{i,0}-\delta_{i,p}\big)T_{p}\big(\frac{X}{2}\big)T_{i}\big(\frac{X}{2}\big)
   \big(2-\delta_{j,0}-\delta_{j,p'}\big)T_{p'-j}\big(\frac{Y}{2}\big),\qquad (\mathrm{mod}\ H_p(X))\nn
 &\equiv&\big(2-\delta_{i,0}-\delta_{i,p}\big)T_{i}\big(\frac{X}{2}\big)
   \big(2-\delta_{j,0}-\delta_{j,p'}\big)T_{p'}\big(\frac{Y}{2}\big)T_{p'-j}\big(\frac{Y}{2}\big),
      \qquad\! (\mathrm{mod}\ H_{p,p'}(X,Y))\nn
&\equiv&\big(2-\delta_{i,0}-\delta_{i,p}\big)T_{i}\big(\frac{X}{2}\big)
   \big(2-\delta_{j,0}-\delta_{j,p'}\big)T_{j}\big(\frac{Y}{2}\big),
      \qquad\qquad\qquad\ (\mathrm{mod}\ H_{p'}(Y))\nn
 &=&\mathrm{pol}_{\Hc_{i,j}}(X,Y)
\eea
In addition, we observe that 
$\mathrm{pol}_{I\ast\Hc_{i,0}}(X)=\mathrm{pol}_I(X)\mathrm{pol}_{\Hc_{i,0}}(X)$,
\bea
 \mathrm{pol}_{C\ast\Hc_{i,0}}(X)&=&\mathrm{pol}_{\Hc_{p-i,0}}(X)\ =\ \big(2-\delta_{i,0}-\delta_{i,p}\big)
   T_{p-i}\big(\frac{X}{2}\big)\ \equiv\ T_p\big(\frac{X}{2}\big)\big(2-\delta_{i,0}-\delta_{i,p}\big)
   T_{i}\big(\frac{X}{2}\big)\nn
 &=&\mathrm{pol}_{C}(X)\mathrm{pol}_{\Hc_{i,0}}(X)
\eea
and
\bea
 \mathrm{pol}_{\Hc_{a,0}\ast\Hc_{a',0}}(X)&=&\big(1+\delta_{a,a'}\big)\mathrm{pol}_{\Hc_{|a-a'|,0}}(X)
  +\big(1+\delta_{a+a',p}\big)\mathrm{pol}_{\Hc_{p-|p-a-a'|,0}}(X)\nn
 &=&2T_{|a-a'|}\big(\frac{X}{2}\big)+2T_{p-|p-a-a'|}\big(\frac{X}{2}\big)
 \ \equiv\ 2T_a\big(\frac{X}{2}\big)2T_{a'}\big(\frac{X}{2}\big)\nn
 &=&\mathrm{pol}_{\Hc_{a,0}}(X)\mathrm{pol}_{\Hc_{a',0}}(X)
\eea
where the congruences are modulo $H_p(X)$. This establishes the homomorphic property of the map on the horizontal component, with the vertical component following similarly.
Finally, the polynomial analogue of the separation property (\ref{HHH}) trivially reads
\be
 \mathrm{pol}_{\Hc_{i,0}(X)}\mathrm{pol}_{\Hc_{0,j}}(Y)\ =\ 
   \mathrm{pol}_{\Hc_{i,j}}(X,Y)
\ee
%
%
$\Box$
\\[.2cm]
We note that $X$ and $Y$ are the isomorphic images of the two
fundamental generators 
\be
 \Hc_{1,0}\ \leftrightarrow\ \frac{1}{1+\delta_{p,1}}X,\qquad \Hc_{0,1}\ \leftrightarrow\ Y
\ee
We also note that the polynomial $H_{p,p'}(X,Y)=T_p(X)-T_{p'}(Y)$ corresponds to equating
the horizontal and vertical ways of realizing the conjugation $C$
\be
 T_p(X)\ \leftrightarrow\ \Hc_{p,0}\ =\ C\ =\ \Hc_{0,p'}\ \leftrightarrow\ T_{p'}(Y)
\ee
%
%
{\bf Proposition \ref{SecQuo}.3}\ \ \ 
Let $\varphi$ be the isomorphism 
\be
 \varphi:\ \mathrm{Fund}[{\cal WLM}(p,p')]\ \to\ \mathbb{C}[X,Y]/\big(P_p(X),P_{p'}(Y),P_{p,p'}(X,Y)\big)
\ee 
in Proposition~\ref{SecFundWLM}.2, and let $\phi$ be the isomorphism
\be
 \phi:\ \big\langle I, \Hc_{1,0}, \Hc_{0,1} \big\rangle_{p,p'}\ \to\ 
   \mathbb{C}[X,Y]/\big(H_p(X),H_{p'}(Y),H_{p,p'}(X,Y)\big)
\ee 
in Proposition~\ref{SecQuo}.2. With $\pi$ being the projection
\bea
 &&\pi:\  \mathbb{C}[X,Y]/\big(P_p(X),P_{p'}(Y),P_{p,p'}(X,Y)\big)\nn
 &&\quad\qquad\to\ 
    \Big(\mathbb{C}[X,Y]/\big(P_p(X),P_{p'}(Y),P_{p,p'}(X,Y)\big)\Big)/\big(H_p(X),H_{p'}(Y),H_{p,p'}(X,Y)\big)
\eea
and $\psi$ the map from $\mathrm{Fund}[{\cal WLM}(p,p')]$ to 
$\langle I, \Hc_{1,0}, \Hc_{0,1} \rangle_{p,p'}$ defined by 
\bea
 \psi\big(\ketw{a,b}\big)&=&\sum_{i}^{a-1}\ 
   \sum_{j}^{b-1}\Hc_{i,j}
    \nn
\psi\big(\ketw{\R_{\kappa p,b}^{\al,0}}\big)&=&\big(2-\delta_{\al,0}\big)
  \sum_{i}^{p+\eps(\al+(\kappa-1)p)-1}\ \ \!\!
  \sum_{j}^{b-1}\Hc_{i,j}
  \nn
 \psi\big(\ketw{\R_{a,\kappa p'}^{0,\beta}}\big)&=&\big(2-\delta_{\beta,0}\big) 
   \sum_{i}^{a-1}\ \ \!\!
   \sum_{j}^{p'+\eps(\beta+(\kappa-1)p')-1}\Hc_{i,j}
  \nn
 \psi\big(\ketw{\R_{\kappa p,p'}^{\al,\beta}}\big)&=&\big(2-\delta_{\al,0}\big) \big(2-\delta_{\beta,0}\big) 
   \sum_{i}^{p+\eps(\al+(\kappa-1)p)-1}\ \ \!\!
   \sum_{j}^{p'+\eps(\beta)-1}\Hc_{i,j}
\label{psiprop3}
\eea
we have the commutative diagram 
\be
 \mbox{
 \begin{picture}(100,150)(40,-10)
    \unitlength=0.9cm
  \thinlines
\put(0,4.5){\vector(1,0){4}}
\put(0,0.5){\vector(1,0){4}}
\put(-0.3,4.2){\vector(0,-1){3.4}}
\put(4.3,4.2){\vector(0,-1){3.4}}
\put(1.85,5){$\varphi$}
\put(1.85,-0.2){$\phi$}
\put(-1.01,2.45){$\psi$}
\put(4.7,2.45){$\pi$}
\put(-3.65,4.45){$\mathrm{Fund}[{\cal WLM}(p,p')]$}
\put(4.2,4.45){$\mathbb{C}[X,Y]/\big(P_p(X),P_{p'}(Y),P_{p,p'}(X,Y)\big)$}
\put(-3.1,0.25){$\big\langle I, \Hc_{1,0}, \Hc_{0,1} \big\rangle_{p,p'}$}
\put(4.2,0.25){$\mathbb{C}[X,Y]/\big(H_p(X),H_{p'}(Y),H_{p,p'}(X,Y)\big)$}
 \end{picture}
}
\label{Hinter}
\ee
\\[.2cm]
{\bf Proof}\ \ \ It follows from the division properties indicated in (\ref{PHPH}),
see (\ref{fP}) and (\ref{Qfff}), that
\bea
 &&\Big(\mathbb{C}[X,Y]/\big(P_{p}(X),P_{p'}(Y),P_{p,p'}(X,Y)\big)\Big)/
    \big(H_p(X),H_{p'}(Y),H_{p,p'}(X,Y)\big)\nn
 &&\qquad\qquad\qquad\simeq\  \mathbb{C}[X,Y]/\big(H_p(X),H_{p'}(Y),H_{p,p'}(X,Y)\big)
\eea
giving rise to the form of the lower-right corner of (\ref{Hinter}).
We also have
\be
 \phi^{-1}\circ\pi\circ\varphi\big(\ketw{a,b}\big)
 \ =\ \phi^{-1}\circ\pi\big(\mathrm{pol}_{\ketw{a,b}}\big)
 \ =\ \phi^{-1}\big(U_{a-1}\big(\frac{X}{2}\big)U_{b-1}\big(\frac{Y}{2}\big)\big)
 \ =\ \sum_{i}^{a-1}\
   \sum_{j}^{b-1}\Hc_{i,j}
\ee
and
\bea
 \phi^{-1}\circ\pi\circ\varphi\big(\ketw{\R_{\kappa p,b}^{\al,0}}\big)&=&\phi^{-1}\circ\pi\big(
  \mathrm{pol}_{\ketw{\R_{\kappa p,b}^{\al,0}}}(X,Y)\big)\nn
  &=&\frac{2-\delta_{\al,0}}{2\kappa}\phi^{-1}\circ\pi\Big(2T_{\al}\big(\frac{X}{2}\big)
    U_{\kappa p-1}\big(\frac{X}{2}\big)U_{b-1}\big(\frac{Y}{2}\big)\Big)\nn
  &=&\big(2-\delta_{\al,0}\big)
  \sum_{i}^{p+\eps(\al+(\kappa-1)p)-1}\ \ \!
  \sum_{j}^{b-1}\Hc_{i,j}
\eea
where we have used Lemma~\ref{AppCheb}.3, in particular. These results are
in accordance with (\ref{psiprop3}), and the remaining two expressions in (\ref{psiprop3})
are recovered similarly.
\\
$\Box$
\\[.2cm]
Recalling (\ref{identi}), we note that
\be
 \psi\big(\ketw{\R_{p,2p'}^{\al,\beta}}\big)\ =\ \psi\big(\ketw{\R_{2p,p'}^{\al,\beta}}\big),\qquad
 \psi\big(\ketw{\R_{2p,2p'}^{\al,\beta}}\big)\ =\ \psi\big(\ketw{\R_{p,p'}^{\al,\beta}}\big)
\ee
are consequences of
\bea
 \sum_{i}^{p+\eps(\al+(\kappa-1)p)-1}\ \ \!\!
   \sum_{j}^{p'+\eps(\beta+(\kappa'-1)p')-1}\Hc_{i,j}
  &=&\sum_{i}^{p+\eps(\al+(\kappa-1)p)-1}\ \ \!\!
   \sum_{j}^{p'+\eps(\beta+(\kappa'-1)p')-1}\Hc_{p-i,p'-j}\nn
  &=&\sum_{i}^{p+\eps(\al+\kappa p)-1}\ \ \!\!
   \sum_{j}^{p'+\eps(\beta+\kappa'p')-1}\Hc_{i,j}
\label{sumid}
\eea
where the two rewritings are due to (\ref{Hcond}) and a change of summation variables, respectively.

\subsubsection{Without disentangling in the lattice approach}
\label{SecDis}

The identities (\ref{sumid}) are indicators of an underlying structure. To appreciate this, we introduce
the linear combinations
\be
 \Gamma^{p-1,0}\ =\ \sum_i^{p-1}\Hc_{i,0},\qquad
 \Gamma^{p,0}\ =\ \sum_i^p\Hc_{i,0},\qquad
 \Gamma^{0,p'-1}\ =\ \sum_j^{p'-1}\Hc_{0,j},\qquad
 \Gamma^{0,p'}\ =\ \sum_j^{p'}\Hc_{0,j}
\label{Gamma}
\ee
satisfying 
\be
 C\ast\Gamma^{p-1,0}\ =\ \begin{cases} \Gamma^{p-1,0},\quad&p\ \mathrm{even}\\
    \Gamma^{p,0},\quad&p\ \mathrm{odd}  \end{cases},\qquad\qquad
 C\ast\Gamma^{p,0}\ =\ \begin{cases} \Gamma^{p,0},\quad&p\ \mathrm{even}\\
    \Gamma^{p-1,0},\quad&p\ \mathrm{odd}  \end{cases}
\ee 
with similar results for the vertical fusions $C\ast\Gamma^{0,p'-1}$ and $C\ast\Gamma^{0,p'}$.
We also find that for $p$ even
\be
 \Gamma^{p,0}\ast\Gamma^{p,0}\ =\ \Gamma^{p-1,0}\ast\Gamma^{p-1,0}\ =\ p\Gamma^{p,0},\qquad
 \Gamma^{p,0}\ast\Gamma^{p-1,0}\ =\ p\Gamma^{p-1,0}
\ee
while for $p$ odd
\be
 \Gamma^{p,0}\ast\Gamma^{p,0}\ =\ \Gamma^{p-1,0}\ast\Gamma^{p-1,0}\ =\ p\Gamma^{p-1,0},\qquad
 \Gamma^{p,0}\ast\Gamma^{p-1,0}\ =\ p\Gamma^{p,0}
\ee
again with similar results for the vertical components.
With $\Gamma^{\ell,\ell'}$ defined by
\be
 \Gamma^{\ell,\ell'}\ =\ \Gamma^{\ell,0}\ast\Gamma^{0,\ell'}\ =\ \sum_i^{\ell}\sum_{j}^{\ell'}
  \Hc_{i,j},\qquad\quad \ell\in\mathbb{Z}_{p-1,p},\quad\ell'\in\mathbb{Z}_{p'-1,p'}
\label{Gll}
\ee
it follows from $C\ast C=I$ that
\be
 \Gamma^{\ell,\ell'}\ =\ \big(C\ast\Gamma^{\ell,0}\big)\ast\big(C\ast\Gamma^{0,\ell'}\big)
\ee
and hence that
\bea
 \Gamma^{p,p'}\ =\ \Gamma^{p,p'-1},\qquad \Gamma^{p-1,p'}\ =\ \Gamma^{p-1,p'-1},\qquad\quad&&
   p\ \mathrm{even},\ p'\ \mathrm{odd}\nn
 \Gamma^{p,p'}\ =\ \Gamma^{p-1,p'},\qquad \Gamma^{p,p'-1}\ =\ \Gamma^{p-1,p'-1},\qquad\quad&&
   p\ \mathrm{odd},\ p'\ \mathrm{even}\nn
 \Gamma^{p,p'}\ =\ \Gamma^{p-1,p'-1},\qquad \Gamma^{p-1,p'}\ =\ \Gamma^{p,p'-1},\qquad\quad&&
   p,p'\ \mathrm{odd}
\label{GG}
\eea
There are thus 6 linearly independent $\Gamma$'s for $p>1$ but only 4 for $p=1$.
Cayley tables of the associated fusion algebras are easily completed and depend
on the parities of $p$ and $p'$.

We now recognize the identities (\ref{sumid}) as the sums behind (\ref{GG}) and furthermore observe
that
\be
 \psi\big(\ketw{\R_{\kappa p,1}^{\al,0}}\big)\ =\ \big(2-\delta_{\al,0}\big)
  \Gamma^{p+\eps(\al+(\kappa-1)p)-1,0},\qquad
  \psi\big(\ketw{\R_{1,\kappa p'}^{0,\beta}}\big)\ =\ \big(2-\delta_{\beta,0}\big)
  \Gamma^{0,p'+\eps(\beta+(\kappa-1)p')-1}
\label{psiRG}
\ee
and, in accordance with (\ref{Gll}), that
\bea
  \psi\big(\ketw{\R_{\kappa p,p'}^{\al,\beta}}\big)&=&\big(2-\delta_{\al,0}\big) \big(2-\delta_{\beta,0}\big) 
   \Gamma^{p+\eps(\al+(\kappa-1)p)-1,p'+\eps(\beta)-1}\nn
  &=&\big(2-\delta_{\al,0}\big) \big(2-\delta_{\beta,0}\big) 
   \Gamma^{p+\eps(\al)-1,p'+\eps(\beta+(\kappa-1)p')-1}
\eea
where $\psi$ is the epimorphism (\ref{psiprop3}) in Proposition~\ref{SecQuo}.3.

A particular lift of the $\Gamma$-system already appeared in~\cite{Ras0805}, as we will discuss 
presently.
{}From the lattice, certain direct sums of ${\cal W}$-indecomposable representations arise naturally
without employing the disentangling procedure. This procedure was actually designed
and introduced exactly to {\em distinguish} between the representations in these direct sums.
Following~\cite{Ras0805}, we thus consider
\bea
 \ketw{\Ec,1}
  &=&\bigoplus_{\al}^{p-2}(p-\al)\Big\{
   \ketw{\R_{p,1}^{\al,0}}\oplus\ketw{\R_{2p,1}^{\al,0}}\Big\}\nn
 \ketw{\Oc,1}
  &=&\bigoplus_{\al}^{p-1}(p-\al)\Big\{
    \ketw{\R_{p,1}^{\al,0}}\oplus\ketw{\R_{2p,1}^{\al,0}}\Big\}
\label{E1even}
\eea
for $p$ even, and
\bea
 \ketw{\Ec,1}&=&\bigoplus_{\al}^{p-2}(p-\al)\ketw{\R_{p,1}^{\al,0}}
   \oplus\bigoplus_{\al}^{p-1}(p-\al)
    \ketw{\R_{2p,1}^{\al,0}}  \nn
 \ketw{\Oc,1}&=&\bigoplus_{\al}^{p-1}(p-\al)
    \ketw{\R_{p,1}^{\al,0}}
  \oplus\bigoplus_{\al}^{p-2}(p-\al)
    \ketw{\R_{2p,1}^{\al,0}}
\label{E1odd}
\eea
for $p$ odd. Direct sums denoted by $\ketw{1,\Ec}$ and $\ketw{1,\Oc}$ are defined similarly.
Using (\ref{psiRG}), it is a straightforward exercise to verify that
\be
 \psi\big(\ketw{\Ec,1}\big)\ =\ \begin{cases} p^2\Gamma^{p-1,0},\quad&p\ \mathrm{even}\\
  p^2\Gamma^{p,0},\qquad&p\ \mathrm{odd}  \end{cases},\qquad\qquad
\psi\big(\ketw{\Oc,1}\big)\ =\ \begin{cases} p^2\Gamma^{p,0},\quad&p\ \mathrm{even}\\
  p^2\Gamma^{p-1,0},\qquad&p\ \mathrm{odd}  \end{cases}
\ee
with similar expressions for the images of $\ketw{1,\Ec}$ and $\ketw{1,\Oc}$.
By comparing these results with those of~\cite{Ras0805}, we observe 
that the six-dimensional (four-dimensional for $p=1$) fusion algebra 
of the $\Gamma$-system is isomorphic to the fusion algebra of the $\Ec,\Oc$ system
\be
 \big\langle \Gamma^{p-1,0},\Gamma^{p,0},\Gamma^{0,p'-1},\Gamma^{0,p'}\big\rangle_{p,p'}
   \ \simeq\ \big\langle \ketw{\Ec,1},\ketw{\Oc,1},\ketw{1,\Ec},\ketw{1,\Oc}\big\rangle_{p,p'}
\ee

On one hand, we expect~\cite{PR0812} that the $\Hc$-algebra, and hence the $\Gamma$-system,
is generated by the Verlinde fusion matrices associated to modular transformations of
projective characters, see the discussion preceding (\ref{Hij}). 
On the other hand, we have just demonstrated that the $\Gamma$-system
corresponds to the fusion algebra of the ${\cal W}$-extended, Yang-Baxter integrable,
boundary conditions prior to an eventual application of the disentangling procedure.
It thus seems natural to conjecture that the class of modular transformations just mentioned is linked
to this set of `fully entangled' boundary conditions. This question, however, will
not be addressed here any further.

\subsection{Minimal models}

In this section, we show how the usual fusion algebra, here denoted by $\mathrm{Fund}[{\cal M}(p,p')]$,
of the {\em rational} minimal model ${\cal M}(p,p')$~\cite{BPZ84,DiFMS} can be obtained from
the fundamental fusion algebra of either ${\cal LM}(p,p')$ or ${\cal WLM}(p,p')$.

First, we recall that the minimal-model fusion algebra is isomorphic to the quotient
polynomial fusion ring
\be
 \mathrm{Fund}[{\cal M}(p,p')]\ \simeq\ \mathbb{C}[X,Y]/\big(M_p(X),M_{p'}(Y),M_{p,p'}(X,Y)\big)
\label{FMCMMM}
\ee
where
\be
 M_n(x)\ =\ U_{n-1}\big(\frac{x}{2}\big),\qquad\quad M_{n,n'}(x,y)\ =\ U_{n-2}\big(\frac{x}{2}\big)
   -U_{n'-2}\big(\frac{y}{2}\big)
\ee
The fusion-algebra generators $\Nc_{a,b}$ are subject to $\Nc_{p-a,p'-b}\equiv\Nc_{a,b}$,
and the isomorphism in (\ref{FMCMMM}) reads
\be
 \Nc_{a,b}\ \leftrightarrow\ U_{a-1}\big(\frac{X}{2}\big)U_{b-1}\big(\frac{Y}{2}\big)
\label{NUU}
\ee
It is stressed that it is the condition $\Nc_{p-a,p'-b}\equiv\Nc_{a,b}$ in $\mathrm{Fund}[{\cal M}(p,p')]$
which on the ring side corresponds to $M_{p,p'}(X,Y)\equiv0$.
We also note that
\be
 P_n(x)\ =\ (x^2-4)U_{n-1}^2\big(\frac{x}{2}\big)M_n(x),\qquad\quad
   P_{n,n'}(x,y)\ =\ \Big(T_n\big(\frac{x}{2}\big)-T_{n'}\big(\frac{y}{2}\big)\Big)M_n(x)M_{n'}(y)
\label{PMPM}
\ee

Considering the set
\be
 \big\{[a,b]\big\}
 \ \simeq\ \Jc^{\mathrm{Fund}}_{p,p'}\setminus\Jc^{\mathrm{Out}}_{p,p'}
 \ \simeq\ \ketw{\Jc^{\mathrm{Fund}}_{p,p'}}\setminus\ketw{\Jc^{\mathrm{Out}}_{p,p'}}
\label{JJab}
\ee
of cardinality $|\{[a,b]\}|=(p-1)(p'-1)$, we introduce the fusion rules
\be
 [a,b]\ast[a',b']\ =\ \sum_{a''=|a-a'|+1,\ \!\mathrm{by}\ \!2}^{p-|p-a-a'|-1}\ 
   \sum_{b''=|b-b'|+1,\ \!\mathrm{by}\ \!2}^{p'-|p'-b-b'|-1}[a'',b'']
\label{abastab}
\ee
defining the fusion algebra $\langle[a,b]\rangle_{p,p'}$.
Since $\mathrm{Out}[{\cal LM}(p,p')]$ and $\mathrm{Out}[{\cal WLM}(p,p')]$
are algebraic ideals of $\mathrm{Fund}[{\cal LM}(p,p')]$ and 
$\mathrm{Fund}[{\cal WLM}(p,p')]$, respectively, we can form and study the quotients
$\mathrm{Fund}[{\cal LM}(p,p')]/\mathrm{Out}[{\cal LM}(p,p')]$ and
$\mathrm{Fund}[{\cal WLM}(p,p')]/\mathrm{Out}[{\cal WLM}(p,p')]$.
\\[.2cm]
{\bf Lemma \ref{SecQuo}.4}\ \ \ The quotients 
$\mathrm{Fund}[{\cal LM}(p,p')]/\mathrm{Out}[{\cal LM}(p,p')]$ and
$\mathrm{Fund}[{\cal WLM}(p,p')]/\mathrm{Out}[{\cal WLM}(p,p')]$ are both isomorphic to the
fusion algebra $\langle[a,b]\rangle_{p,p'}$, that is,
\be
 \mathrm{Fund}[{\cal LM}(p,p')]/\mathrm{Out}[{\cal LM}(p,p')]\ \simeq\ 
 \mathrm{Fund}[{\cal WLM}(p,p')]/\mathrm{Out}[{\cal WLM}(p,p')]\ \simeq\ 
  \big\langle[a,b]\big\rangle_{p,p'}
\ee
{\bf Proof}\ \ \ This readily follows by setting
\be
 \R\ \equiv\ 0,\qquad 
   \R\in\Jc^{\mathrm{Out}}_{p,p'};\qquad\qquad
   \ketw{\R}\ \equiv\ 0,\qquad
   \ketw{\R}\in\ketw{\Jc^{\mathrm{Out}}_{p,p'}}
\ee
in $\mathrm{Fund}[{\cal LM}(p,p')]$ and $\mathrm{Fund}[{\cal WLM}(p,p')]$, respectively, and from
\be
 (a,b)\ \leftrightarrow\ \ketw{a,b}\ \leftrightarrow\ [a,b]
\ee
$\Box$
\\[.2cm]
We note that both quotients in Lemma~\ref{SecQuo}.4 are trivial for $p=1$ since the sets in
(\ref{JJab}) are empty in that case. 
\\[.2cm]
{\bf Lemma \ref{SecQuo}.5}\ \ \  The fusion algebra $\langle[a,b]\rangle_{p,p'}$ is
isomorphic to the quotient polynomial fusion ring 
$\mathbb{C}[X,Y]/(M_p(X),M_{p'}(Y))$. The isomorphism reads
\be
  [a,b]\ \leftrightarrow\   U_{a-1}\big(\frac{X}{2}\big)U_{b-1}\big(\frac{Y}{2}\big)
\ee
{\bf Proof}\ \ \  This follows from (\ref{abastab}) and (\ref{UUU}).
\\
$\Box$
\\[.2cm]
{\bf Proposition \ref{SecQuo}.6}\ \ \ Let $\varphi$ be the isomorphism
\be
 \varphi:\ \mathrm{Fund}[{\cal LM}(p,p')]\ \to\ \mathbb{C}[X,Y]/\big(P_{p,p'}(X,Y)\big)
\ee 
in Proposition~\ref{SecLM}.2, let $\phi$ be the isomorphism
\be
 \phi:\   \big\langle[a,b]\big\rangle_{p,p'}  \ \to\ \mathbb{C}[X,Y]/\big(M_p(X),M_{p'}(Y)\big)
\ee
in Lemma~\ref{SecQuo}.5, and let $\sigma$ be the isomorphism
\be
 \sigma:\ \mathrm{Fund}[{\cal M}(p,p')]\ \to\ \mathbb{C}[X,Y]/\big(M_p(X),M_{p'}(Y),M_{p,p'}(X,Y)\big)
\ee 
in (\ref{FMCMMM}) and (\ref{NUU}).
Let $\pi_1$ be the projection
\be
 \pi_1:\ \mathbb{C}[X,Y]/\big(P_{p,p'}(X,Y)\big)\ \to\ 
    \Big(\mathbb{C}[X,Y]/\big(P_{p,p'}(X,Y)\big)\Big)/\big(M_p(X),M_{p'}(Y)\big)
\ee
and let $\pi_2$ be the projection
\be
 \pi_2:\ \mathbb{C}[X,Y]/\big(M_p(X),M_{p'}(Y)\big)\ \to\ 
    \Big(\mathbb{C}[X,Y]/\big(M_p(X),M_{p'}(Y)\big)\Big)/\big(M_{p,p'}(X,Y)\big)
\ee
With $\psi_1$ being the map from $\mathrm{Fund}[{\cal LM}(p,p')]$ to 
$\langle[a,b]\rangle_{p,p'}$ defined by
\be
 \psi_1\big((a,b)\big)\ =\ [a,b],\qquad\quad \psi_1\big(\R\big)\ =\ 0,\quad \R\in\Jc^{\mathrm{Out}}_{p,p'}
   \subseteq\Jc^{\mathrm{Fund}}_{p,p'}
\label{psi1LM}
\ee
and $\psi_2$ the map from $\langle[a,b]\rangle_{p,p'}$ to 
$\mathrm{Fund}[{\cal M}(p,p')]$ defined by
\be
 \psi_2\big([a,b]\big)\ =\ \Nc_{a,b}
\label{psi2LM}
\ee
the diagram
\be
 \mbox{
 \begin{picture}(150,250)(25,-10)
    \unitlength=0.9cm
  \thinlines
\put(0,4.6){\vector(1,0){4}}
\put(0,0.4){\vector(1,0){4}}
\put(-0.3,4.2){\vector(0,-1){3.4}}
\put(4.3,4.2){\vector(0,-1){3.4}}
\put(0,8.7){\vector(1,0){4}}
\put(-0.3,8.4){\vector(0,-1){3.4}}
\put(4.3,8.4){\vector(0,-1){3.4}}
\put(1.85,9.1){$\varphi$}
\put(1.85,5){$\phi$}
\put(1.85,-0.2){$\sigma$}
\put(-1.01,2.45){$\psi_2$}
\put(4.7,2.45){$\pi_2$}
\put(-1.01,6.55){$\psi_1$}
\put(4.7,6.55){$\pi_1$}
\put(-2,4.5){$\big\langle[a,b]\big\rangle_{p,p'}$}
\put(4.2,4.45){$\mathbb{C}[X,Y]/\big(M_p(X),M_{p'}(Y)\big)$}
\put(-2.9,0.25){$\mathrm{Fund}[{\cal M}(p,p')]$}
\put(4.2,0.25){$\mathbb{C}[X,Y]/\big(M_p(X),M_{p'}(Y),M_{p,p'}(X,Y)\big)$}
\put(4.2,8.65){$\mathbb{C}[X,Y]/\big(P_{p,p'}(X,Y)\big)$}
\put(-3.2,8.65){$\mathrm{Fund}[{\cal LM}(p,p')]$}
 \end{picture}
}
\label{diagLMM}
\ee
is a commutative diagram.
\\[.2cm]
{\bf Proof}\ \ \ It follows from the factorizations (\ref{PMPM})
that
\bea
 \!\!\!\!\!\!\!\!\!\!\Big(\mathbb{C}[X,Y]/\big(P_{p,p'}(X,Y)\big)\Big)/\big(M_p(X),M_{p'}(Y)\big)
  &\simeq&\mathbb{C}[X,Y]/\big(M_p(X),M_{p'}(Y)\big)\nn
  &\simeq&\Big(\mathbb{C}[X,Y]/\big(M_{p}(X)\big)\Big)\times\Big(\mathbb{C}[X,Y]/\big(M_{p'}(Y)\big)\Big)
\eea
This explains the form of the middle fusion ring in (\ref{diagLMM}), 
while the form of the lower fusion ring is obvious.
Also, it follows from (\ref{2TU}) that
\be
 2T_\al\big(\frac{x}{2}\big)U_{kn-1}\big(\frac{x}{2}\big)\ \equiv\ 0\qquad\big(\mathrm{mod}\ M_n(x)\big)
\ee
implying that $\phi^{-1}\circ\pi_1\circ\varphi(\R)=\phi^{-1}(0)=0$ for $\R\in\Jc^{\mathrm{Out}}_{p,p'}$,
while
\be
 \phi^{-1}\circ\pi_1\circ\varphi\big((a,b)\big)\ =\ \phi^{-1}\circ\pi_1\big(U_{a-1}\big(\frac{X}{2}\big)
   U_{b-1}\big(\frac{Y}{2}\big)\big)\ =\ \phi^{-1}\big(U_{a-1}\big(\frac{X}{2}\big)
   U_{b-1}\big(\frac{Y}{2}\big)\big)\ =\ [a,b]
\ee
This is in accordance with (\ref{psi1LM}). Finally, we have
\be
 \sigma^{-1}\circ\pi_2\circ\phi\big([a,b]\big)\ =\ \sigma^{-1}\circ\pi_2\big(U_{a-1}\big(\frac{X}{2}\big)
   U_{b-1}\big(\frac{Y}{2}\big)\big)\ =\ \sigma^{-1}\big(U_{a-1}\big(\frac{X}{2}\big)
   U_{b-1}\big(\frac{Y}{2}\big)\big)\ =\ \Nc_{a,b}
\ee
in accordance with (\ref{psi2LM}). 
\\
$\Box$
\\[.2cm]
{\bf Proposition \ref{SecQuo}.7}\ \ \ Let $\hat{\varphi}$ be the isomorphism
\be
 \hat{\varphi}:\ \mathrm{Fund}[{\cal WLM}(p,p')]\ \to\ 
    \mathbb{C}[X,Y]/\big(P_p(X),P_{p'}(Y),P_{p,p'}(X,Y)\big)
\ee 
in Proposition~\ref{SecFundWLM}.2, let $\phi$ be the isomorphism
\be
 \phi:\   \big\langle[a,b]\big\rangle_{p,p'}  \ \to\ \mathbb{C}[X,Y]/\big(M_p(X),M_{p'}(Y)\big)
\ee
in Lemma~\ref{SecQuo}.5, and let $\sigma$ be the isomorphism
\be
 \sigma:\ \mathrm{Fund}[{\cal M}(p,p')]\ \to\ \mathbb{C}[X,Y]/\big(M_p(X),M_{p'}(Y),M_{p,p'}(X,Y)\big)
\ee 
in (\ref{FMCMMM}) and (\ref{NUU}).
Let $\hat{\pi}_1$ be the projection
\bea
 &&\hat{\pi}_1:\ \mathbb{C}[X,Y]/\big(P_p(X),P_{p'}(Y),P_{p,p'}(X,Y)\big)\nn
  &&\qquad\qquad\ \to\ 
    \Big(\mathbb{C}[X,Y]/\big(P_p(X),P_{p'}(Y),P_{p,p'}(X,Y)\big)\Big)/\big(M_p(X),M_{p'}(Y)\big)
\eea
and let $\pi_2$ be the projection
\be
 \pi_2:\ \mathbb{C}[X,Y]/\big(M_p(X),M_{p'}(Y)\big)\ \to\ 
    \Big(\mathbb{C}[X,Y]/\big(M_p(X),M_{p'}(Y)\big)\Big)/\big(M_{p,p'}(X,Y)\big)
\ee
With $\hat{\psi}_1$ being the map from $\mathrm{Fund}[{\cal WLM}(p,p')]$ to 
$\langle[a,b]\rangle_{p,p'}$ defined by
\be
 \hat{\psi}_1\big(\ketw{a,b}\big)\ =\ [a,b],\qquad\quad 
   \hat{\psi}_1\big(\ketw{\R}\big)\ =\ 0,\quad \ketw{\R}\in\ketw{\Jc^{\mathrm{Out}}_{p,p'}}
   \subseteq\ketw{\Jc^{\mathrm{Fund}}_{p,p'}}
\ee
and $\psi_2$ the map from $\langle[a,b]\rangle_{p,p'}$ to 
$\mathrm{Fund}[{\cal M}(p,p')]$ defined by
\be
 \psi_2\big([a,b]\big)\ =\ \Nc_{a,b}
\ee
the diagram
\be
 \mbox{
 \begin{picture}(150,250)(25,-10)
    \unitlength=0.9cm
  \thinlines
\put(0,4.6){\vector(1,0){4}}
\put(0,0.4){\vector(1,0){4}}
\put(-0.3,4.2){\vector(0,-1){3.4}}
\put(4.3,4.2){\vector(0,-1){3.4}}
\put(0,8.7){\vector(1,0){4}}
\put(-0.3,8.4){\vector(0,-1){3.4}}
\put(4.3,8.4){\vector(0,-1){3.4}}
\put(1.85,9.1){$\hat{\varphi}$}
\put(1.85,5){$\phi$}
\put(1.85,-0.2){$\sigma$}
\put(-1.01,2.45){$\psi_2$}
\put(4.7,2.45){$\pi_2$}
\put(-1.01,6.55){$\hat{\psi}_1$}
\put(4.7,6.55){$\hat{\pi}_1$}
\put(-2,4.5){$\big\langle[a,b]\big\rangle_{p,p'}$}
\put(4.2,4.45){$\mathbb{C}[X,Y]/\big(M_p(X),M_{p'}(Y)\big)$}
\put(-2.9,0.25){$\mathrm{Fund}[{\cal M}(p,p')]$}
\put(4.2,0.25){$\mathbb{C}[X,Y]/\big(M_p(X),M_{p'}(Y),M_{p,p'}(X,Y)\big)$}
\put(4.2,8.65){$\mathbb{C}[X,Y]/\big(P_p(X),P_{p'}(Y),P_{p,p'}(X,Y)\big)$}
\put(-3.65,8.65){$\mathrm{Fund}[{\cal WLM}(p,p')]$}
 \end{picture}
}
\label{diagWLMM}
\ee
is a commutative diagram.
\\[.2cm]
{\bf Proof}\ \ \ The proof is similar to the proof of Proposition~\ref{SecQuo}.6.
\\
$\Box$
\\[.2cm]
In partial summary, the fusion algebra of the rational minimal model ${\cal M}(p,p')$ can
be viewed as
\bea
 \mathrm{Fund}[{\cal M}(p,p')]
   &\simeq& 
     \Big(\mathrm{Fund}[{\cal LM}(p,p')]/\mathrm{Out}[{\cal LM}(p,p')]\Big)/\mathbb{Z}_2\nn
   &\simeq&
     \Big(\mathrm{Fund}[{\cal WLM}(p,p')]/\mathrm{Out}[{\cal WLM}(p,p')]\Big)/\mathbb{Z}_{2}
\eea
where the quotients by $\mathbb{Z}_2$ correspond to the equivalence relations 
identifying $(a,b)$ with $(p-a,p'-b)$ and $\ketw{a,b}$ with $\ketw{p-a,p'-b}$, respectively.

\section{Discussion}
\label{SecDisc}

The logarithmic minimal model ${\cal LM}(p,p')$~\cite{PRZ0607} contains a countably infinite number of 
Virasoro representations whose fusion rules are described in~\cite{RP0706,RP0707}.
These representations 
can be reorganized into a finite number of ${\cal W}$-representations with respect
to the extended Virasoro algebra symmetry ${\cal W}$.
Using a lattice implementation of fusion, we recently~\cite{PRR0803,RP0804,Ras0805} 
determined the fusion algebra of 
these ${\cal W}$-representations and found that it closes, albeit without an identity for $p>1$.
In this paper, we have provided a fusion-matrix realization of this fusion algebra and identified 
a fusion ring isomorphic to it.
We have also considered various extensions of it and quotients thereof, 
and introduced and analyzed commutative
diagrams with morphisms between the involved fusion algebras and the corresponding
fusion rings.
One particular extension is reminiscent of the fundamental fusion algebra of ${\cal LM}(p,p')$
and offers a natural way of introducing the missing identity for $p>1$.
In this example, the complete set of supplementary fusion rules has been posited and shown to 
yield an associative and commutative fusion algebra.
We have discussed how working out explicit fusion matrices 
is facilitated by a further enlargement based on a pair
of mutual Moore-Penrose inverses intertwining between the ${\cal W}$-fundamental and enlarged
fusion algebras. 
We have also detailed how the ${\cal W}$-extended fusion algebras and the corresponding 
fusion rings can be obtained, using quotient constructions, 
from the infinite fusion algebras and corresponding fusion rings in the Virasoro picture.
Two additional quotients have been examined. The first one introduces a fusion algebra of expected
relevance~\cite{PR0812} to modular transformations in the ${\cal W}$-extended picture.
The other one demonstrates how the fusion algebras of the rational minimal 
models~\cite{BPZ84,DiFMS} arise as the algebras of certain equivalence classes in the much richer
logarithmic minimal models.

We stress that the process of forming quotient constructions of fusion algebras 
by collecting representations
in equivalence classes is in general {\em incompatible} with naive character considerations.
Not only can two lifts of a given class have different characters, their conformal weights
may differ by a non-integer fraction.

After the introduction of our logarithmic minimal models~\cite{PRZ0607}, an alternative
lattice-based description of a family of logarithmic conformal field theories has emerged~\cite{RS0701}.
Critical percolation is considered there as a main example and an associated fusion algebra
is presented. As discussed in~\cite{RP0706}, this fusion algebra corresponds to a particular
subalgebra of the vertical fusion algebra 
$\langle (1,1), (1,2)\rangle_{2,3}$ which itself is a subalgebra of the fundamental fusion
algebra $\mathrm{Fund}[{\cal LM}(2,3)]$ of critical percolation. 

The fusion algebra of~\cite{RS0701} can be generalized from critical percolation ${\cal LM}(2,3)$
to ${\cal LM}(p,p')$ as $\mathrm{Fund}[{\cal LM}(p,p')]$ contains a similar fusion subalgebra.
To appreciate this, we focus on the vertical components of $\mathrm{Fund}[{\cal LM}(p,p')]$
and the associated fusion ring by considering
\be
 \big\langle(1,1),(1,2)\big\rangle_{p,p'}\ \simeq\ \mathbb{C}[Y]
\label{1112Y}
\ee
Since the product of two even polynomials in $\mathbb{C}[Y]$ is an even polynomial,
the set of even polynomials in $\mathbb{C}[Y]$ is closed under multiplication.
Keeping the isomorphism (\ref{1112Y}) in mind, the natural description of this set depends on
the parity of $p'$. For $p'$ odd, in particular, the corresponding fusion subalgebra of
$\langle(1,1),(1,2)\rangle_{p,p'}$ reads
\be
 \big\langle(1,2j-1), (1,(2k-1)p'), \R_{1,(2k-1)p'}^{0,2j}, \R_{1,2kp'}^{0,2j-1};\ 
   j\in\mathbb{Z}_{1,\frac{p'-1}{2}},\ k\in\mathbb{N}\big\rangle_{p,p'}
\label{12j1}
\ee
The isomorphism to the set of even polynomials in $\mathbb{C}[Y]$ is then constructed as the
isomorphism given in Proposition~\ref{SecLM}.2 restricted to the generators of (\ref{12j1})
\bea
 (1,2j-1) &\leftrightarrow& U_{2j-2}\big(\frac{Y}{2}\big)\nn
 (1,(2k-1)p')&\leftrightarrow& U_{(2k-1)p'-1}\big(\frac{Y}{2}\big)\nn
 \R_{1,(2k-1)p'}^{0,2j}&\leftrightarrow&2T_{2j}\big(\frac{Y}{2}\big)U_{(2k-1)p'-1}\big(\frac{Y}{2}\big)\nn
 \R_{1,2kp'}^{0,2j-1}&\leftrightarrow&2T_{2j-1}\big(\frac{Y}{2}\big)U_{2kp'-1}\big(\frac{Y}{2}\big)
\label{RSiso}
\eea
where $j\in\mathbb{Z}_{1,\frac{p'-1}{2}}$.
As observed in~\cite{RP0706}, the fusion algebra of critical percolation discussed in~\cite{RS0701}
is equivalent to
\be
  \big\langle(1,1), (1,6k-3), \R_{1,6k-3}^{0,2}, \R_{1,6k}^{0,1};\ 
    k\in\mathbb{N}\big\rangle_{2,3}
\ee
which is exactly the fusion subalgebra (\ref{12j1}) of $\mathrm{Fund}[{\cal LM}(2,3)]$.
We thus conclude that the fusion algebra of~\cite{RS0701} and its extension to ${\cal LM}(p,p')$
are associated to the fusion rings of even polynomials (given in (\ref{RSiso}) for $p'$ odd), 
and we have the particular isomorphism
\be
 (1,1)\ \leftrightarrow\ 1,\quad (1,6k-3)\ \leftrightarrow\ U_{6k-4}\big(\frac{Y}{2}\big),\quad
  \R_{1,6k-3}^{0,2}\ \leftrightarrow\ (Y^2-2)U_{6k-4}\big(\frac{Y}{2}\big),\quad
  \R_{1,6k}^{0,1}\ \leftrightarrow\ YU_{6k-1}\big(\frac{Y}{2}\big)
\ee
in the case of critical percolation.

{}From the perspective of our ${\cal W}$-extensions of ${\cal LM}(p,p')$, there is a natural, algebraic 
candidate for a ${\cal W}$-extension of the fusion algebra of~\cite{RS0701} and its generalization
to ${\cal LM}(p,p')$ as described above. Reorganizing the Virasoro
representations appearing in (\ref{12j1})
in ${\cal W}$-representations, recalling that we have specialized to $p'$ odd, we recognize
\bea
 \ketw{1,p'}&=&\bigoplus_{k\in\mathbb{N}}(2k-1)(1,(2k-1)p')\nn
 \ketw{\R_{1,p'}^{0,2j}}&=&\bigoplus_{k\in\mathbb{N}}(2k-1)\R_{1,(2k-1)p'}^{0,2j}\nn
 \ketw{\R_{1,2p'}^{0,2j-1}}&=&\bigoplus_{k\in\mathbb{N}}2k\R_{1,2kp'}^{0,2j-1}
\eea
discussed in Section~\ref{SecRepCont}. They generate the closed fusion algebra
\be
 \big\langle\ketw{1,p'}, \ketw{\R_{1,p'}^{0,2j}}, \ketw{\R_{1,2p'}^{0,2j-1}};\ j\in\mathbb{Z}_{1,\frac{p'-1}{2}}
   \big\rangle_{p,p'}
\label{1p1p02}
\ee
as a $p'$-dimensional
fusion subalgebra of $\mathrm{Out}[{\cal WLM}(p,p')]$. In the case of critical percolation,
the fusion rules of the three-dimensional fusion algebra (\ref{1p1p02})
\be
 \big\langle\ketw{1,3}, \ketw{\R_{1,3}^{0,2}}, \ketw{\R_{1,6}^{0,1}}\big\rangle_{2,3}
\ee
are given by
\bea
 \ketw{1,3}\fus\ketw{1,3}&=&\ketw{1,3}\oplus\ketw{\R_{1,3}^{0,2}}\nn
 \ketw{1,3}\fus\ketw{\R_{1,3}^{0,2}}\ =\ \ketw{1,3}\fus\ketw{\R_{1,6}^{0,1}}&=&
   2\ketw{1,3}\oplus2\ketw{\R_{1,6}^{0,1}}\nn
 \ketw{\R_{1,3}^{0,2}}\fus\ketw{\R_{1,3}^{0,2}}\ =\ 
  \ketw{\R_{1,3}^{0,2}}\fus\ketw{\R_{1,6}^{0,1}}\nn
 =\ \ketw{\R_{1,6}^{0,1}}\fus\ketw{\R_{1,6}^{0,1}}&=&4\ketw{1,3}\oplus2\ketw{\R_{1,3}^{0,2}}
   \oplus2\ketw{\R_{1,6}^{0,1}}
\eea
It would be interesting to see if this algebraically motivated proposal for a 
${\cal W}$-extension of~\cite{RS0701} actually admits a lattice realization within the
framework of~\cite{RS0701}.  
\vskip.5cm
\subsection*{Acknowledgments}
\vskip.1cm
\noindent
This work is supported by the Australian Research Council. 
The author thanks Paul A. Pearce for helpful discussions and comments.

\appendix

\section{Chebyshev polynomials}
\label{AppCheb}

\subsection{Chebyshev polynomials of the first kind}

Recursion relation:
\be
 T_{n}(x)\ =\ 2xT_{n-1}(x)-T_{n-2}(x),\ \ \ \ \ \ \ n=2,3,\ldots
\label{Trec}
\ee
Initial conditions:
\be
 T_0(x)\ =\ 1,\ \ \ \ \ \ \ T_1(x)\ =\ x
\ee
Examples:
\bea
 T_2(x)\!\!&=&\!\! 2x^2-1\nn
 T_3(x)\!\!&=&\!\! 4x^3-3x\nn
 T_4(x)\!\!&=&\!\! 8x^4-8x^2+1\nn
 T_5(x)\!\!&=&\!\! 16x^5-20x^3+5x
\label{Tex}
\eea
Decomposition of product:
\be
 T_m(x)T_n(x)\ =\ \frac{1}{2}\Big(T_{|m-n|}(x)+T_{m+n}(x)\Big)
\label{TTT}
\ee

\subsection{Chebyshev polynomials of the second kind}

Recursion relation:
\be
 U_{n}(x)\ =\ 2xU_{n-1}(x)-U_{n-2}(x),\ \ \ \ \ \ \ n=2,3,\ldots
\label{Urec}
\ee
Initial conditions:
\be
 U_0(x)\ =\ 1,\ \ \ \ \ \ \ U_1(x)\ =\ 2x
\ee
Examples:
\bea
 U_2(x)\!\!&=&\!\!4x^2-1\nn
 U_3(x)\!\!&=&\!\!8x^3-4x\nn
 U_4(x)\!\!&=&\!\!16x^4-12x^2+1\nn
 U_5(x)\!\!&=&\!\!32x^5-32x^3+6x
\label{Uex}
\eea
Extension:
\be
 U_{-1}(x)\ =\ 0
\ee
Decomposition of product:
\be
 U_m(x)U_n(x)\ =\ \sum_{j=|m-n|,\ \!\!{\rm by}\ \!2}^{m+n}U_j(x)
\label{UUU}
\ee

\subsection{Relating Chebyshev polynomials of the first and second kind}

The basic relation
\be
 2T_n(x)\ =\ U_n(x)-U_{n-2}(x)
\label{TU}
\ee
generalizes to
\be
 2T_{n}(x)U_{m-1}(x)\ =\ \begin{cases}
   U_{n+m-1}(x)-U_{|n-m|-1}(x),\hspace{1cm}&n>m\\[.2cm]
   U_{n+m-1}(x),  &n=m\\[.2cm]
   U_{n+m-1}(x)+U_{|n-m|-1}(x),  &n<m
 \end{cases}
\label{2TU}
\ee
Also, inverting (\ref{TU}) yields
\be
 U_m(x)\ =\ \sum_{i}^m\big(2-\delta_{i,0}\big)T_i(x)
\label{UT}
\ee
where we recall the convention (\ref{sum2a}), from which we deduce that
\bea
 2T_n(x)U_{m-1}(x)&=&
  \sum_{i}^{m-1}\big(2-\delta_{i,0}\big)\big(T_{|n-i|}(x)+T_{n+i}(x)\big)\nn
 &=& \begin{cases} 
   {\displaystyle \sum_{i=|n-m|+1,\ \!\!\mathrm{by}\ \!2}^{n+m-1}\!\!\!2T_i(x)},  &n\geq m\\
   {\displaystyle \sum_{i=|n-m|+1,\ \!\!\mathrm{by}\ \!2}^{n+m-1}\!\!\!2T_i(x)}
   +{\displaystyle \sum_{i}^{m-n-1}\!\!2\big(2-\delta_{i,0}\big)T_i(x)}
   ,\qquad &n<m
  \end{cases}
\label{2TUT}
\eea
Some other useful relations are
\be
 T_n^2(x)\ =\ 1+(x^2-1)U_{n-1}^2(x),\qquad\quad
 T_n(x)T_{n+1}(x)\ =\ x+(x^2-1)U_{n-1}(x)U_{n}(x)
\ee
and hence
\bea
 T_{2n}(x)-1&=&2\big(T_n^2(x)-1\big)\ =\ 2(x^2-1)U_{n-1}^2(x)\nn
 T_{2n+1}(x)-x&=&2\big(T_n(x)T_{n+1}(x)-x\big)\ =\ 2(x^2-1)U_{n-1}(x)U_n(x)
\label{T2n}
\eea

\subsection{Congruences modulo Chebyshev polynomials}

{\bf Lemma \ref{AppCheb}.1}\ \ \  Modulo the polynomial $P_n(x)$ defined in
(\ref{Pn}), we have
\be
 U_{(2k-1)n-1}\big(\frac{x}{2}\big)\ \equiv\ (2k-1)U_{n-1}\big(\frac{x}{2}\big),\qquad\quad
 U_{2kn-1}\big(\frac{x}{2}\big)\ \equiv\ kU_{2n-1}\big(\frac{x}{2}\big) 
\ee
{\bf Proof}\ \ \ This follows by induction in $k$. Simple evaluations verify the statements
for small $k$, while the induction steps read
\bea
 U_{(2k+1)n-1}\big(\frac{x}{2}\big)&=&2T_{2n}\big(\frac{x}{2}\big)U_{(2k-1)n-1}\big(\frac{x}{2}\big)
  -U_{(2k-3)n-1}\big(\frac{x}{2}\big)\nn 
 &\equiv&2T_{2n}\big(\frac{x}{2}\big)(2k-1)U_{n-1}\big(\frac{x}{2}\big)
   -(2k-3)U_{n-1}\big(\frac{x}{2}\big)\nn
 &=&(2k-1)\Big(U_{3n-1}\big(\frac{x}{2}\big)-U_{n-1}\big(\frac{x}{2}\big)\Big)
  -(2k-3)U_{n-1}\big(\frac{x}{2}\big)\nn 
 &\equiv&(2k+1)U_{n-1}\big(\frac{x}{2}\big) 
\eea
and subsequently
\bea
 U_{2(k+1)n-1}\big(\frac{x}{2}\big)&=&2T_n\big(\frac{x}{2}\big)U_{(2k+1)n-1}\big(\frac{x}{2}\big)
  -U_{2kn-1}\big(\frac{x}{2}\big)\nn
 &\equiv&2T_n\big(\frac{x}{2}\big)(2k+1)U_{n-1}\big(\frac{x}{2}\big)
  -kU_{2n-1}\big(\frac{x}{2}\big)\nn
 &=&(2k+1)U_{2n-1}\big(\frac{x}{2}\big) -kU_{2n-1}\big(\frac{x}{2}\big)\nn
 &=&(k+1)U_{2n-1}\big(\frac{x}{2}\big)
\eea
$\Box$
\\[.2cm]
{\bf Lemma \ref{AppCheb}.2}\ \ \ For $i\in\mathbb{Z}_{0,n}$ and modulo the polynomial
$H_n(x)$ defined in (\ref{Hn}), we have
\be
 T_n\big(\frac{x}{2}\big)T_i\big(\frac{x}{2}\big)\ \equiv\ T_{n-i}\big(\frac{x}{2}\big),\qquad\quad
   T_{n+i}\big(\frac{x}{2}\big)\ \equiv\ T_{n-i}\big(\frac{x}{2}\big)
\label{TTTTT}
\ee
In particular, 
$T_{2n}\big(\frac{x}{2}\big)\equiv T_n\big(\frac{x}{2}\big)T_n\big(\frac{x}{2}\big)\equiv1$.
\\[.2cm]
{\bf Proof}\ \ \ Both congruences in (\ref{TTTTT}) follow by induction in $i$. They are readily verified
for small $i$, while the induction steps read
\bea
 T_n\big(\frac{x}{2}\big)T_{i+1}\big(\frac{x}{2}\big)
   &=&T_n\big(\frac{x}{2}\big)\Big(xT_i\big(\frac{x}{2}\big)
   -T_{i-1}\big(\frac{x}{2}\big)\Big)\ \equiv\ xT_{n-i}\big(\frac{x}{2}\big)-T_{n-i+1}\big(\frac{x}{2}\big)
     \ =\ T_{n-i-1}\big(\frac{x}{2}\big)\nn
 T_{n+i+1}\big(\frac{x}{2}\big)&=&xT_{n+i}\big(\frac{x}{2}\big)-T_{n+i-1}\big(\frac{x}{2}\big)
   \ \equiv\ xT_{n-i}\big(\frac{x}{2}\big)-T_{n-i+1}\big(\frac{x}{2}\big)\ =\ T_{n-i-1}\big(\frac{x}{2}\big)
\eea
$\Box$
\\[.2cm]
{\bf Lemma \ref{AppCheb}.3}\ \ \ For $n\in\mathbb{Z}_{0,m-1}$ and modulo the polynomial
$H_m(x)$ defined in (\ref{Hn}), we have
\be
 2T_n\big(\frac{x}{2}\big)U_{\kappa m-1}\big(\frac{x}{2}\big)\ \equiv
  \sum_{i}^{m+\eps(n+(\kappa-1)m)-1}\!\!\!
   2\kappa\big(2-\delta_{i,0}-\delta_{i,m}\big) T_i\big(\frac{x}{2}\big)
\ee
{\bf Proof}\ \ \ The congruence for $\kappa=1$ follows by application of Lemma~\ref{AppCheb}.2 
to (\ref{2TUT}). The congruence for $\kappa=2$ is subsequently obtained by applying
the result for $\kappa=1$ to
\be
 2T_n\big(\frac{x}{2}\big)U_{2m-1}\big(\frac{x}{2}\big)\ =\ 2\Big(T_{n+m}\big(\frac{x}{2}\big)
   +T_{m-n}\big(\frac{x}{2}\big)\Big)U_{m-1}\big(\frac{x}{2}\big)\ \equiv\ 
    4T_{m-n}\big(\frac{x}{2}\big)U_{m-1}\big(\frac{x}{2}\big)
\ee
$\Box$

\section{Quotient polynomial rings}
\label{AppQuo}

Here, we present our notation for quotient polynomial rings and recall a couple
of their simple properties. 
By
\be
 \mathbb{K}[x_1,\ldots,x_{\ell}]/(f_1,\ldots,f_m)
\label{Kff}
\ee
we denote a quotient of the polynomial ring $\mathbb{K}[x_1,\ldots,x_{\ell}]$ over the
field $\mathbb{K}$ where the ideal, with respect to which the quotient is formed,
is specified by a collection of generators (polynomials) $f_i\in\mathbb{K}[x_1,\ldots,x_{\ell}]$.
A further quotient construction by the ideal $(g_1,\ldots,g_{m'})$, whose generators are elements of
$\mathbb{K}[x_1,\ldots,x_{\ell}]/(f_1,\ldots,f_m)$, can be written
\be
 \big(\mathbb{K}[x_1,\ldots,x_{\ell}]/(f_1,\ldots,f_m)\big)/(g_1,\ldots,g_{m'})
 \ \simeq\ \mathbb{K}[x_1,\ldots,x_{\ell}]/(f_1,\ldots,f_m,\tilde{g}_1,\ldots,\tilde{g}_{m'})
\label{Kfg}
\ee
Here, $\tilde{g}_i$ is a lift of $g_i$, that is, it is any element in $\mathbb{K}[x_1,\ldots,x_{\ell}]$
whose class in (\ref{Kff}) is $g_i$.
With a slight but natural abuse of notation, we will denote $\tilde{g}_i$ simply by $g_i$.

The characterization of the `combined' ideal in (\ref{Kfg})
can often be simplified. Of particular interest here,
are the cases where $m'=m$ and $g_i$ is a (not necessarily proper) divisor of $f_i$ for
$\forall i\in\mathbb{Z}_{1,m}$. In such a case, we simply have
\be
 (f_1,\ldots,f_m,g_1,\ldots,g_{m})\ \simeq\ (g_1,\ldots,g_{m})
\ee
and hence
\be
 \big(\mathbb{K}[x_1,\ldots,x_{\ell}]/(f_1,\ldots,f_m)\big)/(g_1,\ldots,g_{m})
 \ \simeq\ \mathbb{K}[x_1,\ldots,x_{\ell}]/(g_1,\ldots,g_{m})
\label{Kgg}
\ee

We are mainly interested in the case where $\mathbb{K}=\mathbb{C}$ and $\ell=2$.
For simplicity, we may thus write $x=x_1$ and $y=x_2$.
The number of linearly independent generators of $\mathbb{C}[x,y]/(f_1,\ldots,f_m)$
is equal to the number of inequivalent monomials in $x$ and $y$ (including $x^0y^0=1$).
First, for $m=1$ with a single non-trivial polynomial $f_1$, the number of linearly
independent generators of $\mathbb{C}[x,y]/(f_1)$ is infinite.
Second, we let $m=2$ and assume that the two non-trivial polynomials 
are of the form $f_1(x)$ and $f_2(y)$. 
The quotient ring (\ref{Kff}) then corresponds to the tensor product of the two quotient polynomial rings
$\mathbb{C}[x]/(f_1)$ and $\mathbb{C}[y]/(f_2)$.
The number of linearly independent generators of this ring $\mathbb{C}[x,y]/(f_1,f_2)$ 
is given by $\mathrm{deg}(f_1)\mathrm{deg}(f_2)$.

Partly to illustrate how sensitive a quotient polynomial ring is to a simple change of its ideal, we 
now consider the two cases
\be
 Q_\ell\ =\ \mathbb{C}[x,y]/\big(f(x),g_\ell(y),h(x,y)\big),\qquad\quad \ell=1,2 
\ee
where
\be
 f(x)\ =\ x^2-1,\qquad g_1(y)\ =\ y^3-1,\qquad g_2(y)\ =\ y^3-y,\qquad h(x,y)\ =\ xy-y^2
\ee
In $Q_1$, we have
\be
 0\ \equiv\ (x+y)h(x,y)\ =\ (x^2y-xy^2)+(xy^2-y^3)\ \equiv\ y-1
\ee
from which it follows that
\be
 0\ \equiv\ xh(x,y)\ =\ x^2y-xy^2\ \equiv\ 1-x
\ee
and hence
\be
 Q_1\ \simeq\ \mathbb{C}
\ee
In $Q_2$, on the other hand, $(x+y)h(x,y)\equiv0$ is trivially satisfied, and $Q_2$
has four linearly independent generators which we can choose as $1$, $x$, $y$ and $xy$.
This pair of examples also demonstrates that counting the number of linearly independent generators,
which is of interest to us, can be a delicate business.
Even though $g_1(y)$ and $g_2(y)$ have the common factor $y-1$, this analysis furthermore illustrates 
that care has to be exercised when considering situations like the left-hand side of (\ref{Kgg})
if $g_i\nmid f_i$ for some $i\in\mathbb{Z}_{1,m}$.

\section{${\cal W}$-extended fusion algebra $\mathrm{Out}[{\cal WLM}(p,p')]$}
\label{AppFus}

Here, we summarize the fusion rules, obtained in~\cite{Ras0805}, underlying the 
fusion algebra $\mathrm{Out}[{\cal WLM}(p,p')]$ as given in (\ref{WfusOut}).
The fusion of two ${\cal W}$-indecomposable rank-1 representations is given by
\bea
 \ketw{\kappa p,s}\fus\ketw{\kappa'p,s'}&=&\bigoplus_{\al}^{p-1}\Big\{
  \!\bigoplus_{j=|s-s'|+1,\ \!{\rm by}\ \!2}^{p'-|p'-s-s'|-1}
  \!\!\!\ketw{\R_{(\kappa\cdot\kappa')p,j}^{\al,0}}
    \oplus\!\bigoplus_{\beta}^{s+s'-p'-1}
    \!\ketw{\R_{\kappa p,\kappa'p'}^{\al,\beta}}
    \Big\}   \nn
 \ketw{\kappa p,s}\fus\ketw{r,\kappa'p'}&=&\bigoplus_{\al}^{r-1}
  \Big\{\bigoplus_{\beta}^{s-1}
     \ketw{\R_{\kappa p,\kappa'p'}^{\al,\beta}}\Big\}\nn
 \ketw{r,\kappa p'}\fus\ketw{r',\kappa'p'}&=&\bigoplus_{\beta}^{p'-1}\Big\{
   \!\bigoplus_{j=|r-r'|+1,\ \!{\rm by}\ \!2}^{p-|p-r-r'|-1}
  \!\!\!\ketw{\R_{j,(\kappa\cdot\kappa')p'}^{0,\beta}}
  \oplus\!\bigoplus_{\al}^{r+r'-p-1}
    \!\ketw{\R_{\kappa p,\kappa'p'}^{\al,\beta}}
    \Big\}
\label{fus11}
\eea
The fusion of a ${\cal W}$-indecomposable rank-1 representation with a 
${\cal W}$-indecomposable rank-2 representation is given by
\bea
 \ketw{\kappa p,s}\fus\ketw{\R_{\kappa'p,s'}^{a,0}}
   &=&\!\bigoplus_{j=|s-s'|+1,\ \!{\rm by}\ \!2}^{p'-|p'-s-s'|-1}
  \!\!\Big\{\bigoplus_{\al}^{p-a-1}\!\!2\ketw{\R_{(\kappa\cdot\kappa')p,j}^{\al,0}}\oplus
    \bigoplus_{\al}^{a-1}2\ketw{\R_{(2\cdot\kappa\cdot\kappa')p,j}^{\al,0}}\Big\}  \nn
  &\oplus&\bigoplus_{\beta}^{s+s'-p'-1}
  \!\!\Big\{\bigoplus_{\al}^{p-a-1}\!\!2\ketw{\R_{\kappa p,\kappa'p'}^{\al,\beta}}\oplus
    \bigoplus_{\al}^{a-1}2\ketw{\R_{\kappa p,(2\cdot\kappa')p'}^{\al,\beta}}\Big\}  \nn
 \ketw{\kappa p,s}\fus\ketw{\R_{r,\kappa'p'}^{0,b}}
  &=&\bigoplus_{\al}^{r-1}\Big\{
    \!\bigoplus_{\beta=|b-s|+1,\ \!{\rm by}\ \!2}^{p'-|p'-s-b|-1}
      \!\!\!\!\ketw{\R_{\kappa p,\kappa'p'}^{\al,\beta}}
     \oplus\!\bigoplus_{\beta}^{s-b-1}\!2\ketw{\R_{\kappa p,\kappa'p'}^{\al,\beta}}\oplus
     \!\!\!\bigoplus_{\beta}^{b+s-p'-1}
        \!\!\!2\ketw{\R_{\kappa p,(2\cdot\kappa')p'}^{\al,\beta}}\Big\}   \nn
 \ketw{r,\kappa p'}\fus\ketw{\R_{\kappa'p,s}^{a,0}}
  &=&\bigoplus_{\beta}^{s-1}\Big\{
    \!\bigoplus_{\al=|a-r|+1,\ \!{\rm by}\ \!2}^{p-|p-r-a|-1}
      \!\!\!\!\ketw{\R_{\kappa p,\kappa'p'}^{\al,\beta}}
    \oplus\!\bigoplus_{\al}^{r-a-1}\!2\ketw{\R_{\kappa p,\kappa'p'}^{\al,\beta}}\oplus 
     \!\!\!\bigoplus_{\al}^{a+r-p-1}
        \!\!\!2\ketw{\R_{\kappa p,(2\cdot\kappa')p'}^{\al,\beta}}\Big\}   \nn
 \!\ketw{r,\kappa p'}\fus\ketw{\R_{r',\kappa'p'}^{0,b}}
  &=&\!\bigoplus_{j=|r-r'|+1,\ \!{\rm by}\ \!2}^{p-|p-r-r'|-1}
  \!\!\Big\{\bigoplus_{\beta}^{p'-b-1}\!\!2\ketw{\R_{j,(\kappa\cdot\kappa')p'}^{0,\beta}}\oplus
    \bigoplus_{\beta}^{b-1}2\ketw{\R_{j,(2\cdot\kappa\cdot\kappa')p'}^{0,\beta}}\Big\}  \nn
  &\oplus&\bigoplus_{\al}^{r+r'-p-1}
  \!\!\Big\{\bigoplus_{\beta}^{p'-b-1}\!\!2\ketw{\R_{\kappa p,\kappa'p'}^{\al,\beta}}\oplus
    \bigoplus_{\beta}^{b-1}2\ketw{\R_{\kappa p,(2\cdot\kappa')p'}^{\al,\beta}}\Big\}   
\label{fus12}
\eea
The fusion of a ${\cal W}$-indecomposable rank-1 representation with a 
${\cal W}$-indecomposable rank-3 representation is given by
\bea
 \ketw{\kappa p,s}\fus\ketw{\R_{p,\kappa'p'}^{a,b}}
  &=&\!\!\!\bigoplus_{\al}^{p-a-1}\!\!\Big\{
     \!\bigoplus_{\beta=|b-s|+1,\ \!{\rm by}\ \!2}^{p'-|p'-s-b|-1}
      \!\!\!\!2\ketw{\R_{\kappa p,\kappa'p'}^{\al,\beta}}
   \oplus\!\bigoplus_{\beta}^{s-b-1}\!\!4\ketw{\R_{\kappa p,\kappa'p'}^{\al,\beta}}
     \oplus\!\!\!\!\bigoplus_{\beta}^{b+s-p'-1}
        \!\!\!\!4\ketw{\R_{\kappa p,(2\cdot\kappa')p'}^{\al,\beta}}\Big\}\nn
 &\oplus&\!\!\!\bigoplus_{\al}^{a-1}\!\Big\{
     \!\!\bigoplus_{\beta=|b-s|+1,\ \!{\rm by}\ \!2}^{p'-|p'-s-b|-1}
      \!\!\!\!2\ketw{\R_{\kappa p,(2\cdot\kappa')p'}^{\al,\beta}}
   \oplus\!\bigoplus_{\beta}^{s-b-1}\!\!4\ketw{\R_{\kappa p,(2\cdot\kappa')p'}^{\al,\beta}}
     \oplus\!\!\!\!\!\bigoplus_{\beta}^{b+s-p'-1}
        \!\!\!\!\!4\ketw{\R_{\kappa p,\kappa'p'}^{\al,\beta}}\Big\}  \nn
 \ketw{r,\kappa p'}\fus\ketw{\R_{p,\kappa'p'}^{a,b}}
  &=&\!\!\!\bigoplus_{\beta}^{p'-b-1}\!\!\!\Big\{
     \!\bigoplus_{\al=|a-r|+1,\ \!{\rm by}\ \!2}^{p-|p-r-a|-1}
      \!\!\!\!2\ketw{\R_{\kappa p,\kappa'p'}^{\al,\beta}}
  \oplus\!\!\bigoplus_{\al}^{r-a-1}\!\!4\ketw{\R_{\kappa p,\kappa'p'}^{\al,\beta}}
     \oplus\!\!\!\!\bigoplus_{\al}^{a+r-p-1}
        \!\!\!\!4\ketw{\R_{\kappa p,(2\cdot\kappa')p'}^{\al,\beta}}\Big\}\nn
 &\oplus&\!\!\!\bigoplus_{\beta}^{b-1}\!\Big\{
     \!\bigoplus_{\al=|a-r|+1,\ \!{\rm by}\ \!2}^{p-|p-r-a|-1}
      \!\!\!\!2\ketw{\R_{\kappa p,(2\cdot\kappa')p'}^{\al,\beta}}
  \oplus\!\!\bigoplus_{\al}^{r-a-1}\!\!\!4\ketw{\R_{\kappa p,(2\cdot\kappa')p'}^{\al,\beta}}
     \oplus\!\!\!\!\!\bigoplus_{\al}^{a+r-p-1}
        \!\!\!\!\!4\ketw{\R_{\kappa p,\kappa'p'}^{\al,\beta}}\Big\}
     \nn
\label{fus13}
\eea
The fusion of two ${\cal W}$-indecomposable rank-2 representations is given by
\bea
 \ketw{\R_{\kappa p,s}^{a,0}}\fus\ketw{\R_{\kappa'p,s'}^{a',0}}
  &=&\!\bigoplus_{j=|s-s'|+1,\ \!{\rm by}\ \!2}^{p'-|p'-s-s'|-1}
   \!\!\Big\{\bigoplus_{\al}^{p-|a-a'|-1}
     \!\!2\ketw{\R_{(\kappa\cdot\kappa')p,j}^{\al,0}}
     \oplus\!\bigoplus_{\al}^{|p-a-a'|-1}
     \!\!2\ketw{\R_{(\kappa\cdot\kappa')p,j}^{\al,0}}\nn
   &&\qquad\oplus\bigoplus_{\al}^{p-|p-a-a'|-1}
     \!\!\!2\ketw{\R_{(2\cdot\kappa\cdot\kappa')p,j}^{\al,0}}
     \oplus\bigoplus_{\al}^{|a-a'|-1}
     \!2\ketw{\R_{(2\cdot\kappa\cdot\kappa')p,j}^{\al,0}}\Big\}\nn
&\oplus&\bigoplus_{\beta}^{s+s'-p'-1}   \!\!\Big\{\bigoplus_{\al}^{p-|a-a'|-1}
     \!\!2\ketw{\R_{\kappa p,\kappa'p'}^{\al,\beta}}
     \oplus\bigoplus_{\al}^{|p-a-a'|-1}
     \!\!2\ketw{\R_{\kappa p,\kappa'p'}^{\al,\beta}}\nn
   &&\qquad\oplus\bigoplus_{\al}^{p-|p-a-a'|-1}
     \!\!\!2\ketw{\R_{\kappa p,(2\cdot\kappa')p'}^{\al,\beta}}
     \oplus\bigoplus_{\al}^{|a-a'|-1}
     \!2\ketw{\R_{\kappa p,(2\cdot\kappa')p'}^{\al,\beta}}\Big\}\nn
 \ketw{\R_{\kappa p,s}^{a,0}}\fus\ketw{\R_{r,\kappa'p'}^{0,b}}
  &=&
   \!\!\bigoplus_{\al=|a-r|+1,\ \!{\rm by}\ \!2}^{p-|p-r-a|-1}
   \!\Big\{\bigoplus_{\beta=|b-s|+1,\ \!{\rm by}\ \!2}^{p'-|p'-s-b|-1}
    \!\!\!\!\ketw{\R_{\kappa p,\kappa'p'}^{\al,\beta}}\Big\}
    \nn
 &\oplus&  
     \!\!\bigoplus_{\al=|a-r|+1,\ \!{\rm by}\ \!2}^{p-|p-r-a|-1}
   \!\Big\{  \bigoplus_{\beta}^{s-b-1}2\ketw{\R_{\kappa p,\kappa'p'}^{\al,\beta}}\Big\}
    \oplus\bigoplus_{\beta=|b-s|+1,\ \!{\rm by}\ \!2}^{p'-|p'-s-b|-1}
    \!\Big\{  \bigoplus_{\al}^{r-a-1}2\ketw{\R_{\kappa p,\kappa'p'}^{\al,\beta}}\Big\}
    \nn
  &\oplus&\!\bigoplus_{\al}^{r-a-1}\!\Big\{ 
     \bigoplus_{\beta}^{s-b-1}4\ketw{\R_{\kappa p,\kappa'p'}^{\al,\beta}}\Big\}
      \oplus\!\bigoplus_{\al}^{a+r-p-1}\!\Big\{
     \bigoplus_{\beta}^{b+s-p'-1}\!4\ketw{\R_{\kappa p,\kappa'p'}^{\al,\beta}}\Big\}
     \nn
 &\oplus&\!\bigoplus_{\al}^{a+r-p-1}\!\Big\{
   \bigoplus_{\beta=|b-s|+1,\ \!{\rm by}\ \!2}^{p'-|p'-s-b|-1}
     \!\!2\ketw{\R_{\kappa p,(2\cdot\kappa')p'}^{\al,\beta}}\oplus
      \bigoplus_{\beta}^{s-b-1}\!4\ketw{\R_{\kappa p,(2\cdot\kappa')p'}^{\al,\beta}}\Big\}
      \nn
 &\oplus&\!\bigoplus_{\beta}^{b+s-p'-1}\!\Big\{
  \bigoplus_{\al=|a-r|+1,\ \!{\rm by}\ \!2}^{p-|p-r-a|-1}
    \!\!2\ketw{\R_{\kappa p,(2\cdot\kappa')p'}^{\al,\beta}}\oplus
     \bigoplus_{\beta}^{r-a-1}\!4\ketw{\R_{\kappa p,(2\cdot\kappa')p'}^{\al,\beta}}\Big\}
     \nn
     \nn
 \ketw{\R_{r,\kappa p'}^{0,b}}\fus\ketw{\R_{r',\kappa'p'}^{0,b'}}
  &=&\!\!\bigoplus_{j=|r-r'|+1,\ \!{\rm by}\ \!2}^{p-|p-r-r'|-1}
   \!\!\Big\{\bigoplus_{\beta}^{p'-|b-b'|-1}
     \!\!2\ketw{\R_{j,(\kappa\cdot\kappa')p'}^{0,\beta}}
     \oplus\bigoplus_{\beta}^{|p'-b-b'|-1}
     \!\!2\ketw{\R_{j,(\kappa\cdot\kappa')p'}^{0,\beta}}\nn
   &&\qquad\oplus\bigoplus_{\beta}^{p'-|p'-b-b'|-1}
     \!\!\!2\ketw{\R_{j,(2\cdot\kappa\cdot\kappa')p'}^{0,\beta}}
     \oplus\bigoplus_{\beta}^{|b-b'|-1}
     \!2\ketw{\R_{j,(2\cdot\kappa\cdot\kappa')p'}^{0,\beta}}\Big\}\nn
&\oplus&\bigoplus_{\al}^{r+r'-p-1}   \!\Big\{\bigoplus_{\beta}^{p'-|b-b'|-1}
     \!\!2\ketw{\R_{\kappa p,\kappa'p'}^{\al,\beta}}
     \oplus\!\bigoplus_{\beta}^{|p'-b-b'|-1}
     \!\!2\ketw{\R_{\kappa p,\kappa'p'}^{\al,\beta}}\nn
   &&\qquad\oplus\!\bigoplus_{\beta}^{p'-|p'-b-b'|-1}
     \!\!\!2\ketw{\R_{\kappa p,(2\cdot\kappa')p'}^{\al,\beta}}
     \oplus\bigoplus_{\beta}^{|b-b'|-1}
     \!2\ketw{\R_{\kappa p,(2\cdot\kappa')p'}^{\al,\beta}}\Big\}  
\label{fus22}
\eea
The fusion of a ${\cal W}$-indecomposable rank-2 representation with a 
${\cal W}$-indecomposable rank-3 representation is given by
\bea
 \ketw{\R_{\kappa p,s}^{a,0}}\fus\ketw{\R_{p,\kappa'p'}^{a',b'}}
  &=&\!\!\bigoplus_{\beta=|b'-s|+1,\ \!{\rm by}\ \!2}^{p'-|p'-s-b'|-1}\!\!\Big\{
   \bigoplus_{\al}^{p-|a-a'|-1}\!\!\!2\ketw{\R_{\kappa p,\kappa'p'}^{\al,\beta}}
   \oplus\!\bigoplus_{\al}^{|p-a-a'|-1}\!\!\!2\ketw{\R_{\kappa p,\kappa'p'}^{\al,\beta}}\Big\}
  \nn
  &\oplus&\bigoplus_{\beta}^{s-b'-1}\!\Big\{
   \bigoplus_{\al}^{p-|a-a'|-1}\!\!\!4\ketw{\R_{\kappa p,\kappa'p'}^{\al,\beta}}
   \oplus\!\bigoplus_{\al}^{|p-a-a'|-1}\!\!\!4\ketw{\R_{\kappa p,\kappa'p'}^{\al,\beta}}\Big\}
  \nn
  &\oplus&\bigoplus_{\beta}^{b'+s-p'-1}\!\Big\{
   \bigoplus_{\al}^{p-|p-a-a'|-1}\!\!\!4\ketw{\R_{\kappa p,\kappa'p'}^{\al,\beta}}\oplus
   \bigoplus_{\al}^{|a-a'|-1}\!4\ketw{\R_{\kappa p,\kappa'p'}^{\al,\beta}}\Big\}
 \nn
 &\oplus&\!\!\bigoplus_{\beta=|b'-s|+1,\ \!{\rm by}\ \!2}^{p'-|p'-s-b'|-1}\!\!\Big\{
   \bigoplus_{\al}^{p-|p-a-a'|-1}\!\!\!2\ketw{\R_{\kappa p,(2\cdot\kappa')p'}^{\al,\beta}}
   \oplus\bigoplus_{\al}^{|a-a'|-1}\!2\ketw{\R_{\kappa p,(2\cdot\kappa')p'}^{\al,\beta}}\Big\}
 \nn
 &\oplus&\bigoplus_{\beta}^{s-b'-1}\!\Big\{
   \bigoplus_{\al}^{p-|p-a-a'|-1}\!\!\!4\ketw{\R_{\kappa p,(2\cdot\kappa')p'}^{\al,\beta}}\oplus
   \bigoplus_{\al}^{|a-a'|-1}\!4\ketw{\R_{\kappa p,(2\cdot\kappa')p'}^{\al,\beta}}\Big\}
 \nn
 &\oplus&\bigoplus_{\beta}^{b'+s-p'-1}\!\Big\{
   \bigoplus_{\al}^{p-|a-a'|-1}\!\!\!4\ketw{\R_{\kappa p,(2\cdot\kappa')p'}^{\al,\beta}}
   \oplus\!\bigoplus_{\al}^{|p-a-a'|-1}\!\!\!4\ketw{\R_{\kappa p,(2\cdot\kappa')p'}^{\al,\beta}}\Big\} 
 \nn 
 \ketw{\R_{r,\kappa p'}^{0,b}}\fus\ketw{\R_{p,\kappa'p'}^{a',b'}}
  &=&\!\!\bigoplus_{\al=|a'-r|+1,\ \!{\rm by}\ \!2}^{p-|p-r-a'|-1}\!\!\Big\{
   \bigoplus_{\beta}^{p'-|b-b'|-1}\!\!\!2\ketw{\R_{\kappa p,\kappa'p'}^{\al,\beta}}
   \oplus\!\bigoplus_{\beta}^{|p'-b-b'|-1}\!\!\!2\ketw{\R_{\kappa p,\kappa'p'}^{\al,\beta}}\Big\}
  \nn
  &\oplus&\bigoplus_{\al}^{r-a'-1}\!\Big\{
   \bigoplus_{\beta}^{p'-|b-b'|-1}\!\!\!4\ketw{\R_{\kappa p,\kappa'p'}^{\al,\beta}}
   \oplus\!\bigoplus_{\beta}^{|p'-b-b'|-1}\!\!\!4\ketw{\R_{\kappa p,\kappa'p'}^{\al,\beta}}\Big\}
  \nn
  &\oplus&\bigoplus_{\al}^{a'+r-p-1}\!\Big\{
   \bigoplus_{\beta}^{p'-|p'-b-b'|-1}\!\!\!4\ketw{\R_{\kappa p,\kappa'p'}^{\al,\beta}}\oplus
   \bigoplus_{\beta}^{|b-b'|-1}\!4\ketw{\R_{\kappa p,\kappa'p'}^{\al,\beta}}\Big\}
 \nn
 &\oplus&\!\!\bigoplus_{\al=|a'-r|+1,\ \!{\rm by}\ \!2}^{p-|p-r-a'|-1}\!\!\Big\{
   \bigoplus_{\beta}^{p'-|p'-b-b'|-1}\!\!\!2\ketw{\R_{\kappa p,(2\cdot\kappa')p'}^{\al,\beta}}
   \oplus\bigoplus_{\beta}^{|b-b'|-1}\!2\ketw{\R_{\kappa p,(2\cdot\kappa')p'}^{\al,\beta}}\Big\}
 \nn
 &\oplus&\bigoplus_{\al}^{r-a'-1}\!\Big\{
   \bigoplus_{\beta}^{p'-|p'-b-b'|-1}\!\!\!4\ketw{\R_{\kappa p,(2\cdot\kappa')p'}^{\al,\beta}}\oplus
   \bigoplus_{\beta}^{|b-b'|-1}\!4\ketw{\R_{\kappa p,(2\cdot\kappa')p'}^{\al,\beta}}\Big\}
 \nn
 &\oplus&\bigoplus_{\al}^{a'+r-p-1}\!\Big\{
   \bigoplus_{\beta}^{p'-|b-b'|-1}\!\!\!4\ketw{\R_{\kappa p,(2\cdot\kappa')p'}^{\al,\beta}}
   \oplus\!\bigoplus_{\beta}^{|p'-b-b'|-1}\!\!\!4\ketw{\R_{\kappa p,(2\cdot\kappa')p'}^{\al,\beta}}\Big\}  
\label{fus23}
\eea
Finally, the fusion of two ${\cal W}$-indecomposable rank-3 representations is given by
\bea
 \ketw{\R_{\kappa p,p'}^{a,b}}\fus\ketw{\R_{p,\kappa'p'}^{a',b'}}
 &=&\!\bigoplus_{\al}^{p-|a-a'|-1}\!\!\Big\{
  \bigoplus_{\beta}^{p'-|b-b'|-1}\!\!\!4\ketw{\R_{\kappa p,\kappa'p'}^{\al,\beta}}\Big\}
   \oplus\!\!\bigoplus_{\al}^{|p-a-a'|-1}\!\!\Big\{
  \bigoplus_{\beta}^{|p'-b-b'|-1}\!\!\!4\ketw{\R_{\kappa p,\kappa'p'}^{\al,\beta}}\Big\}
 \nn
 &\oplus&\!\bigoplus_{\al}^{p-|a-a'|-1}\!\!\Big\{
  \bigoplus_{\beta}^{|p'-b-b'|-1}\!\!\!4\ketw{\R_{\kappa p,\kappa'p'}^{\al,\beta}}\Big\}
   \oplus\!\!\bigoplus_{\al}^{|p-a-a'|-1}\!\!\Big\{
  \bigoplus_{\beta}^{p'-|b-b'|-1}\!\!\!4\ketw{\R_{\kappa p,\kappa'p'}^{\al,\beta}}\Big\}
 \nn
 &\oplus&\!\bigoplus_{\al}^{p-|p-a-a'|-1}\!\!\Big\{
  \bigoplus_{\beta}^{p'-|p'-b-b'|-1}\!\!\!4\ketw{\R_{\kappa p,\kappa'p'}^{\al,\beta}}\Big\}
   \oplus\!\!\bigoplus_{\al}^{|a-a'|-1}\!\!\Big\{
  \bigoplus_{\beta}^{|b-b'|-1}\!\!\!4\ketw{\R_{\kappa p,\kappa'p'}^{\al,\beta}}\Big\}
 \nn
 &\oplus&\!\bigoplus_{\al}^{p-|p-a-a'|-1}\!\!\Big\{
  \bigoplus_{\beta}^{|b-b'|-1}\!\!\!4\ketw{\R_{\kappa p,\kappa'p'}^{\al,\beta}}\Big\}
   \oplus\!\!\bigoplus_{\al}^{|a-a'|-1}\!\!\Big\{
  \bigoplus_{\beta}^{p'-|p'-b-b'|-1}\!\!\!4\ketw{\R_{\kappa p,\kappa'p'}^{\al,\beta}}\Big\}
 \nn
 &\oplus&\!\bigoplus_{\al}^{p-|a-a'|-1}\!\!\Big\{
  \bigoplus_{\beta}^{p'-|p'-b-b'|-1}\!\!\!4\ketw{\R_{\kappa p,(2\cdot\kappa')p'}^{\al,\beta}}
  \oplus\!\!\bigoplus_{\beta}^{|b-b'|-1}\!\!\!4\ketw{\R_{\kappa p,(2\cdot\kappa')p'}^{\al,\beta}}\Big\}
 \nn
 &\oplus&\!\bigoplus_{\al}^{|p-a-a'|-1}\!\!\Big\{
    \bigoplus_{\beta}^{p'-|p'-b-b'|-1}\!\!\!4\ketw{\R_{\kappa p,(2\cdot\kappa')p'}^{\al,\beta}}
  \oplus\!\!\bigoplus_{\beta}^{|b-b'|-1}\!\!\!4\ketw{\R_{\kappa p,(2\cdot\kappa')p'}^{\al,\beta}}\Big\}
 \nn
 &\oplus&\!\bigoplus_{\beta}^{p'-|b-b'|-1}\!\!\Big\{
  \bigoplus_{\al}^{p-|p-a-a'|-1}\!\!\!4\ketw{\R_{\kappa p,(2\cdot\kappa')p'}^{\al,\beta}}
  \oplus\!\!\bigoplus_{\al}^{|a-a'|-1}\!\!\!4\ketw{\R_{\kappa p,(2\cdot\kappa')p'}^{\al,\beta}}\Big\}
 \nn
 &\oplus&\!\bigoplus_{\beta}^{|p'-b-b'|-1}\!\!\Big\{
    \bigoplus_{\al}^{p-|p-a-a'|-1}\!\!\!4\ketw{\R_{\kappa p,(2\cdot\kappa')p'}^{\al,\beta}}
  \oplus\!\!\bigoplus_{\al}^{|a-a'|-1}\!\!\!4\ketw{\R_{\kappa p,(2\cdot\kappa')p'}^{\al,\beta}}\Big\}
\label{fus33}
\eea



\begin{thebibliography}{99}


\bib{PRZ0607} P.A.~Pearce, J.~Rasmussen, J.-B.~Zuber, 
 {\em Logarithmic minimal models}, 
 J. Stat. Mech. (2006) P11017,
 arXiv:hep-th/0607232.

\bib{RP0706} J.~Rasmussen, P.A.~Pearce, 
 {\em Fusion algebra of critical percolation}, J. Stat. Mech. P09002 (2007), 
 arXiv:0706.2716 [hep-th].

\bib{RP0707} J.~Rasmussen, P.A.~Pearce, 
{\em Fusion algebras of logarithmic minimal models},
 J. Phys. {\bf A40} (2007) 13711--13733,
 arXiv:0707.3189 [hep-th].

\bib{RP0709} J.~Rasmussen, P.A.~Pearce, 
 {\em Polynomial fusion rings of logarithmic minimal models}, 
 J. Phys. A: Math. Theor. {\bf 41} (2008) 175210,
 arXiv:0709.3337 [hep-th].

\bib{Gep91} D.~Gepner,
 {\em Fusion rings and geometry}, 
 Commun. Math. Phys. {\bf 141} (1991) 381--411.

\bib{WLogCFT} M.~Flohr,
{\em On modular invariant partition functions of conformal field theories with 
 logarithmic operators}, 
 Int. J. Mod. Phys. {\bf A11} (1996) 4147--4172,  
 arXiv:hep-th/9509166;\\
%
 M.~Flohr, M.R.~Gaberdiel, 
 {\em Logarithmic torus amplitudes}, 
 J. Phys. {\bf A39} (2006) 1955--1968,
 arXiv:hep-th/0509075;\\
%
 M.R.~Gaberdiel, I.~Runkel,
 {\em The logarithmic triplet theory with boundary},
 J. Phys. {\bf A39} (2006) 14745--14780,
 arXiv:hep-th/0608184.

\bib{GK9606} M.R.~Gaberdiel, H.G.~Kausch, 
 {\em A rational logarithmic conformal field theory},
 Phys. Lett. {\bf B386} (1996) 131--137,
 arXiv:hep-th/9606050.
 
\bib{Walgebra} A.B.~Zamolodchikov, 
 {\em Infinite additional symmetries in two-dimensional
 conformal quantum field theory},
 Theor. Math. Phys. {\bf 65} (1985) 1205--1213;\\
%
 P.~Bouwknegt, K.~Schoutens, 
 {\em W-symmetry in conformal field theory},
 Phys. Rept. {\bf 223} (1993) 183--276,
 arXiv:hep-th/9210010.

\bibitem{LogCFT} V.~Gurarie, 
 {\em Logarithmic operators in conformal field theory},
 Nucl. Phys. {\bf B410} (1993) 535--549,
 arXiv:hep-th/9303160;\\
%
 M.~Flohr, 
 {\em Bits and pieces in logarithmic conformal field theory}, 
 Int. J. Mod. Phys. {\bf A18} (2003) 4497--4592,
 arXiv:hep-th/0111228;\\
%
 M.R.~Gaberdiel, 
 {\em An algebraic approach to logarithmic conformal field theory},
 Int. J. Mod. Phys. {\bf A18} (2003) 4593--4638,
 arXiv:hep-th/0111260;\\
%
 S.~Kawai, 
 {\em Logarithmic conformal field theory with boundary}, 
 Int. J. Mod. Phys. {\bf A18} (2003) 4655--4684,
 arXiv:hep-th/0204169.

\bib{WLM1p} J.~Fuchs, S.~Hwang, A.M.~Semikhatov, I.Yu.~Tipunin, 
 {\em Nonsemisimple fusion algebras and the Verlinde formula}, 
 Commun. Math. Phys. {\bf 247} (2004) 713--742,
 arXiv:hep-th/0306274;\\
%
 B.~Feigin, A.M.~Gainutdinov, A.M.~Semikhatov, I.Yu.~Tipunin, 
 {\em Modular group representations and fusion in logarithmic conformal field theories 
 and in the quantum group center}, 
 Commun. Math. Phys. {\bf 265} (2006) 47--93,
 arXiv:hep-th/0504093;\\
%
 M.R.~Gaberdiel, I.~Runkel, 
 {\em From boundary to bulk in logarithmic CFT},
 J. Phys. {\bf A41} (2008) 075402,
 arXiv:0707.0388 [hep-th];\\
%
 A.M.~Gainutdinov, I. Yu.~Tipunin,
 {\em Radford, Drinfeld and Cardy boundary states in $(1,p)$ logarithmic conformal field models},
 arXiv:0711.3430 [hep-th].

\bib{PRR0803} P.A.~Pearce, J.~Rasmussen, P.~Ruelle, 
 {\em Integrable boundary conditions and ${\cal W}$-extended fusion in the logarithmic minimal
 models ${\cal LM}(1,p)$}, 
 J. Phys. A: Math. Theor. {\bf 41} (2008) 295201,
 arXiv:0803.0785 [hep-th].

\bib{FGST0606a} B.L.~Feigin, A.M.~Gainutdinov, A.M.~Semikhatov, I.Yu.~Tipunin, 
 {\em Kazhdan-Lusztig dual quantum group for logarithmic extensions of Virasoro minimal models}, 
 J. Math. Phys. {\bf 48} (2007) 032303,
 arXiv:math.QA/0606506.

\bib{FGST0606b} B.L.~Feigin, A.M.~Gainutdinov, A.M.~Semikhatov, I.Yu.~Tipunin, 
 {\em Logarithmic extensions of minimal models: characters and modular transformations},
 Nucl. Phys. {\bf B757} (2006) 303--343,
 arXiv:hep-th/0606196.

\bib{RP0804} J.~Rasmussen, P.A.~Pearce, 
 {\em ${\cal W}$-extended fusion algebra of critical percolation},
 J. Phys. A: Math. Theor. {\bf 41} (2008) 295208,
 arXiv:0804.4335 [hep-th].

\bib{Ras0805} J.~Rasmussen, 
 {\em ${\cal W}$-extended logarithmic minimal models}, 
 Nucl. Phys. {\bf B807} (2009) 495--533, 
 arXiv:0805.2991 [hep-th].

\bib{Symplectic} H.G.~Kausch, 
 {\em Curiosities at $c = -2$}, arXiv:hep-th/9510149;
 {\em Symplectic fermions}, 
 Nucl. Phys. {\bf B583} (2000) 513--541,
 arXiv:hep-th/0003029.

\bib{Polymers} 
 P.G.~de~Gennes, 
 {\em Scaling concepts in polymer physics}, Cornell University, Ithaca (1979);\\
 %
 J.~des~Cloizeaux, G.~Jannink, 
 {\em Polymers in solution: their modelling and structure}, 
 Clarendon Press (1990);\\
 %
 H.~Saleur, 
 {\em New exact exponents for the two-dimensional self-avoiding walks}, 
 J. Phys. {\bf A19} (1986) L807--L810;
 {\em Magnetic properties of the two-dimensional $n=0$ vector model}, 
 Phys. Rev. {\bf B35} (1987) 3657--3660;\\
 %
 B.~Duplantier, 
 {\em Exact critical exponents for two-dimensional dense polymers},
 J. Phys. {\bf A19} (1986) L1009--L1014;\\
 %
 P.A.~Pearce, J.~Rasmussen, 
 {\em Solvable critical dense polymers}, 
 J. Stat. Mech. P02015 (2007),
 arXiv:hep-th/0610273.

\bib{Penrose} R.~Penrose, 
 {\em A generalized inverse for matrices}, 
 Proc. Cambridge Philos. Soc. {\bf 51} (1955) 406--413;\\
 %
 A.~Ben-Israel, T.N.E.~Greville, 
 {\em Generalized inverses: theory and applications},
 Springer (2003).

\bib{PR0812} P.A.~Pearce, J.~Rasmussen, 
 {\em Verlinde formulas for logarithmic minimal models},
 in preparation.

\bib{BPZ84} A.A.~Belavin, A.M.~Polyakov, A.B.~Zamolodchikov, 
 {\em Infinite conformal symmetry in two-dimensional quantum field theory},
 Nucl. Phys. {\bf B214} (1984) 333--380.

\bib{DiFMS} P.~Di~Francesco, P.~Mathieu, D.~S\'en\'echal, 
 {\em Conformal field theory}, Springer (1996).

\bib{RS0701} N. Read, H. Saleur, {\em Enlarged symmetry algebras of spin chains, 
 loop models,  and S-matrices}, Nucl. Phys. {\bf B777} (2007) 263--315,
 arXiv:cond-mat/0701259;
%
{\em Associative-algebraic approach to logarithmic 
 conformal field theories}, Nucl. Phys. {\bf B777} (2007) 316--351,
 arXiv:hep-th/0701117.
 
\bib{FSZ87} P.~Di~Francesco, H.~Saleur, J.-B.~Zuber, {\em Modular invariance in nonminimal
 two-dimensional conformal theories},
 Nucl. Phys. {\bf B285} (1987) 454--480.

\bib{Embedding} M.R.~Gaberdiel, H.G.~Kausch, 
 {\em Indecomposable fusion products}, 
 Nucl. Phys. {\bf B477} (1996) 293--318,
 arXiv:hep-th/9604026;\\
%
 H.~Eberle, M.~Flohr, 
 {\em Virasoro representations and fusion for general augmented minimal models}, 
 J. Phys. {\bf A39} (2006) 15245--15286,
 arXiv:hep-th/0604097. 

\bib{FusionPotential} P.~Di~Francesco, J.-B.~Zuber, 
 {\em Fusion potentials I}, 
 J. Phys. A: Math. Gen. {\bf 26} (1993) 1441--1454, 
 arXiv:hep-th/9211138;\\
%
 O.~Aharony, 
 {\em Generalized fusion potentials}, 
 Phys. Lett. {\bf B306} (1993) 276--282, 
 arXiv:hep-th/9301118.


\end{thebibliography}
\end{document}